\documentclass[fleqn,usenatbib,useAMS]{mnras}

\usepackage{graphicx}	
\usepackage{amsmath}	
\usepackage{multicol}        
\usepackage{bm}		

\usepackage{hyperref}
\usepackage{appendix}
\usepackage{lscape}
\usepackage{ulem}
\usepackage[detect-none]{siunitx}
\usepackage{tabularx}
\usepackage{lineno}
\usepackage[usestackEOL]{stackengine}

\usepackage[T1]{fontenc}
\usepackage{ae,aecompl}

\usepackage{newtxtext,newtxmath}

\newcommand{\emerlin}{\textit{e}{-MERLIN}}
\newcommand{\microJybm}{\,\mathrm{\umu Jy\,beam^{-1}}}
\newcommand{\microJy}{\,\mathrm{\umu Jy}}

\newcommand{\plus}{$\pm$}

\title[Quasars in the radio-quiet regime]{The Quasar Feedback Survey: Revealing the importance of sensitive radio imaging for AGN identification deeper into the radio-quiet regime}

\author[A. Njeri et al.]{Ann Njeri,$^{1}$\thanks{E-mail: ann.njeri@newcastle.ac.uk}
Chris M. Harrison,$^{1}$\thanks{E-mail: christopher.harrison@newcastle.ac.uk}
Preeti Kharb,$^{2}$
David M. Alexander,$^{3}$
Vincenzo Mainieri,${^4}$
Chiara Circosta,$^{5}$
\newauthor
Victoria A. Fawcett,$^{1}$
Darshan Kakkad,${^6}$
Dipanjan Mukherjee,$^7$
Stephen Molyneux$^{8,9}$
and Silpa Sasikumar${^{10}}$
\\
$^{1}$School of Mathematics, Statistics and Physics, Newcastle University, Newcastle upon Tyne, NE1 7RU, UK\\
$^{2}$National Centre for Radio Astrophysics (NCRA) - Tata Institute of Fundamental Research (TIFR), S. P. Pune University Campus, Post Bag 3, Ganeshkhind,\\ 
411007 Pune, India\\
$^{3}$Centre for Extragalactic Astronomy, Department of Physics, Durham University, South Road, Durham, DH1 3LE, UK \\
$^{4}$European Southern Observatory (ESO), Karl-Schwarzschild-Stra{\ss}e 2, 85748 Garching bei M\"{u}nchen, Germany\\  
$^{5}$Institut de Radioastronomie Millimétrique (IRAM), 300 rue de la Piscine, 38400 Saint-Martin-d’Hères, France\\
$^{6}$Centre for Astrophysics Research, Department of Physics, Astronomy and Mathematics, University of Hertfordshire, Hatfield, AL10 9AB, UK\\
$^{7}$Inter-University Centre for Astronomy and Astrophysics, Post Bag - 4, Pune University, Ganeshkhind, Pune - 411007, India \\
$^{8}$School of Physics and Astronomy, University of Southampton, Southampton, SO17 1BJ, UK\\
$^{9}$Institute of Theoretical Astrophysics, University of Oslo, P.O. Box 1029, Blindern, 0315 Oslo, Norway\\
$^{10}$Departamento de Astronom\`{i}a, Universidad de Concepci\'{o}n, Concepci\'{o}n, Chile\\
}

\date{Accepted 2026 January 9. Received 2025 January 7; in original form 2025 October 6}

\pubyear{2025}

\begin{document}
\label{firstpage}
\pagerange{\pageref{firstpage}--\pageref{lastpage}}
\maketitle

\begin{abstract}

We present new sub-arcsecond ($\sim$0.3-1 arcsec; $\sim$1--3\,kpc) VLA imaging at 1.4\,GHz and 6\,GHz of 29 optically-selected, [O~{\sc iii}] luminous ($L_{\rm [O III]}$ > 10$^{42.1}$\,erg\,s$^{-1}$), $z<0.2$ quasars drawn from the expanded Quasar Feedback Survey (QFeedS; with $L_\mathrm{1.4\,GHz} = 10^{22.6}$--10$^{26.3}$\,W\,Hz$^{-1}$). These 29 new objects occupy the low end of the radio-power distribution ($L_\mathrm{1.4\,GHz}$=$10^{22.63}$--10$^{23.45}$\,W\,Hz$^{-1}$) in the QFeedS sample and are nominally `radio quiet'. Despite this, we find widespread evidence of AGN-driven synchrotron activity. Nearly $\sim 31\,$per\,cent exhibit resolved radio structures on $\sim$0.1--20\,kpc scales consistent with compact jets or wind-driven outflows, and $\sim 90\,$per\,cent display steep spectra ($\alpha \lesssim -1$) indicative of optically thin synchrotron emission. Combining morphology, spectral index and brightness-temperature diagnostics, at least $\sim38\,$per\,cent of the sample show clear AGN signatures that cannot be explained by star formation alone. These constitute the first results from the expanded QFeedS (now 71 quasars spanning $\approx 4$ dex in radio power) and demonstrate that compact, low-power jets and AGN shocks are common deep inside the radio-quiet regime. A thorough understanding of feedback processes from quasars, deep into the `radio-quiet' regime, will be obtained by connecting these high resolution radio observations with multi-wavelength observations.

\end{abstract}




\section{Introduction}
The radio-quiet quasar population is the predominant sample in optically-selected quasar samples, such as those from the Sloan Digital Sky Survey (SDSS; \citeauthor{Kellerman2016} \citeyear{Kellerman2016}). Despite being among the most luminous optical sources, radio-quiet quasars remain challenging to study in the radio due to their intrinsically weak emission, compact morphology, and ambiguous emission mechanisms \citep[e.g.][]{Padovani2016,Panessa2019}. These luminous active galactic nuclei (AGN) remain largely unresolved at the spatial resolution of all sky radio surveys such as FIRST and NVSS\footnote{These are the Faint Images of the Radio Sky at Twenty-cm survey (FIRST) and the NRAO VLA Sky Survey (NVSS) \citep[][]{Becker1995,Condon1998}}\citep[e.g.][]{Tadhunter2016,Kellerman2016,McCaffrey2022}, meaning that the radio emission is typically confined within the host galaxies, i.e., on scales smaller than a few kiloparsecs. However, extended radio structures with an AGN origin may exist on smaller scales if produced by jets or shocks due to AGN-driven outflows, that are young and/or frustrated by a dense interstellar medium \citep[e.g.][]{Panessa2019,Klindt2019,Fawcett2020}.

Despite the lack of powerful radio jets, it is becoming increasingly clear that studying radio emission in detail is very important for understanding the feedback processes from radio-quiet quasars. For example, hydrodynamic simulations have demonstrated that low-power radio jets, characteristic of radio-quiet AGN, can still exert significant influence on their host galaxies. This is particularly through their interactions with the interstellar medium (ISM) when the jets are closely inclined relative to the galactic disk and can drive strong shocks and accelerate the gas (e.g., \citealt{Meenakshi2022,Tanner2022,Mukherjee2025}). Importantly, jets do not need to escape the host galaxy to affect star formation. Even when confined within the inner kiloparsecs, these jets can cause turbulence, heat cold gas, and regulate the local gas reservoir. These theoretical insights are particularly relevant for compact low-luminosity radio sources, whose sub-kpc sizes and steep spectral indices are consistent with jet-ISM interactions occurring at $<2\,$kpc scales, potentially in early or frustrated phases of feedback \citep[e.g.][]{Chilufya2024,Njeri2025}.
Furthermore, quasar-driven winds could cause similar signatures in the extended radio emission, since they propagate as multi-phase outflows through the ISM, shocking the material to produce synchrotron emission (e.g., \citealt{Zakamska2016,Nims2015,Meenaksh2024}). While these processes can mimic starburst-driven outflows in the optical line kinematics, careful radio morphological and spectral analysis, such as identifying steep-spectrum, compact structures, and high brightness temperatures, can distinguish AGN from star formation origins \citep[e.g.][]{Jarvis2021,Morabito2022,Njeri2025}. 

A key open question in galaxy evolution is the contribution to feedback from low-power and compact radio jets and quasar-driven winds in radio-quiet quasars. While classical radio-loud quasars with powerful jets are well established as efficient regulators of gas cooling for massive galaxies \citep[e.g.][]{HeckmanBest2014,Fabian2012,Harrison2017}, the impact of faint or compact structures such as frustrated jets and disk winds, remains less clear \citep[e.g.][]{Laor2008,Panessa2019,Baldi2023}. Recent studies suggest that even in radio-quiet quasars, small-scale jets and outflows can efficiently couple to the ISM on small scales, driving turbulence and modest outflows \citep[e.g.][]{Venturi2021,Jarvis2019,Kharb2023,Cresci2023,Morganti2023,Audibert2023,Ulivi2024,Mukherjee2025}. Determining whether such processes represent a significant feedback mode, and how efficiently they regulate star formation, is crucial for building a complete picture of AGN-galaxy co-evolution. 

A related challenge, is how can we robustly separate star formation from AGN-driven radio emission, and establishing what this reveals about the spectrum of feedback modes in quasars.  This complicates efforts to assess the star-formation rates at quasar luminosities and raises questions about how feedback scales across the radio luminosity function. The emerging view is that quasars span a continuum of radio modes, from weak, AGN-contaminated synchrotron at low luminosities to powerful jet-dominated systems at high luminosities \citep[e.g.][]{Zakamska2014,Sabater2019,Jarvis2021,Girdhar2022,Slob2022}. This spectrum challenges the traditional binary classification of `radio-loud' versus `radio-quiet' AGN and motivates a more continuum between accretion states, production of radio emission, and efficiency of feedback \citep[e.g.][]{Padovani2017,Hardcastle2020,Mingo2022}. 

Towards addressing the outstanding issues outlined above, we have been undertaking a systematic survey of low redshift Type 1 and Type 2 ($z<0.2$) quasars, called the ``Quasar Feedback Survey'' (QFeedS; \citealt{Jarvis2021})\footnote{A survey website, including results, information on targets and open access to our data products/images can be found here: (\url{https://blogs.ncl.ac.uk/quasarfeedbacksurvey/})}. The primary objectives of this survey include: (1) understanding the origin of radio emission in sources with moderate-to-low radio luminosities; (2) characterizing the properties of multi-phase outflows in the quasar host galaxies; and, (3) assessing the potential long-term impact of the central quasar on the host galaxies. 

In this work we introduce our first results on the radio properties from a new sample, exploring deeper into the radio-quiet regime ($L_\mathrm{1.4\,GHz}<10^{23.4}$\,W\,Hz$^{-1}$) than our earlier work ($L_\mathrm{1.4\,GHz}>10^{23.4}$\,W\,Hz$^{-1}$). In Section~\ref{sec:QFeedS} we introduce the Quasar Feedback Survey (QFeedS) and describe the new sample for this current work (QFeedS-2). In Section~\ref{sec:VLA_obs} we describe the observations and imaging methods for the new radio data, before discussing the results in Section~\ref{sec:results}. In Section~\ref{sec:Discussions} we discuss our new results in the context of the wider QFeedS sample and other AGN populations and present our main conclusions in Section~\ref{sec:conclusion}. 
We adopt $H_{0}$ = 70\,km\,s$^{-1}$\,Mpc$^{-1}$, $\Omega_{M}$ = 0.3, $\Omega_{\Lambda}$ = 0.7
throughout. In this cosmology, 300\,mas (i.e., our typical resolution at $\sim 6.0\,$GHz), corresponds to $\sim 0.8$\,kpc for the median redshift of the sample ($z$=0.15). We define the radio spectral index, $\nu$, using $S_{\nu}\propto\nu^{\alpha}$.

\section{The Quasar Feedback Survey}\label{sec:QFeedS}
In this section we present an overview of QFeedS and  the motivation for the new sample (Section~\ref{sec:survey_description}), before describing in more detail the sample used in this work (Section~\ref{sec:sample_selection} and Section~\ref{sec:sample_overview}).

\subsection{Survey description}\label{sec:survey_description}
The initial sample of 42 quasars (QFeedS-1) was fully introduced in \cite{Jarvis2021}, although this followed multiple pilot studies (e.g., \citealt{Harrison2014,Harrison2015,Lansbury2018,Jarvis2019}). The goal of QFeedS is to build up a multi-wavelength picture of feedback from quasars; e.g., covering radio imaging, and multi-phase outflow measurements. Quasars for QFeedS are selected to be low redshift ($z<0.2$) luminous [O~{\sc iii}] emitters ($L_{\rm[O III]}\gtrsim10^{42}$\,erg\,s$^{-1}$) from the AGN sample presented in \cite{Mullaney2013}, who used spectroscopy from the SDSS, Data Release 7 \citep[][]{SDSS72009}. This redshift and luminosity selection criteria enables high spatial resolution observations (reaching sub-kiloparsec resolution) of sources that have quasar-like AGN luminosities ($L_{\rm AGN}\gtrsim10^{45}$\,erg\,s$^{-1}$). These luminosities are equivalent to the knee of the luminosity function around cosmic noon (see \citealt{Jarvis2020,Jarvis2021}). Therefore, this survey is complementary to surveys of more nearby radiatively luminous AGN, which cover typically lower bolometric powers or smaller samples \citep[e.g.,][]{Riffel2017,Rose2018,Davies2020,Venturi2021,GarciaBurillo2021,Fluetsch2021}, or those that focus on radio galaxies \citep[e.g.,][]{Morganti2005,Santoro2018,Speranza2021}. It also complements those that focus on higher redshifts, which reach higher luminosities but with more limited sensitivity and spatial resolution \citep[e.g.,][]{Nesvadba2017,Circosta2018,Roy2025,Ilha2025}. 

Similar surveys at the redshift range of QFeedS, also cover multi-wavelength approaches to understanding AGN feedback in quasars \citep[e.g.,][]{Karouzos2016,RamosAlmeida2022,Bessiere2024}. However, unlike most other surveys of radiatively luminous AGN, a key part of QFeedS has been to start by obtaining complete coverage with high spatial resolution radio imaging. Specifically, we have focused on the C-band (4--8\,GHz) and L-band (1--2\,GHz) from both the Very Large Array (VLA; \citealt{Jarvis2019,Jarvis2021}) and \emerlin{} (\citealt{Njeri2025}). Combining these new observations, with data from NVSS and FIRST, this has enabled us to create multi-frequency radio images of these quasars, which are sensitive to structures on scales of $\sim 0.05$\,arcseconds (i.e., $\sim$0.1\,kpc) to several arcseconds (i.e., 10s\,kpc).

\begin{figure}
\centering
\includegraphics[width=\columnwidth]{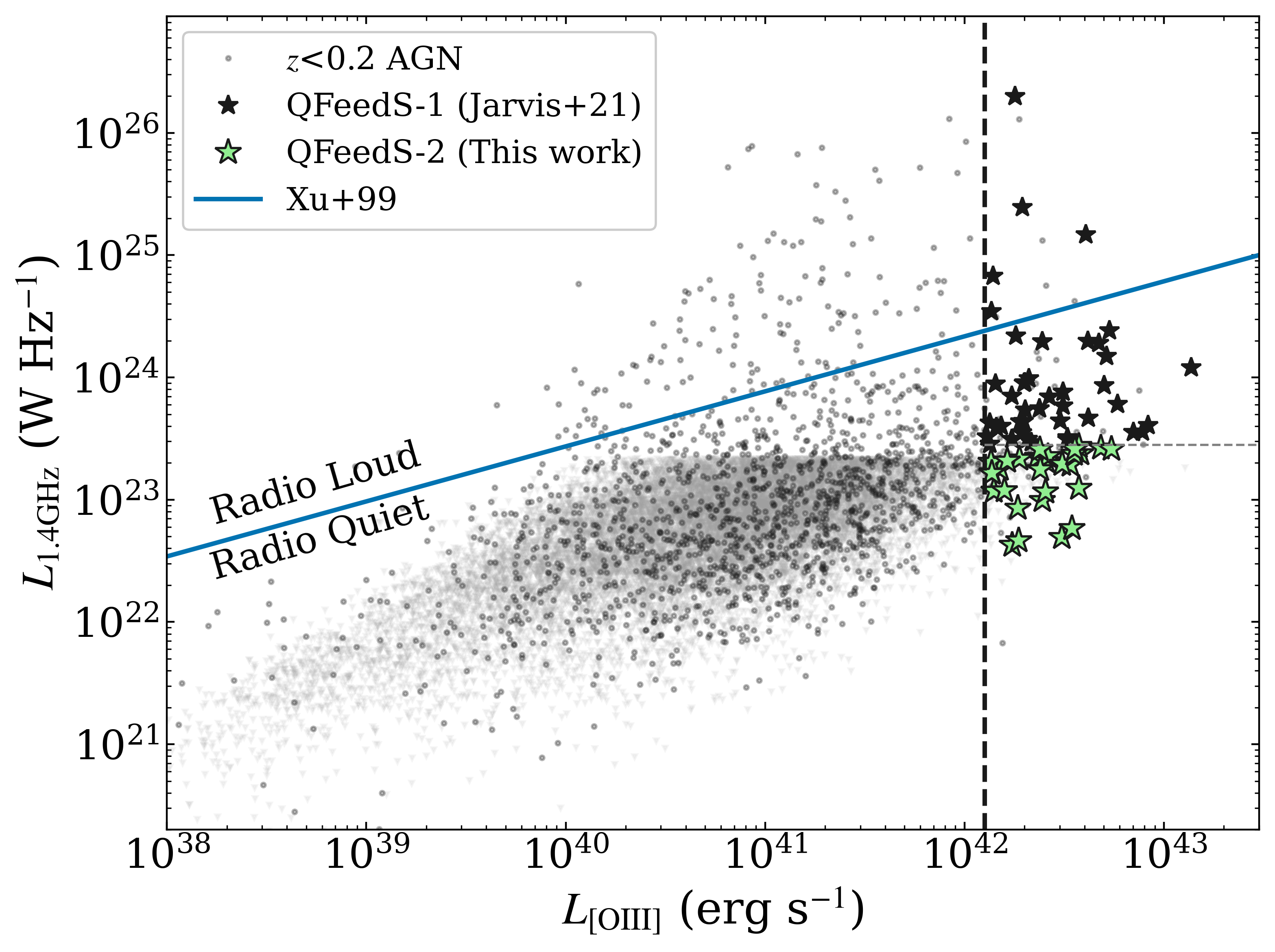}
\includegraphics[width=\columnwidth]{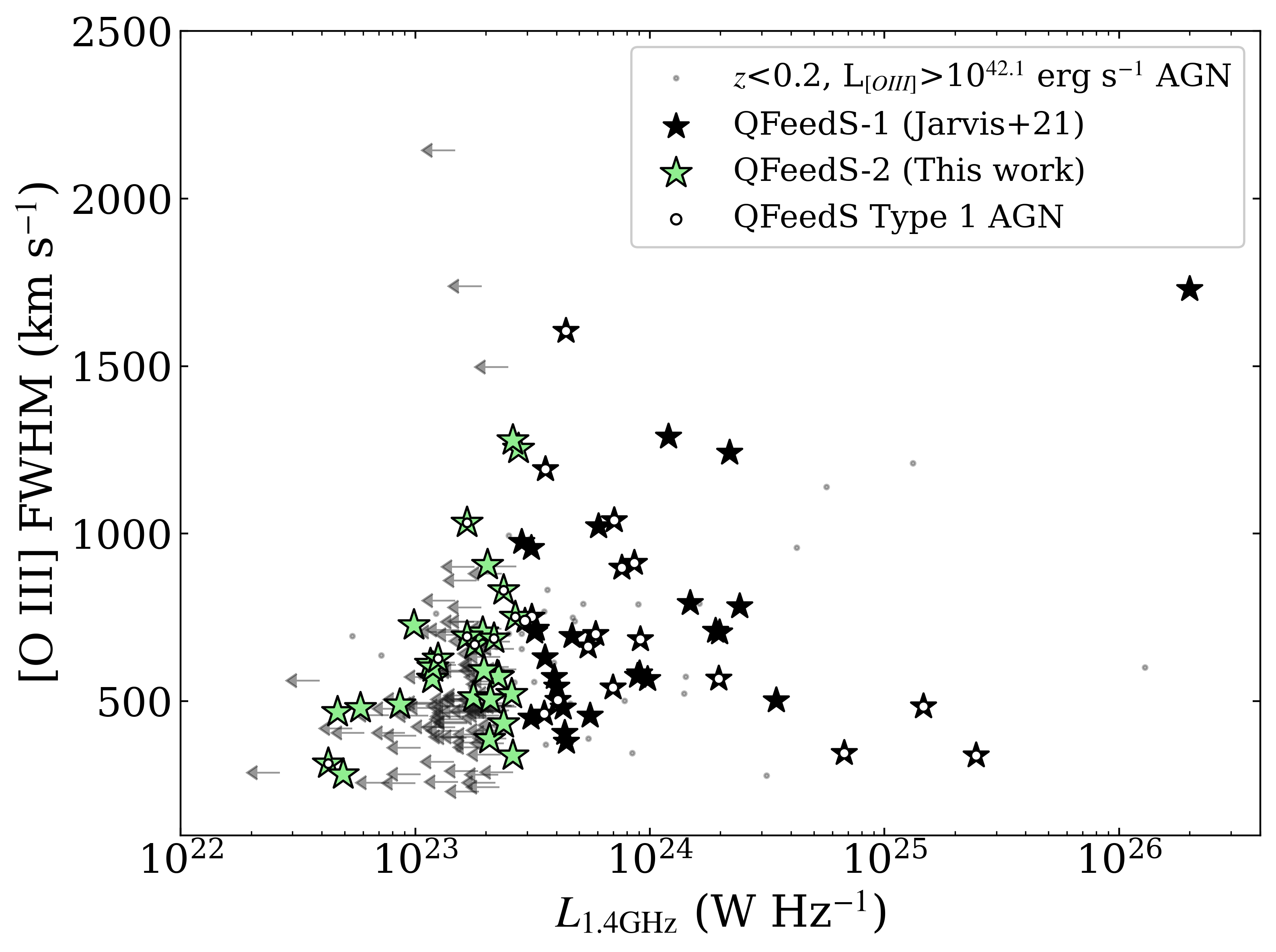}
\caption{{\textit Top:} 1.4\,GHz radio luminosity versus [O~{\sc III}] luminosity for the parent sample of $z<0.2$ AGN (grey circles, with radio upper limits represented as lighter grey triangles; \citealt{Mullaney2013}). The dashed vertical line shows our quasar luminosity threshold for selection ($L_{\rm [O III]}>10^{42.11}$\,erg\,s$^{-1}$). {\textit Bottom:} [O~{\sc iii}] FWHM versus 1.4\,GHz radio luminosity for the parent sample above the [O~{\sc III}] luminosity cut (grey dots, with upper limits represented as arrows). In both panels the initial 42 QFeedS targets are represented with black stars (QFeedS-1), for which we imposed a minimum radio luminosity threshold ($L_{\rm 1.4GHz}>10^{23.45}$\,W\,Hz$^{-1}$; dashed horizontal line in top panel). The 29 new targets, which are selected to be below this radio luminosity (QFeedS-2), are represented with green stars. Type 1 quasars are highlighted with white circles in the bottom panel.}
\label{fig:selection}
\end{figure}

Our initial sample of 42 targets (QFeedS-1) was selected to have radio luminosities of $L_\mathrm{1.4\,GHz} > 10^{23.45}\,$W\,Hz$^{-1}$. The sample is presented in an [O~{\sc iii}] luminosity versus 1.4\,GHz luminosity plane in Fig.~\ref{fig:selection}. It can be seen that, despite a minimum radio luminosity cut, the majority ($\sim$90\,per\,cent) are still consistent with traditional definitions of `radio quiet' (see discussion in \citealt{Jarvis2021}). However, based on on a variety of radio-AGN selection criteria (i.e., radio morphology, radio--infrared excess parameter, core spectral indices and brightness temperatures), $86^{+4.6}_{-6.2}$\,per\,cent reveal a radio-AGN (\citealt{Jarvis2021,Njeri2025}). Furthermore, radio sizes are 0.03--65\,kpc, covering the same size range of archetypal compact radio galaxy populations \citep[e.g.][]{An_Baan2012}, but with 1--3 orders-of-magnitude lower radio luminosities. Two-thirds of the sample show distinct radio structures, often indicative of jets and lobes (e.g., see Fig.~4 in \citealt{Jarvis2021} and Fig.~7 in \citealt{Njeri2025}). These results highlight the important contribution of AGN to the radio emission of quasars even with moderate radio luminosities (i.e., for `radio-quiet' quasars; also see e.g., \citealt{White2015,Zakamska2016,White2017}). 

Follow-up multi-wavelength observations of QFeedS-1 targets give insight into feedback mechanisms in these sources. For example, spatially unresolved sub-mm emission-line observations suggest little immediate impact on the global molecular gas reservoirs in the host galaxies by the quasars in these systems \citep[][]{Jarvis2020,Molyneux2024}. However, spatially-resolved observations show a strong localized connection between the radio structures and high levels of turbulence and outflows, revealed through optical integral field spectroscopy \citep[][]{Harrison2015,Jarvis2019,Girdhar2022}; sub-mm interferometry \citep[][]{Girdhar2022,Girdhar2024}; and X-rays \citep[]{Lansbury2018}. Follow-up polarisation-sensitive radio imaging has revealed a high fractional polarization (10--30 per cent) in the sources with the largest radio lobes, which provides further evidence of an interplay of jets or winds and emission-line gas as an important contribution to the nature of radio outflows in radio-quiet AGN \citep[][]{Silpa2022}. These observations are broadly consistent with simulations that show low power jets, confined within galaxy disks can drive outflows and have a feedback effect on their hosts, with potential additional contributions from outflow-driven shocks \citep[e.g.][]{Mukherjee2018,Tanner2022,Meenakshi2022,Meenaksh2024}.


The results from QFeedS-1 reveal important information on the prevalence of radio AGN and how feedback works in a sample of mostly traditionally radio-quiet quasars. However, we are unable to address how representative these results are to the wider quasar population, reaching lower radio luminosities. Therefore in this work, we present a new sample, QFeedS-2, reaching an order of magnitude lower in radio luminosity. With this, we can investigate the radio properties and feedback processes over several orders of magnitude in radio power and place them even more completely into the context of other AGN populations \citep[following e.g.,][]{Jarvis2021,Hardcastle2020,Baldi2023}. 


\subsection{Sample selection for this work}\label{sec:sample_selection}
Our goal is to expand the survey selection for QFeedS presented in \citet{Jarvis2021}, to quasars with lower radio luminosities.\footnote{We calculate the radio luminosities using the 1.4\,GHz flux densities for each source provided by \cite{Mullaney2013} (i.e., those taken from NVSS with preference, or FIRST otherwise) and assuming a spectral index of $\alpha=-0.7$. For the non radio detected sources, we use a flux density of 2.5\,mJy to derive luminosity upper limits, corresponding to the 50\% completeness of the NVSS survey (\citealt{Condon1998})} Therefore, we follow \citet{Jarvis2021}, and choose $z<0.2$ AGN from \cite{Mullaney2013} with a luminosity $L_{\rm [O III]}>10^{42.11}$\,erg\,s$^{-1}$. This leaves a potential sample of 226 targets. We then exclude any sources with a declination between 24 and 44\,degrees, to avoid observing targets that transit with very high elevations above the VLA.\footnote{We note that for the initial sample we also only considered targets in the RA range of 10--360\,degrees, for more efficient observation scheduling.}

For the initial sample of 42 targets we applied a radio luminosity selection of $L_{\rm 1.4GHz}>10^{23.45}$\,W\,Hz$^{-1}$. These 42 sources, are labeled as ``QFeedS-1'' in a $L_{\rm 1.4GHz}$ versus $L_{\rm [O III]}$ diagram in the top panel of Fig.~\ref{fig:selection}. To create the new sample, we observed the remaining 29 sources that have a 1.4\,GHz detection (in NVSS or FIRST) but with luminosities below the $L_{\rm 1.4GHz}=10^{23.45}$\,W\,Hz$^{-1}$ luminosity threshold. The lower luminosity sample, which is new for this work, is labeled as ``QFeedS-2'' in Figure~\ref{fig:selection}. We list key properties of the new QFeedS-2 sample in Table~\ref{Tab:sources1}.

\begin{table*}																	
\caption[]{Key properties for the 29 QFeedS-2 targets. (1) Source ID; (2\--3) optical positions (J2000); (4) redshifts from SDSS DR7; (5) AGN Type from the optical spectra; (6--7) [O\,{\sc iii}] luminosity and FWHM from \cite{Mullaney2013}; (8--9) integrated 1.4\,GHz flux density and peak flux density from FIRST; (10) 1.4\,GHz radio luminosity (based on NVSS; \citealt{Mullaney2013}).}

\label{Tab:sources1}														\begin{tabular}{lccccccccc}																			
\hline																			
Source ID	&	RA	&	DEC	&	z	&	Type	&	 log($L_\mathrm{[O III]}$)  	&	 FWHM$_\mathrm{[O III ]}$  	&	$S_\mathrm{FIRST}$	&	$P_\mathrm{FIRST}$	&	 log($L_\mathrm{1.4\,GHz}$)    	\\
	&		&		&		&		&	(/erg\,s$^{-1}$)	&	(km\,s$^{-1}$) 	&	($\mathrm{mJy}$)	&	($\mathrm{mJy\,beam^{-1}}$)	&	(/W Hz$^{-1}$)	\\
(1) & (2) & (3) & (4) & (5) & (6) & (7) & (8) & (9) & (10) \\
    
\hline																			
J0213$+$0042 	&	02:13:59.78	&	$+$00:42:26.8	&	0.182	&	1	&	42.68	&	752	&	2.90$\pm{0.12}$ 	&	3.16$\pm{0.07}$ 	&	23.43	\\
J0232$-$0811 	&	02:32:24.25	&	$-$08:11:40.2	&	0.100	&	2	&	42.15	&	567	&	3.48\plus{0.12} 	&	3.77\plus{0.08} 	&	23.07	\\
J0827$+$2233 	&	08:27:11.22	&	$+$22:33:24.2	&	0.173	&	2	&	42.37	&	576	&	2.03\plus{0.13} 	&	2.03\plus{0.07} 	&	23.35	\\
J0841$+$0101 	&	08:41:35.09	&	$+$01:01:56.3	&	0.111	&	 2		&	42.41	&	388	&	3.51\plus{0.19} 	&	2.50\plus{0.09} 	&	23.32	\\
J0924$+$1504 	&	09:24:35.36	&	$+$15:04:10.0	&	0.125	&	 2		&	42.39	&	727	&	3.43\plus{0.09} 	&	3.22\plus{0.05} 	&	22.99	\\
J0947$+$1005 	&	09:47:33.22	&	$+$10:05:08.8	&	0.139	&	1	&	42.18	&	669	&	1.41\plus{0.10} 	&	1.21\plus{0.05} 	&	23.25	\\
J1034$+$6001 	&	10:34:08.60	&	$+$60:01:52.2	&	0.051	&	 2		&	42.41	&	612	&	18.98\plus{0.19} 	&	16.78\plus{0.10} 	&	23.06	\\
J1110$+$5848 	&	11:10:15.25	&	$+$58:48:46.0	&	0.143	&	 2		&	42.42	&	574	&	4.11\plus{0.21} 	&	3.90\plus{0.12}	&	23.35	\\
J1141$+$2156 	&	11:41:16.16	&	$+$21:56:21.8	&	0.063	&	1	&	42.24	&	314	&	2.68\plus{0.45} 	&		1.80\plus{0.20} 	&	22.63	\\
J1152$+$1016 	&	11:52:45.66	&	$+$10:16:23.8	&	0.070	&	 2		&	42.27	&	468	&	3.57\plus{0.08} 	&		3.39\plus{0.05} 	&	22.67	\\
J1203$+$1624 	&	12:03:00.20	&	$+$16:24:43.8	&	0.166	&	 2		&	42.74	&	339	&	2.91\plus{0.13} 	&		2.90\plus{0.07} 	&	23.42	\\
J1300$+$5454 	&	13:00:38.10	&	$+$54:54:36.9	&	0.088	&	 2		&	42.49	&	282	&	2.16\plus{0.13} 	&		1.96\plus{0.07} 	&	22.69	\\
J1316$+$4452 	&	13:16:39.75	&	$+$44:52:35.1	&	0.091	&	2	&	42.20	&	602	&	4.30\plus{0.13} 	&		3.96\plus{0.07} 	&	23.07	\\
J1325$+$1137 	&	13:25:52.16	&	$+$11:37:09.8	&	0.161	&	 1		&	42.58	&	831	&	2.84\plus{0.10} 	&		2.93\plus{0.06} 	&	23.38	\\
J1338$+$1503 	&	13:38:06.53	&	$+$15:03:56.1	&	0.185	&	 2		&	42.53	&	705	&	2.22\plus{0.09} 	&	 	1.91\plus{0.05} 	&	23.29	\\
J1355$+$5612 	&	13:55:16.55	&	$+$56:12:44.7	&	0.122	&	1	&	42.28	&	688	&	6.04\plus{0.15} 	&		5.88\plus{0.08} 	&	23.34	\\
J1410$+$2233 	&	14:10:41.50	&	$+$22:33:37.1	&	0.173	&	1	&	42.14	&	1033	&	3.11\plus{0.19} 	&		2.23\plus{0.09} 	&	23.22	\\
J1419$+$1144 	&	14:19:43.79	&	$+$11:44:26.2	&	0.124	&	 2		&	42.27	&	491	&	2.94\plus{0.13} 	&		2.54\plus{0.07} 	&	22.93	\\
J1426$+$1040 	&	14:26:14.62	&	$+$10:40:13.0	&	0.137	&	 2		&	42.38	&	513	&	1.96\plus{0.12} 	&		2.03\plus{0.07} 	&	23.25	\\
J1426$+$1949 	&	14:26:26.93	&	$+$19:49:54.4	&	0.175	&	 2		&	42.13	&	906	&	1.15\plus{0.06} 	&		1.44\plus{0.04} 	&	23.31	\\
J1434$+$5016 	&	14:34:11.18	&	$+$50:16:40.8	&	0.199	&	 2		&	42.38	&	522	&	2.75\plus{0.13} 	&	 	2.84\plus{0.08} 	&	23.41	\\
J1440$+$5303 	&	14:40:38.10	&	$+$53:30:15.9	&	0.038	&	2	&	42.49	&	593	&	59.21\plus{0.30} 	&	 	57.20\plus{0.17} 	&	23.29	\\
J1450$+$0713 	&	14:50:34.12	&	$+$07:31:32.3	&	0.154	&	 2		&	42.22	&	507	&	1.71\plus{0.27} 	&		1.56\plus{0.15} 	&	23.32	\\
J1507$+$0029 	&	15:07:19.94	&	$+$00:29:05.0	&	0.182	&	 2		&	42.57	&	1254	&	4.36\plus{0.11} 	&		4.40\plus{0.06} 	&	23.44	\\
J1529$+$5616 	&	15:29:07.46	&	$+$56:16:06.7	&	0.100	&	 1		&	42.57	&	627	&	4.26\plus{0.13} 	&		4.31\plus{0.08} 	&	23.10	\\
J1543$+$1148 	&	15:43:03.83	&	$+$11:48:38.5	&	0.098	&	2	&	42.54	&	479	&	2.09\plus{0.11} 	&	 2.60\plus{0.07} 	&	22.77	\\
J1653$+$2349 	&	16:53:15.05	&	$+$23:49:43.0	&	0.103	&	 2		&	42.55	&	435	&	7.14\plus{0.12} 	&		6.72\plus{0.07} 	&	23.38	\\
J1713$+$5729 	&	17:13:50.32	&	$+$57:29:54.9	&	0.113	&	2	&	42.55	&	1278	&	7.38\plus{0.09} 	&		7.29\plus{0.05} 	&	23.42	\\
J2304$-$0841 	&	23:04:43.48	&	$-$08:41:08.6	&	0.047	&	1	&	42.14	&	693	&	21.51\plus{0.75} 	&		17.85\plus{0.38}	&	23.22	\\
\hline																			
\end{tabular}
\end{table*}															

\subsection{New sample overview}\label{sec:sample_overview}

This new work brings our total sample in the QFeedS to 71 $z<0.2$ targets, all of which have been selected to have high [O~{\sc iii}] luminosities, corresponding to bolometric luminosities of $L_{\rm bol}\gtrsim10^{45}$\,erg\,s$^{-1}$ (see \citealt{Jarvis2020}). Of the 29 targets in QFeedS-2, eight are classified as Type 1 and 21 are classified as Type 2 based on their optical spectroscopy (\citealt{Mullaney2013}). Across the combined sample (QFeedS-1 and QFeedS-2), 25 (35\,per\,cent) are classified as Type 1 and 46 (65\,per\,cent) are classified as Type 2. This is representative of the parent sample of AGN from \citealt{Mullaney2013}, which meet the redshift and [O~{\sc iii}] luminosity cuts, for which 88/226 (i.e., 39\,per\,cent) are confirmed as Type 1 AGN. 

For our [O~{\sc iii}] luminosity and redshift selection criteria (see Fig.~\ref{fig:selection}); $52^{+3.3}_{-3.3}$\,per\,cent of the parent sample have radio detections in FIRST/NVSS from \cite{Mullaney2013}. Our combined sample still does not represent the targets from the parent sample that only have upper limits on their 1.4\,GHz radio luminosities based on FIRST and NVSS. The non radio detected sources have a mean upper limit of 1.9$\times$10$^{23}$\,W\,Hz$^{-1}$, with a range of (0.27--2.7)$\times$10$^{23}$\,W\,Hz$^{-1}$. Nonetheless, the combined QFeedS sample covers nearly four orders of magnitude in radio luminosity from $L_{\rm 1.4GHz}$=10$^{22.6}$ to 10$^{26.3}$\,W\,Hz$^{-1}$. Indeed, the sample expands deep into the `radio-quiet' regime following the definition of \cite{Xu1999} (see blue line in Fig.~\ref{fig:selection}). 

All targets in the QFeedS-2 sample are classified as `radio quiet', with at least one order of magnitude lower radio luminosities than the `radio-loud' cut off. For the Type 1 targets, we additionally investigated the radio-loudness parameter, $R$, which compares radio to optical continuum luminosity as a means to define sources as `radio loud'. Following \cite{Kellerman1989,Ivezic2002}, $R = 0.4(m_i - t)$, where $m_i$ is the $i-$band magnitude from SDSS DR16 \citep[][]{Ahumada2020} and $t$ is the `AB radio magnitude' given by $t = -2.5\,\mathrm{log} \left(\frac{S_{1.4\,\mathrm{GHz}}}{{3631\mathrm{Jy}}}\right)$, none of the Type 1 targets in QFeedS-2 pass the radio-loudness criteria of $R\gtrsim 1.0$. In contrast, in the initial sample (QFeedS-1), 9 targets were classified as `radio loud' following both criteria. Furthermore, many targets in the QFeedS-1 sample could be classified as `radio intermediate', due to their proximity to the radio-loud/quiet boundary (\citealt{Jarvis2021}). 

In the bottom panel of Figure~\ref{fig:selection} we show [O~{\sc iii}] FWHM as a function of radio luminosity for the parent sample in the redshift and [O~{\sc iii}] luminosity range of our selection, and highlight the QFeedS targets with larger symbols. In this figure the plotted FWHM is the flux-weighted average of the two components that are fitted to the [O~{\sc iii}] emission lines in the SDSS spectra, following \cite{Mullaney2013}. It can be seen that the combined QFeedS sample covers the full range of FWHM parameter space of the parent sample. Furthermore, the figure re-highlights the previously identified increased fraction of extreme kinematics ($\gtrsim$800\,km\,s$^{-1}$) for radio luminosities greater than $\sim$2$\times$10$^{23}$\,W\,Hz$^{-1}$ (\citealt{Mullaney2013}; also see \citealt{Zakamska2014}, \citealt{Woo2016}, \citealt{Jarvis2021}). 

The new QFeedS-2 sample provides the opportunity to study the radio properties of quasars over multiple orders of magnitude in radio luminosity, and explore deep into the `radio-quiet' regime. Building on the work with QFeedS-1, future multi-wavelength follow-up observations will enable an assessment of the connection between radio emission and multi-phase outflows across the quasar population. In this work, we present new L-band and C-band VLA observations of the 29 QFeedS-2 targets, following similar approaches to the equivalent data presented for the initial 42 QFeedS-1 targets in \citep{Jarvis2019,Jarvis2021} and with C-band \emerlin{} data in \cite{Njeri2025}. We then combine and compare the radio properties of the full sample (QFeedS-1 and QFeedS-2), and explore the possible origin of the radio emission across the full radio luminosity range.


\begin{figure*}
\centerline{\includegraphics[width=1\linewidth]{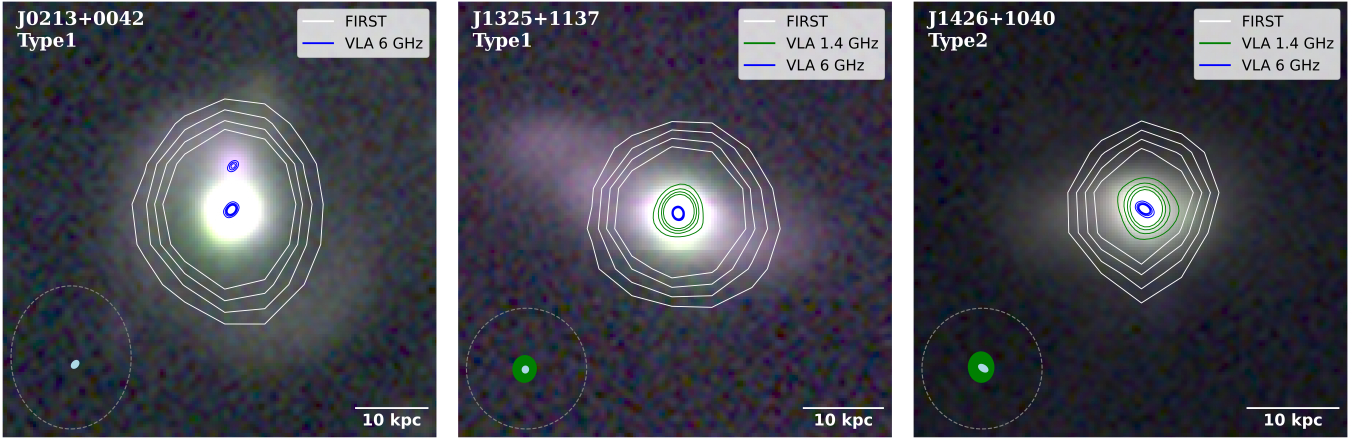}}
\caption[]{Examples of our radio data for sources classified as compact based on radio morphology in the QFeedS-2 sample (see Section~\ref{sec:radio_morph}). The rgb images come from the DESI Legacy Imaging Survey in the $z,r,g$ bands. The contours overlaid represent: FIRST maps in white with levels $1\sigma \times [2,3,4,5]$; our new VLA 1.4\,GHz maps in green with levels $1\sigma \times [2,4,6,8]$ (except for J0213$+$0042, which was not observed) ; and our new VLA 6\,GHz maps in blue with levels $1\sigma \times [2,4,6,8]$. Ellipses represent the synthesised beam for the corresponding radio map contours; $\sim$5\,arcsec for FIRST { (in dashed grey ellipses)}, $\sim$1\,arcsec for VLA 1.4\,GHz (in green) and $\sim$0.3\,arcsec for VLA 6\,GHz ({ in light blue}). The scale bar highlights the physical size scales. Equivalent figures for all sources are provided (see the Data Availability section). For J0213+0042, we note the target was classified as a `compact' source because the second radio component, which coincides with a second optical companion, was ignored.}
\label{fig:compact_sources}
\end{figure*}

\begin{figure*}
\centerline{\includegraphics[width=1\linewidth]{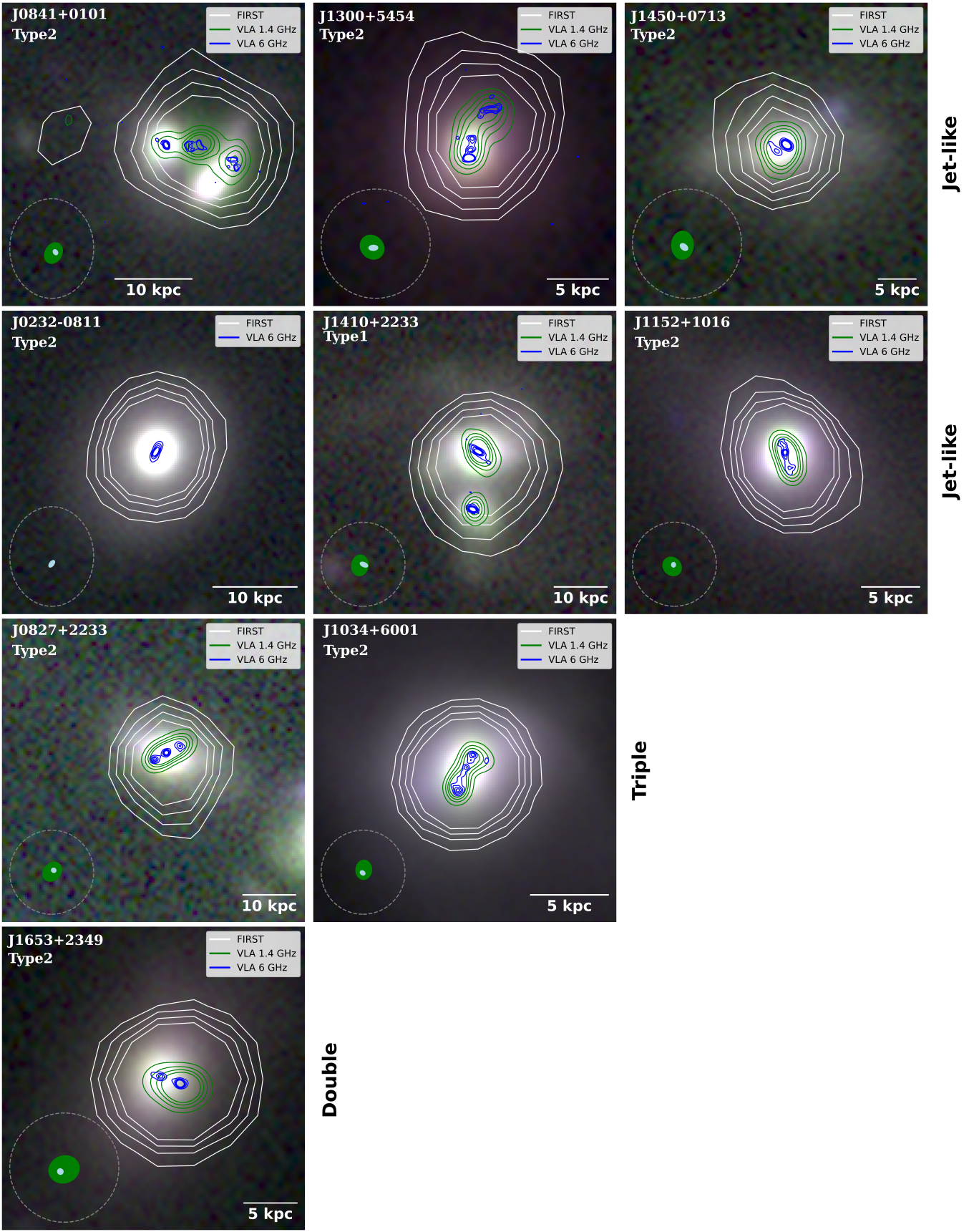}}
\caption[]{The same as Figure~\ref{fig:compact_sources}, but presenting the nine targets identified as having extended 'jet-like' structures. These targets are classified as radio-AGN based on radio morphology. We note that the target J0232$-$0811 was not observed in the L-band.}
\label{fig:jet_like}
\end{figure*}

\section{Observations and data reduction}\label{sec:VLA_obs}
Observations of the 29 quasars (QFeedS-2) were conducted with VLA in the L-band (1.2 -- 1.7\,GHz) and C-band (5.25 -- 7.20\,GHz), under the proposal ID 23A-214 [PI. Njeri]. These observations were carried out between July and September 2023 using the A array, providing a representative resolution of 1.0 and 0.3 arcsec, in the L-band and C-band, respectively. Whilst all targets were covered by the C-band observations, three targets were not observed in the L-band (J0213$+$0042, J0232$-$0811, and J2304$-$0841), due to incompleted observations. We observed for 10.5\,hours in L-band (5–30 min per source) and 9.5\,hours in C-band (5–30 min per source). Each target was observed in at least two scans spread across 1 -- 2 hour observing blocks. At the start of each observing block, a standard calibrator (3C286, 3C48 or 3C147) was observed for $\sim$10\,minutes. A phase calibrator was observed for $\sim$3\,minutes per scan (including the slewing time) followed by the target.

We used the VLA Calibration Pipeline 2024.1.0.8 for \textsc{casa} version 6.6.1 to reduce our data. The calibrated measurement set was then inspected using \textsc{casa} task {PLOTMS} and any remaining bad data was visually identified and flagged using `tfcrop' mode in \textsc{casa} task {FLAGDATA}. Each target was split into their own measurement set and imaged using \textsc{casa} task {TCLEAN}. Maps of $2500 \times 2500$ cells were constructed, with a cell size of 0.2 arcsecs per pixel for L-band, and 0.06 arcsecs per pixel for C-band. To produce a consistent set of images to QFeedS-1 we followed \cite{Jarvis2021} and used a Briggs weighting of $r=0.5$ for all radio maps. This provides a suitable compromise between sensitivity to larger scale structures and higher spatial resolution. This uniform approach is most suitable for this systematic study, but we note that a source-by-source imaging optimisation can sometimes reveal additional structures or detail (e.g., comparing different approaches for the same targets in \citealt{Jarvis2021} and \citealt{Jarvis2019}). We generated radio maps with typical median root-mean square (rms) sensitivities of $\sim24\,\microJybm$ in L-band and $\sim 10\,\microJybm$ in C-band. In comparison, the QFeedS-1 radio maps were produced at typical median rms sensitivities of $\sim 47\,\microJy$ in L-band and $\sim 12\,\microJybm$ in C-band.

Examples of these radio maps are presented in Figure~\ref{fig:compact_sources} and Figure~\ref{fig:jet_like}, which showcase a range of the morphological structures observed (see Section~\ref{sec:radio_morph}). The background optical images are from the archival 3-colour (RGB) image from the Dark Energy Spectroscopic Instrument (DESI) Legacy Imaging Survey using the $z,r,g$ bands \citep[][]{Dey2019}. Overlaid, are contour maps from the VLA FIRST survey in white, and our VLA L-band and C-band maps in green and blue, respectively. We release all the equivalent figures for the whole sample and all the radio maps, which include all of the relevant metadata (e.g., the exact beam sizes for each map; see Data Availability section). 

To extract the radio properties such as the total flux density, peak flux density, and the linear size of radio components, the targets were fitted using the \textsc{casa} task  {IMFIT}. These measured values are presented in Table~\ref{Tab:sources2}. Where distinct multiple components were identified in an individual source, these properties have been extracted separately (i.e., those classified as doubles or triples; see Section~\ref{sec:radio_morph}).


We also produced spectral index  ($\alpha$) maps using the C-band data. We applied the two Taylor terms to model the frequency dependence of the sky emission and generated the in-band (5.25 -- 7.20\, GHz) spectral index maps and  corresponding error maps. To extract the spectral indices of the core regions, and the associated errors, we took the median value of the $\alpha$ image within a region that covered the core only and obtained the equivalent error from the respective error maps (see Section~\ref{sec:spectral_index}). These $\alpha$ values and the corresponding errors are presented in Table~\ref{Tab:sources2}. 

\section{Results}\label{sec:results}
Building on the work presented in \cite{Jarvis2021} and \cite{Njeri2025}, one of the primary goals of QFeedS is to establish the fraction of moderate-to-low radio luminosity quasars (average $\sim L_\mathrm{1.4\,GHz}\approx 10^{23.5}\,$W\,Hz$^{-1}$) that show radio emission associated with AGN processes. In this work, we present the results from the new QFeedS-2 sample, which comprises of 29 targets in a luminosity range of $L_\mathrm{1.4\,GHz}\sim10^{22.63}$--$10^{23.45}$\,W\,Hz$^{-1}$, pushing further into the radio-quiet regime than in the initial QFeedS-1 sample (see Fig.~\ref{fig:selection}). 

We present three sets of radio maps, those from the archival FIRST data, and our new L-band and C-band VLA maps (Section~\ref{sec:VLA_obs}). Accounting for the redshift range of the sample, these maps cover a spatial resolution of $\sim 5$--20\,kpc in the 1.4\,GHz FIRST images, $\sim 1$--20\,kpc in our new L-band (1.4\,GHz) maps and $\sim 0.3$--1\,kpc in our new C-band (6\,GHz) maps. Example maps are shown in Figure~\ref{fig:compact_sources} and Figure~\ref{fig:jet_like}. The measured radio properties are presented in Table~\ref{Tab:sources2}.

In Section~\ref{sec:radio_morph}, we present the radio morphology classification based on visual inspection and we present the results of largest linear size (LLS) measurements in Section~\ref{sec:linearsizes}.  In Section~\ref{sec:spectral_index} and Section~\ref{sec:TB}, we present the spectral index and brightness temperature measurements, respectively. In Section~\ref{sec:radio_AGN}, we summarize how the radio measurements (morphology, brightness temperature and spectral index) were used for the final identification of `radio-AGN'.

\subsection{Detected radio structures and their morphology}\label{sec:radio_morph}
Detecting and characterising radio structures across the quasar sample provides crucial insights into the physical mechanisms driving radio emission and its potential role in feedback. Following the nomenclature used in \cite{Jarvis2021}, we classify our targets as:\\
(i) Compact (C): when only a single compact structure is identified. \\
(ii) Double (D): the target has two clearly distinct radio peaks.\\
(iii) Triple (T): the target has three clearly distinct radio peaks. \\
(iv) Jet-like (J): when the targets show a single central radio component and are visibly spatially extended in one or two directions. In this category, we also include two sources (J0841$+$0101 and J1300$+$5454) with some more irregular structures for which two or three individual components are not well defined, but are generally collimated and jet-like.

Since the targets remained largely compact and unresolved in FIRST and only a few sources (J0841$+$0101, J1034$+$6001, J1300$+$5454 and J1316$+$4452) showed visibly extended emission in L-band maps, we used the C-band radio maps for the radio morphology classification. There are two sources (J0213$+$0042 and J1410$+$2233) that show radio emission associated with secondary optical counterparts. Consequently, the radio structures are likely associated with companion galaxies and we ignored these radio structures in the classifications.

We find that 20/29 sources were classified as compact, with three examples shown in Figure~\ref{fig:compact_sources}. For the nine sources that show extended radio structures from the visual classification, 6/9 are classified as jet-like, 1/9 is classified as a double, and 2/9 are classified as triples. All nine of these sources are presented in Figure~\ref{fig:jet_like}. For simplicity, and for the remainder of this work, all extended emission and well-collimated multiple components (double or triple radio peaks) are considered as jet-like (following our radio morphology classification presented in \citeauthor{Njeri2025}\citeyear{Njeri2025}). However, we note that this purely defines the radio morphology and does not necessarily indicate the presence of a traditional AGN-driven jet (further discussion in Section~\ref{sec:radio_origin}). 

In summary, a total of 9/29 targets ($\sim 31\,$per\,cent) show extended radio structures in the QFeedS-2 sample. This is in contrast to QFeedS-1, for which $67^{+6.8}_{-7.6}$ per\,cent showed extended structures using equivalent datasets, and we discuss the variations across the whole of QFeedS further in Section~\ref{sec:QFS1_QFS2} \citep[also see similar examples from other samples in][]{Middelberg2007,Baldi2018,Odea2021}.

\subsubsection{Comparison with FIRST and implications for diffuse emission}
To assess how much large-scale, low-surface-brightness radio emission might be missed by our higher-resolution L-band data, we compare the integrated flux density measured in our L-band maps with the FIRST integrated flux densities (both at 1.4\,GHz). We report the per-source flux ratio:

\textbf{\begin{equation}   
R = \frac{S_\mathrm{1.4\,GHz}}{S_\mathrm{FIRST}}
\end{equation}}

in Table~\ref{Tab:sources2}. The ratios of integrated flux between our L-band maps and FIRST range from $0.34 \-- 1.57$ with a median value of $\sim 1.0$. Values of $R\,<\,1$ indicate cases where FIRST records more integrated flux than our high-resolution L-band maps, which is consistent with diffuse emission resolved out by the higher-resolution observations. We only find 5/26 of the sources (J1141+2156, J1410+2233, J1316+4452, J0924+1504 and J1419+1144; i.e., 19\,per\,cent of the sample) have $R\,<\,1$ values ($R = 0.34 \--0.85$) beyond the uncertainties. In all of these cases the $R$ values are smaller than $R=1$ at a significance which is greater than three times the uncertainties. In one of these cases, J1410$+$2233, the difference can be attributed to the secondary optical source in the field, which contaminates the FIRST flux. Therefore, only the remaining four cases show strong evidence that $\approx$15--66\,per\,cent of the 1.4\,GHz flux density is resolved out in our L-band maps. These results indicate that our L-band imaging recovers the bulk of the $1.4\,$GHz emission measured by FIRST for the large majority of the sample.

Conversely, $R>1$ values indicate that our L-band maps contain more flux than measured by FIRST. This might indicate source variability between FIRST and our observations, minor variations due to slightly different frequency coverage, or other calibration and systematic errors \citep[e.g.][]{Vries2004,Ofek2011,Perley2013,Mooley2016,Radcliffe2019}. We find 8/26 sources (J0827+2233, J1110+5848, J1300+5454, J1325+1137, J1426+1040, J1426+1949, J1434+5016 and J1543+1148; $R = 1.10 \-- 1.57$) have values of $R>1$ beyond the uncertainties. However, in all but two of these cases the differences are consistent with $R>1$ only at three times the errors. Therefore, there is only weak evidence for values of $R>1$ and we cannot rule out potential calibration/measurement issues. For the remaining two sources (J1426$+$1949 and J1543$+$1148, with R=1.57$\pm$0.08 and R=1.30$\pm$0.07, respectively) there may have been variability of the sources between the two observing epochs at the $\sim$30--60\,per\,cent level.

We stress that additional causes, particularly time variability (FIRST epochs are typically years earlier than our new observations), different flux-extraction methods, and small frequency offsets combined with spectral curvature, can all produce deviations from $R=1$. We therefore treat this comparison only as a rough indicator that, for the majority of our sources, there is not a significant fraction of large-scale gigahertz emission which is missed by our L-band VLA observations. Single-dish data or lower frequency radio observations could better quantify any additional large-scale diffuse components missed by our observations \citep[e.g.][]{Condon1992,Godfrey2016,Morabito2022}.

\subsection{Largest linear sizes in the radio}\label{sec:linearsizes}
To estimate the radio largest linear size (LLS) of each quasar, we used the L-band images under the assumption that the detected radio emission is fully confined within the L-band restoring beam. For compact sources, the LLS was taken directly from the deconvolved full width at half‐maximum (FWHM) of the major axis returned by the two-dimensional Gaussian fitting performed with the \textsc{casa} task IMFIT. These values are presented in Table~\ref{Tab:sources2}. The quoted uncertainties correspond to the formal {IMFIT} errors on the deconvolved major axis, these incorporate the statistical uncertainties from the fit but do not explicitly propagate additional uncertainties in the restoring beam size and shape, potentially resulting in a underestimate of the true uncertainties.\footnote{{\textsc{casa} IMFIT computes uncertainties on the deconvolved Gaussian parameters by propagating the covariance matrix of the best-fitting convolved model. The reported errors reflect statistical uncertainties from the fit (including pixel noise and parameter covariances), but they do not explicitly include systematic uncertainties due to the restoring beam model, residual calibration errors, or departures from a true Gaussian source structure. As a result, deconvolved sizes may have small formal uncertainties even when the intrinsic source size is smaller than the beam (see CASA Documentation: Image Analysis—{IMFIT},  CASA v6.5 from \citealt{casa_imfit_2022} and \citealt{mcmullin2007}). However, IMFIT will not return a size measurement if the source can not be de-convolved from the beam, and is consistent with an unresolved point source.}}

For sources exhibiting multiple components or clearly extended structures in the L-band images, we adopted the same LLS fitting procedure as \cite{Jarvis2021}. In this approach, the LLS is defined as the projected separation between the outermost significant radio peaks. These peaks are identified in the cleaned image using intensity thresholds matched to the local rms, and their angular separation is converted to a linear scale using the source redshift. \cite{Jarvis2021} demonstrated that this method robustly captures the extent of jets, lobes, or knotty emission even when the morphology departs from a simple Gaussian and when individual components cannot be well-described by single-component fits. In cases where we detected a secondary radio component coincident with an optical galaxy (J0213+0042 and J1410+2233), we interpreted that component as an unrelated companion galaxy (see Section~\ref{sec:radio_morph}). Therefore, these secondary components were excluded from our LLS analysis. For the primary quasar component in these two systems, the LLS was taken from the deconvolved major-axis FWHM provided by \textsc{casa} task IMFIT (as decribed above).

For the 3 targets (J0213$+$0042, J0232$-$0811 and J2304$-$0841) that were not observed in L-band, we were forced to consider the {IMFIT} deconvolved major axis in C-band as the LLS for these targets.

\begin{figure*}
\centerline{\includegraphics[width=1\linewidth]{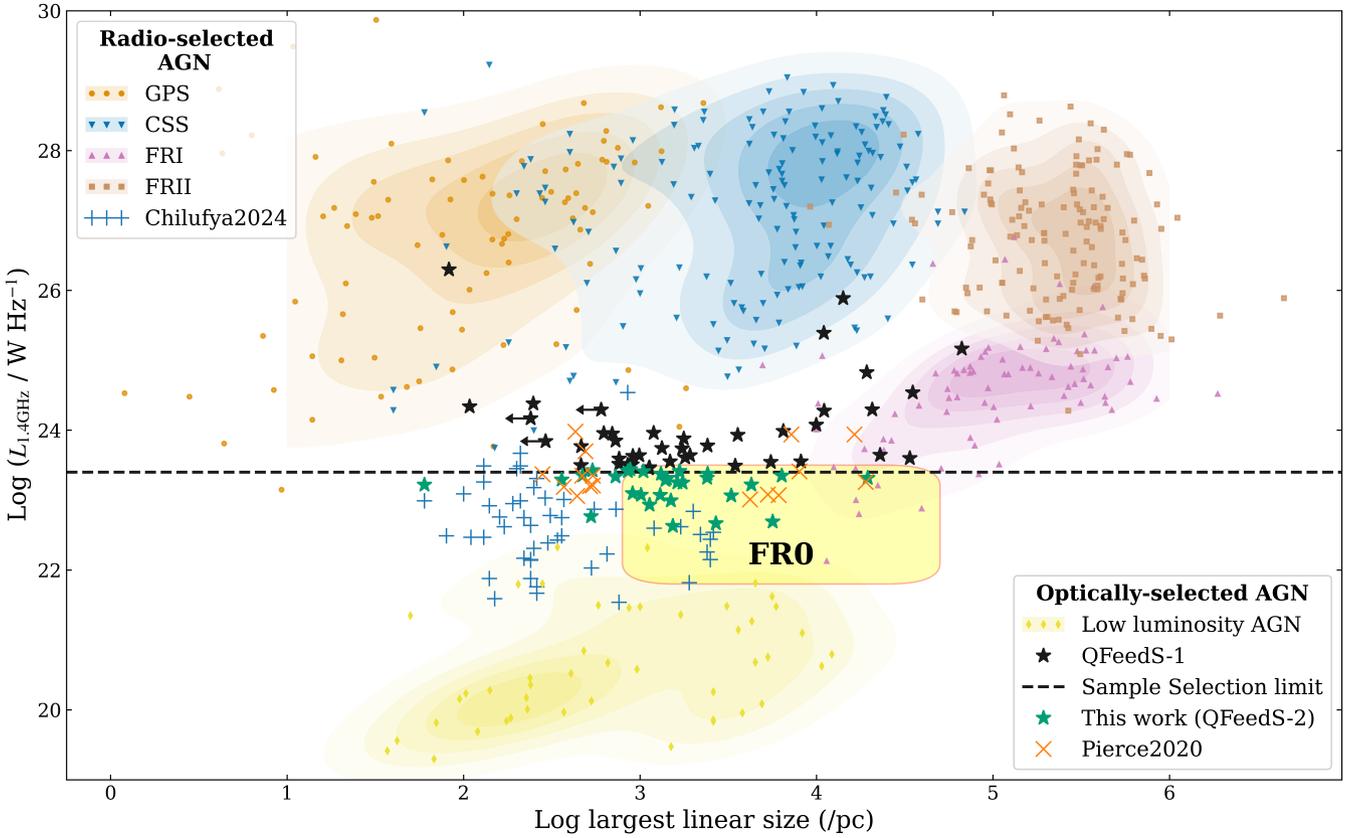}}
\caption[]{Radio size versus radio luminosity for the QFeedS quasars compared to classical radio-selected AGN populations. Coloured points and contours mark the loci of GPS, CSS, FRI nad FRII sources, while the shaded yellow region highlights the FR0 domain as defined in \cite{Baldi2023}. Black stars show the QFeedS-1 sample \citep[][]{Jarvis2021}, and green stars show the new QFeedS-2 sample presented in this work. The dashed line marks the QFeedS-1 radio luminosity selection limit. The majority of QFeedS-2 remain compact at sub-arcsecond to arcsecond scales ($\sim3\,$kpc), and broadly consistent with FR0-like systems. This distribution closely parallels the compact steep-spectrum sources reported in \cite{Chilufya2024} (plus symbols), though QFeedS quasars are optically selected and radiatively efficient, in contrast to many radio-selected samples. By contrast, QFeedS-1 shows a higher fraction of extended sources with higher radio luminosities, bridging towards CSS and FRI-like populations. At the same time, the QFeedS-2 sample overlaps in luminosity with the intermediate-power HERGs of \cite{Pierce2020} (cross symbols), but stands out for its strong {\sc[O iii]} emission despite compact radio morphologies. Taken together, QFeedS maps a broad continuum of radio properties in optically-selected quasars and highlights the role of QFeedS in connecting low-luminosity FR0-like AGN with the more extended radio populations, while demonstrating the diversity of radio output among radiatively-efficient quasars across several orders of magnitude in $L_{1.4\,\mathrm{GHz}}$.}
\label{fig:luminosity_plot}
\end{figure*}

In Figure~\ref{fig:luminosity_plot}, we plot $L_\mathrm{1.4\,GHz}$ (derived from the NVSS measurements) as a function of the LLS for the whole QFeedS sample, where the new QFeedS-2 sample is represented with green stars, and QFeedS-1 with black stars. The LLS values of QFeedS-2 range from 0.4--6.4\,kpc, with two outliers at $\sim$0.1\,kpc (J2304$-$0841; but we note that this is missing the L-band data) and $\sim$20\,kpc (J0841$+$0101; see Fig.~\ref{fig:jet_like}). Our new results reveal that the gigahertz radio emission is typically compact on host galaxy scales, i.e., confined within a few kiloparsecs. Indeed, visual inspection of the radio images shows that the radio emission is confined to the spatial extent of the optical emission (with the exception of J0841$+$0101). 

The QFeedS-2 sources have a very similar distribution of LLS sizes to the majority of QFeedS-1 (see Fig.~\ref{fig:luminosity_plot}). The exception is that ten (i.e., 24\,per\,cent) of the QFeedS-1 sources have LLS values $\gtrsim$10\,kpc, with only one source (i.e., 3\,per\,cent) of QFeedS-2 reaching these scales. The QFeedS-1 targets with the largest sizes also tend to be those with the highest luminosities, i.e.,$L_\mathrm{1.4\,GHz}\gtrsim10^{24}$\,W\,Hz$^{-1}$. We cannot rule out that even deeper radio observations would reveal lower surface brightness extended structures. However, for the majority of our sample the detected radio luminosity is dominated by the compact components recovered in our L-band maps. 

This result highlights a limitation of wide, low-resolution surveys such as FIRST, with a beam of $\sim5\farcs0$, they cannot resolve the sub-kpc to few-kpc structures revealed by higher-resolution VLA imaging. Indeed, all QFeedS-2 targets remained compact, i.e., without any visibly extended radio features in FIRST (except a marginal detection of a secondary component in J0841$+$0101; see top-left of Fig.~\ref{fig:jet_like}). Our sensitive and high spatial resolution radio imaging shows that the radio emission is generally confined to the optical extent of the host galaxy, and that FIRST therefore often blends compact cores, small jets and nearby companions into a single unresolved detection. Consequently, FIRST-scale surveys do not fully characterise the key spatial scales of radio activity in these radio-quiet quasars, and, therefore sensitive sub-arcsecond imaging is required to separate compact AGN components from genuinely diffuse emission and to robustly interpret the physical origin of the radio emission. In Section~\ref{sec:low_luminosity} we discuss QFeedS in the context of the other AGN populations presented in Figure~\ref{fig:luminosity_plot}

\begin{landscape}									\begin{table}					\caption{The derived radio properties for the 29 QFeedS-2 targets, including multiple radio components per source when relevant. The columns are as follows: Column (1) gives the source ID. Columns (2) and (3) give the source positions from C-band data in J2000. Columns (4--5) give the total flux density and the peak brightness from the L-band maps. Column (6) is the largest linear size, derived from the L-band; Column (7) gives the integrated flux ratios ($R$) between our L-band and FIRST data; Columns (8--9) give the total flux density and peak brightness for the components observed in the C-band maps. Column (10) is the de-convolved linear size from the fits to the components observed in the C-band data (Section~\ref{sec:VLA_obs}). Column (11) is the core spectral index ($\alpha$, over 5.25\--7.20 GHz; Section~\ref{sec:spectral_index}). Column (12) is the brightness temperature of the cores (Section~\ref{sec:TB}). Column (13) is the radio morphology from visual inspection (Section~\ref{sec:radio_morph}). Column (14) highlights the sources classified as a Radio-AGN (Section~\ref{sec:radio_AGN}). N/O indicates that the target was not observed in L-band.}																									
 \label{Tab:sources2}																									
\small																									
\centering																									
\begin{tabular}{lccccccccccccc}																									
\hline																									
Source ID	&	RA	&	DEC	&	 $S_\mathrm{1.4\,GHz}$	&	$P_\mathrm{1.4\,GHz}$	&	LLS	 & $S_\mathrm{1.4\,GHz}/$ & 	$S_\mathrm{6.3\,GHz}$	&	$P_\mathrm{6.3\,GHz}$	&	Component	&	SI	&	log($T_\mathrm{B}$)	&	Radio	&	Class	\\

	&		&		&	  ($\mathrm{mJy}$) 	&	 ($\mathrm{mJy\,beam^{-1}}$)  	&	 size(kpc)  	& $S_\mathrm{FIRST}$	&  ($\mathrm{mJy}$) 	&	 ($\mathrm{mJy\,beam^{-1}}$) 	&	(kpc) 	&	 $\alpha$ 	&	(K)	&	 Morph. 	&		\\

 (1) & (2) & (3) & (4) & (5) & (6) & (7) & (8) & (9) & (10) & (11) & (12) & (13) & (14) \\   
\hline																									
J0213$+$0042 	&	 02:13:59.79 	&	 $+$00:42:26.75 	&	 N/O 	&		N/O		&	 N/O 	&	N/O & 0.83$\pm{0.01}$ 	&	 0.69\plus{0.01} 	&	 0.54\plus{0.05} 	&	 -0.71\plus{0.15} 	&	2.95	&	 Compact  	&		\\

J0232$-$0811 	&	 02:32:24.25 	&	 -08:11:40.19 	&	 N/O		&	 N/O 	&	 N/O 	&	N/O &  1.14\plus{0.07} 	&		0.55\plus{0.02} 	&	 1.30\plus{0.11} 	&	 -1.36\plus{0.37} 	&	2.11	&	 Jet-like  	&	 R-AGN	\\

J0827$+$2233 	&	 08:27:11.22 	&	 $+$22:33:24.21  	&	 2.48\plus{0.12} 	&	 1.25\plus{0.04} 	&	 6.35\plus{0.40} 	& 1.22\plus{0.10}	&  0.27\plus{0.01} 	&	 0.20\plus{0.01} 	&	 5.0\plus{0.07} 	&	 -2.15\plus{0.74} 	&	2.48	&	 Triple   	&	 R-AGN 	\\

           	&	 08:27:11.27 	&	 $+$22:33:24.76 	&	$-$	&	$-$	&	$-$	& $-$	& 0.25\plus{0.01} 	&	 0.14\plus{0.01} 	&	 0.96\plus{0.07} 	&	 -1.07\plus{0.36} 	&	1.74	&	$-$	&	$-$	\\
           	&	 08:27:11.16 	&	 $+$22:33:24.64 	&	$-$	&	$-$	&	$-$	& $-$	& 0.21\plus{0.01} 	&	 	0.1\plus{0.002} 	&	 1.12\plus{0.05} 	&	 -1.15\plus{0.30}  	&	1.44	&	$-$	&	$-$	\\
            
J0841$+$0101 	&	 08:41:35.08 	&	 $+$01:01:56.20 	&	 3.65\plus{0.58} 	&		0.68\plus{0.09} 	&	 19.50\plus{0.06} 	& 1.04\plus{0.17}	& 0.26\plus{0.01} 	&		0.19\plus{0.01} 	&	  0.54\plus{0.07} 	&	 -0.91\plus{0.94} 	&	2.16	&	 Jet-like  	&	 R-AGN	\\

J0924$+$1504 	&	 09:24:35.35 	&	 $+$15:04:10.00 	&	 2.71\plus{0.09} 	&		2.33\plus{0.04} 	&	 1.50\plus{0.13} 	& 0.79\plus{0.03}	&	0.65\plus{0.02} 	&	 0.51\plus{0.08} 	&	 0.48\plus{0.03} 	&	 -1.19\plus{0.27} 	&	3.0	&	 Compact	&	 	\\

J0947$+$1005 	&	 09:47:33.22 	&	 $+$10:05:08.75 	&	  1.31\plus{0.07} 	&	 1.06\plus{0.04} 	&	 1.63\plus{0.36}  	& 0.93\plus{0.08}	& 0.47\plus{0.01} 	&	 0.46\plus{0.01} 	&	 $<0.80$ 	&	 -1.15\plus{0.20} 	&	2.21	&	 Compact 	&	 	\\

J1034$+$6001 	&	 10:34:08.52 	&	 $+$60:01:53.00 	&	 17.60\plus{2.40} 	&	 5.41\plus{0.57} 	&	  3.28\plus{0.52} 	& 0.93\plus{0.13}		& 1.72\plus{0.12} 	&		0.57\plus{0.03} 	&	 2.54\plus{0.05} 	&	 -1.56\plus{0.36} 	&	1.95	&	 Triple	&	 R-AGN	\\

           	&	  10:34:08.577	&	$ +$60.01.52.159	&	$-$	&	$-$	&	$-$	& $-$	& 0.85\plus{0.07} 	&	 0.46\plus{0.03} 	&	 0.32\plus{0.06} 	&	 -1.36\plus{0.26} 	&	2.27	&	$-$	&	$-$	\\
            
           	&	 10:34:08.65 	&	 $+$60:01:50.72 	&	$-$	&	$-$	&	$-$	& $-$ & 0.51\plus{0.01} 	&	 0.39\plus{0.06} 	&	 0.20\plus{0.01} 	&	 -1.43\plus{0.22} 	&	2.7	&	$-$	&	$-$	\\
            
J1110$+$5848 	&	 11:10:15.24 	&	 $+$58:48:46.02 	&	  4.61\plus{0.05} 	&	 4.55\plus{0.03} 	&	 0.47\plus{0.17} 	& 1.12\plus{0.06}	& 1.46\plus{0.02} 	&		1.33\plus{0.01} 	&	 0.31\plus{0.03} 	&	 -1.17\plus{0.12} 	&	3.67	&	 Compact 	&		\\

J1141$+$2156 	&	 11:41:16.16 	&	 $+$21:56:21.67 	&	 0.92\plus{0.10} 	&	 0.52\plus{0.04} 	&		1.53\plus{0.24} 	& 0.34\plus{0.07}	& 0.39\plus{0.01} 	&		0.35\plus{0.01} 	&	 0.14\plus{0.03} 	&	 -0.86\plus{0.66} 	&	3.11	&	 Compact	&		\\

J1152$+$1016 	&	 11:52:45.66 	&	 $+$10:16:23.83 	&	 3.50\plus{0.18} 	&	 1.81\plus{0.06} 	&	 {2.69\plus{0.12}} 	& 1.0\plus{0.06}	& 0.44\plus{0.02} 	&		0.27\plus{0.08} 	&	 0.42\plus{0.03} 	&	 -1.06\plus{0.52} 	&	2.24	&	 Jet-like 	&	 R-AGN	\\

J1203$+$1624 	&	 12:03:00.20 	&	 $+$16:24:43.72 	&	 3.01\plus{0.04} 	&	 2.66\plus{0.02} 	&	 1.67\plus{0.08} 	& 1.04\plus{0.05}	& 0.83\plus{0.02} 	&		0.57\plus{0.01} 	&	 0.87\plus{0.03} 	&	 -1.13\plus{0.16} 	&	3.02	&	 Compact	&		\\

J1300$+$5454 	&	 13:00:38.11 	&	 $+$54:54:36.39 	&	 2.61\plus{0.34} 	&	 0.70\plus{0.07} 	&	 5.64\plus{0.8}	 	& 1.21\plus{0.17}	& 0.14\plus{0.06} 	&		0.12\plus{0.01} 	&	 0.24\plus{0.09} 	&	 -2.61\plus{1.46} 	&	2.3	&	 Jet-like  	&	 R-AGN	\\

J1316$+$4452 	&	 13:16:39.75 	&	 $+$44:52:35.03 	&	 3.06\plus{0.15} 	&	 2.42\plus{0.07} 	&	 1.02\plus{0.17}	 	& 0.71\plus{0.04}	& 0.93\plus{0.02} 	&		0.81\plus{0.01} 	&	  0.34\plus{0.03} 	&	 -1.13\plus{0.18} 	&	3.11	&	 Compact 	&		\\

J1325$+$1137 	&	 13:25:52.16 	&	 $+$11:37:09.79 	&	 3.14\plus{0.08} 	&	 2.76\plus{0.04} 	&	 1.31\plus{0.21} 	& 1.11\plus{0.05}	& 0.79\plus{0.01} 	&	 0.68\plus{0.01} 	&	 0.49\plus{0.02} 	&	 -0.97\plus{0.18} 	&	3.7	&	 Compact	&	 	\\

J1338$+$1503 	&	 13:38:06.53 	&	 $+$15:03:56.07 	&	 2.16\plus{0.04} 	&	 1.93\plus{0.02} 	&		 1.39\plus{0.17} 	& 0.97\plus{0.04}	& 0.52\plus{0.01} 	&		0.38\plus{0.01} 	&	 0.68\plus{0.04} 	&	 -1.19\plus{0.21} 	&	2.68	&	 Compact	&		\\

J1355$+$5612 	&	 13:55:16.56 	&	 $+$56:12:44.56 	&	 5.84\plus{0.09} 	&	 5.51\plus{0.05} 	&	 0.73\plus{0.16} 	& 0.97\plus{0.03}	&  1.74\plus{0.02} 	&		1.42\plus{0.01} 	&	 0.42\plus{0.02} 	&	 -0.97\plus{0.12} 	&	3.4	&	 Compact  	&	  	\\

J1410$+$2233 	&	 14:10:41.51 	&	 $+$22:33:37.10 	&	 1.87\plus{0.09}		&	 1.16\plus{0.04} 	&	 4.27\plus{0.30} 	& 0.60\plus{0.05}	& 0.38\plus{0.02} 	&		0.23\plus{0.10} 	&	 1.61\plus{0.18} 	&	 -1.9\plus{0.20} 	&	1.98	&	 Jet-like 	&	 R-AGN	\\

J1419$+$1144 	&	 14:19:43.79 	&	 $+$11:44:26.24 	&	 2.49\plus{0.04} 	&	 2.26\plus{0.02} 	&	 1.13\plus{0.08} 	& 0.85\plus{0.04}	&  0.75\plus{0.02} 	&		0.52\plus{0.01} 	&	 0.79\plus{0.07} 	&	 -1.15\plus{0.34} 	&	2.5	&	 Compact	&		\\

J1426$+$1040 	&	 14:26:14.62 	&	 $+$10:40:12.86 	&	 2.16\plus{0.06} 	&	 1.72\plus{0.03} 	&	 1.74\plus{0.20} 	& 1.10\plus{0.07}	&  0.58\plus{0.02} 	&		0.46\plus{0.01} 	&	 0.48\plus{0.08} 	&	 -1.19\plus{0.27} 	&	2.62	&	 Compact 	&	  	\\

J1426$+$1949 	&	 14:26:26.93 	&	 $+$19:49:54.22 	&	 1.81\plus{0.03} 	&	 1.66\plus{0.02} 	&	 1.42\plus{0.17} 	& 1.57\plus{0.08}	& 1.7\plus{0.01} 	&		1.66\plus{0.01} 	&	 0.19\plus{0.02} 	&	 -0.4\plus{0.06} 	&	4.25	&	 Compact   	&	 R-AGN 	\\

J1434$+$5016 	&	 14:34:11.19 	&	 $+$50:16:40.78 	&	 3.15\plus{0.03} 	&	 2.99\plus{0.02} 	&	 1.04\plus{0.09} 	& 1.14\plus{0.06}	& 0.53\plus{0.01} 	&		0.46\plus{0.01} 	&	 0.58\plus{0.03} 	&	 -1.97\plus{0.37} 	&	3.41	&	 Compact	&		\\

J1440$+$5303 	&	 14:40:38.20 	&	 $+$53:30:15.92 	&	 59.0\plus{1.20} 	&	 52.05\plus{0.65} 	&	 0.36\plus{0.05}		& 1.0\plus{0.02}	& 17.34\plus{0.45} 	&		14.81\plus{0.23} 	&	 0.12\plus{0.01} 	&	 -0.93\plus{0.02} 	&	4.49	&	 Compact	&		\\

J1450$+$0713 	&	 14:50:34.10 	&	 $+$07:31:32.48 	&	 1.81\plus{0.07} 	&	 1.26\plus{0.03} 	&	 2.40\plus{0.29} 	& 1.06\plus{0.17}	& 0.51\plus{0.04} 	&		0.48\plus{0.02} 	&	 0.25\plus{0.22} 	&	 -1.13\plus{0.45} 	&	3.38	&	 Jet-like                        	&	 R-AGN  \\

J1507$+$0029 	&	 15:07:19.93 	&	 $+$00:29:04.95 	&	 4.40\plus{0.05} 	&	 4.39\plus{0.03} 	&	 0.88\plus{0.08} 	& 1.0\plus{0.03}	& 1.14\plus{0.05} 	&		0.82\plus{0.02} 	&	 0.88\plus{0.08} 	&	 -1.17\plus{0.28} 	&	3.35	&	 Compact	&		\\

J1529$+$5616 	&	 15:29:07.45 	&	 $+$56:16:06.67 	&	 4.45\plus{0.14} 	&	 3.90\plus{0.08} 	&	 0.91\plus{0.22}		& 1.04\plus{0.05}	& 1.19\plus{0.03} 	&		0.9\plus{0.02} 	&	 0.35\plus{0.03} 	&	 -1.06\plus{0.17} 	&	2.99	&	 Compact	&		\\

J1543$+$1148 	&	 15:43:03.83 	&	 $+$11:48:38.45 	&	 2.72\plus{0.03} 	&	 2.65\plus{0.02} 	&	 0.53\plus{0.11}		& 1.30\plus{0.07}	& 0.88\plus{0.01} 	&		0.73\plus{0.01} 	&	 0.36\plus{0.14} 	&	 -0.94\plus{0.17} 	&	3.18	&	 Compact	&	 	\\

J1653$+$2349 	&	 16:53:14.99 	&	 $+$23:49:42.55 	&	 6.81\plus{0.31} 	&	 5.14\plus{0.15} 	&	  2.41\plus{0.22} 	& 0.95\plus{0.05}	& 1.28\plus{0.06} 	&		0.90\plus{0.03} 	&	 2.09\plus{0.04} 	&	 -0.35\plus{0.34} 	&	3.0	&	 Double  	&	 R-AGN	\\

           	&	 16:53:15:06 	&	 $+$23:49:42.91 	&	$-$	&	$-$	&	$-$	&	$-$	 & 0.68\plus{0.07} 	&	 0.32\plus{0.02} 	&	 0.91\plus{0.11} 	&	 -0.71\plus{0.22} 	&	2.15	&	$-$	&	$-$	\\
            
J1713$+$5729 	&	 17:13:50.31 	&	 $+$57:29:54.92 	&	 7.39\plus{0.11} 	&	 6.72\plus{0.06} 	&	 0.85\plus{0.12} 	&	1.0\plus{0.02} & 1.74\plus{0.03} 	&		1.60\plus{0.02} 	&	 0.27\plus{0.05} 	&	 -1.34\plus{0.19} 	&	3.77	&	 Compact	&	 	\\

J2304$-$0841 	&	 23:04:43.48 	&	 -08:41:08.65 	&	 N/O  	&	 N/O 	&	N/O &  N/O &  13.25\plus{0.23} 	&	 12.88\plus{0.13} 	&	 0.06\plus{0.02} 	&	  0.64\plus{0.02} 	&	5.3	&	 Compact 	&	 R-AGN 	\\
\hline																									
\end{tabular}																									
\end{table}																									
\end{landscape}															

\subsection{Spectral Indices}\label{sec:spectral_index}

Spectral indices provide a powerful diagnostic tool for identifying AGN cores, lobes, jets, and studying AGN feedback and lifecycles \citep[e.g.][]{Laing1980,Begelman1984,Hardcastle2020,Harwood2022}. Extragalactic radio sources follow the power-law spectra where the observed flux density follows $S_{\nu} \propto \nu^{\alpha}$. A flat or inverted spectrum $\alpha \geq -0.5$ is indicative of an AGN core and a steep spectrum $-0.5 \leq \alpha \lesssim -2$ is indicative of large scale emission from processes such as jets, lobes, shocks from outflows and diffuse emission from star formation. 

We measured the spectral index of radio cores identified in our C-band radio maps ($5.25 \-- 7.20\,$GHz; see Section \ref{sec:VLA_obs}). For the multi-component sources, the brightest radio peak was taken to be the core (although not necessarily a true `core'). The spectral indices are listed in Table~\ref{Tab:sources2} and are presented in Figure~\ref{fig:TB_plot}. 

Only 3/29 (J1426$+$1949, J1653$+$2349 and J2304$-$0841), i.e., 10\,per\,cent, of the QFeedS-2 targets show a flat spectrum ($\alpha \geq -0.5$). In comparison, 7/42 showed a flat spectrum (i.e., 17\,per\,cent) in the QFeedS-1 sample \citep{Jarvis2021}. Overall, only a minority of the sources (14\,per\,cent) across the whole of QFeedS show flat or inverted cores (see Fig.~\ref{fig:TB_plot}). Indeed, the majority show very steep spectra with values around $-1$. The relevance of these measurements across the combined samples is discussed further in Section~\ref{sec:QFS1_QFS2}.

\subsection{Brightness temperature}\label{sec:TB}
Brightness temperature ($T_\mathrm{B}$) is defined as the temperature of a blackbody that would produce the observed radio surface brightness (flux density per solid angle) at the given frequency \citep[e.g.][]{Morabito2022}. In practice, $T_\mathrm{B}$ serves as a proxy for radio surface brightness, whereby compact, synchrotron-emitting AGN cores (the bases of jets) often show very high $T_\mathrm{B}$ because a large flux density is concentrated into a very small solid angle. By contrast, purely star-forming regions have a physical limit on the radio surface brightness they can produce. In fact, models of compact starbursts predict an upper bound of order $T_B\sim10^5$\,K even in the most extreme systems \citep[][]{Condon1992}. Thus, any unresolved radio core with a measured $T_\mathrm{B}$ above this starburst limit is a strong indicator of AGN activity.

\begin{figure}
\centerline{\includegraphics[width=1\linewidth]{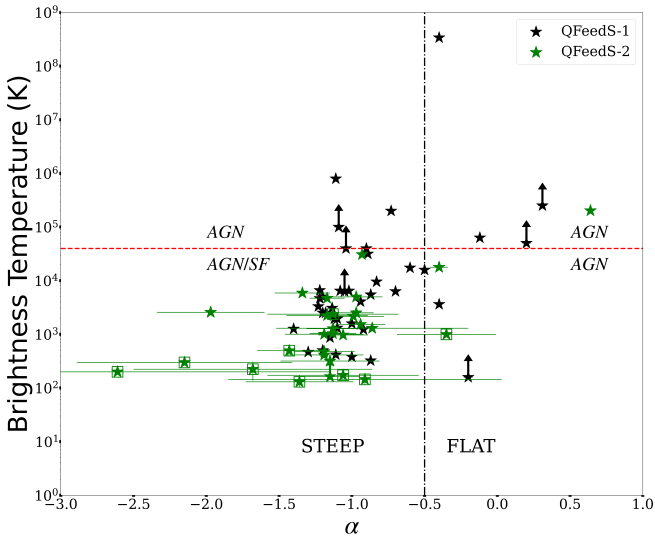}}
\caption[]{Brightness temperature ($T_\mathrm{B}$) as a function of spectral index, $\alpha$, ($\sim$5--7\,GHz) for the whole Quasar Feedback Survey sample. Black stars highlight the QFeedS-1 sample of 42 targets as presented in \cite{Jarvis2021}. The new sample of 29 targets presented in this work (QFeedS-2) is highlighted by green stars. For the multi-component sources, we plotted the component with the highest $T_\mathrm{B}$ value. The dashed red line highlights the lower limit of brightness temperature $T_\mathrm{B} = 10^{4.6}\,$K for selected radio AGN. The vertical dot-dashed line separates flat and steep sources. For sources in the bottom-left quadrant, these radio measurements can not distinguish between AGN and star formation (SF). However, the targets highlighted by green squares indicate the QFeedS-2 sources that were classified as radio-AGN based on their morphology (Section \ref{sec:radio_morph}).}
\label{fig:TB_plot}
\end{figure}

To derive the brightness temperatures, we use our measurements from the high-resolution C-band data ($\theta_\mathrm{res}\sim 0\farcs3$) and apply the standard equation  \citep[e.g.][]{Condon1982,Ulvestad2005,Njeri2025}, following:
\begin{equation}\label{TBequation}
T_\mathrm{B} = 1.22 \times 10^{12}(1 + z) \left (\frac{S_\nu}{1\,\mathrm{Jy}}\right)\left(\frac{\nu}{1\,\mathrm{GHz}}\right)^{-2}\left(\frac{\theta_\mathrm{maj}\theta_\mathrm{min}}{1\,\mathrm{mas^2}}\right)^{-1}~\mathrm{K},
\end{equation}
where $\mathrm{\theta_{maj}}$ and $\mathrm{\theta_{\mathrm min}}$ are the deconvolved major and minor axes from the elliptical Gaussian model, $S_{\nu}$ is the observed peak flux density, and $\nu$ is the observing frequency. As noted by \cite{Morabito2022}, this means $T_\mathrm{B}$ depends inversely on the observing frequency and beam size: at lower frequencies or higher resolution, the same flux density corresponds to a higher $T_\mathrm{B}$. Using our $\sim0\farcs3$ beam at $\sim6$ GHz (Equation 2), we derive $T_\mathrm{B}$ for each source. These values are presented in Figure~\ref{fig:TB_plot} and are tabulated in Table~\ref{Tab:sources2}.

To distinguish AGN from star-formation, we adopt the same criteria used by \cite{Jarvis2021}, namely $T_\mathrm{B}>10^{4.6}\,$K. \citeauthor{Jarvis2021} derive this threshold by scaling the \cite{condon1991} starburst limit to C-band. At our central frequency of $\sim6$\,GHz and redshift ($z\sim 0.1$), starbursts are not expected to exceed this value, so any brighter core is expected to be AGN-dominated. The $T_\mathrm{B}$ values for the QFeedS-2 sample range from $1.3 \times 10^{2} - 2.0 \times 10^{5}\,$K, with only one target (i.e., $\sim 3\,$per\,cent), J2304$-$0841, meeting this criterion (see Fig.~\ref{fig:TB_plot}). For comparison, about $\sim17\,$per\,cent of the QFeedS-1 cores exceed $10^{4.6}$\,K (black stars in Fig.~\ref{fig:TB_plot}), reflecting that those sources are on average more radio-luminous and host more compact, high $T_\mathrm{B}$ cores.

It is important to note that $T_\mathrm{B}$ is a very conservative radio-AGN diagnostic and a measurement below $T_\mathrm{B}<10^{4.6}$\,K does not prove the absence of a radio emission associated with AGN. Many QFeedS-2 sources have lower peak fluxes and slightly resolved low-surface brightness cores, which lead to beam dilution and thus could underestimate the $T_\mathrm{B}$ measurements. Additionally, $T_\mathrm{B}$ measurements depend strongly on the resolution and can underestimate $T_\mathrm{B}$ for extended or partially resolved sources, potentially missing the weaker AGN cores \citep[e.g.][]{Radcliffe2018,Morabito2022,Njeri2023,Njeri2025,Morabito2025b}. Indeed, high spatial resolution and high sensitive radio imaging can be more effective at isolating true AGN cores, with high $T_\mathrm{B}$, from dilution from other processes (see discussion of QFeedS-1 using higher resolution radio data in \citealt{Njeri2025}). In general, one can only securely classify a source as AGN if $T_\mathrm{B}$ exceeds the starburst limit \citep[e.g.][]{Kewley2000}. If $T_\mathrm{B}$ is below the limit, the source may still host an AGN core, but any bright AGN contribution would need confirmation via other means (see ``AGN/SF'' region in Fig.~\ref{fig:TB_plot}). For this reason, we use $T_\mathrm{B}$ as one diagnostic among several. As in previous work, we supplement this diagnostic with spectral index and morphology diagnostics to identify AGN-related radio emission \citep[][]{Jarvis2019,Njeri2025}. As discussed in the following sub-section, this multi-criteria approach, is essential for identifying radio emission associated with AGN.

\subsection{Radio-AGN identification}\label{sec:radio_AGN}
No QFeedS-2 target is classified as `radio loud' based either on their radio to [O~{\sc iii}] luminosities (applicable for all sources) or radio to optical continuum luminosities (applicable for the Type 1 sources; see Section~\ref{sec:sample_overview}). Radio-quiet quasars exhibit radio luminosities that are orders of magnitude fainter than their radio-loud counterparts, typically in the range $L_{\mathrm{1.4\,GHz}} \sim 10^{21-25}$\,W\,Hz$^{-1}$ \citep[][]{Zakamska2004,Padovani2016}. These faint signals are easily lost due to sensitivity limitations, or could be overwhelmed by star formation contributions, especially in shallow or low-resolution surveys such as NVSS or FIRST. Thus detecting and characterising radio emission from radio-quiet quasars is extremely challenging. Using sensitive high-resolution radio imaging obtained in this work, we probed for the presence of radio emission associated with AGN based on the radio morphology (Section~\ref{sec:radio_morph}), the spectral index (Section~\ref{sec:spectral_index}) and the brightness temperature (Section~\ref{sec:TB}):
\begin{itemize}
    \item[i)] Radio morphology: high resolution radio imaging can identify structures on a few kiloparsec to sub-kiloparsec scales. Elongated emission, double-lobed structures, well-collimated multiple components and linear features extending from a central source are indicative of low-power AGN jets, possibly compact due to confinement or young age, or from shocks caused by quasar winds \citep[e.g.][]{Nims2015,Baldi2021a,Njeri2025}. Nine targets (i.e., $\sim31\,$per\,cent) show such morphologies (see Fig.~\ref{fig:jet_like}) and are therefore classified as radio-AGN.

    \item[ii)] Spectral indices: our high resolution radio imaging enabled us to identify compact nuclear components of these radio-quiet quasars targets. Three targets (J1426$+$1949, J1653$+$2349 and J2304$-$0841; i.e., $\sim10\,$per\,cent) show a flat spectrum ($\alpha \geq -0.5$), which is likely indicative of an AGN `core', and are therefore classified as radio-AGN. One of these sources, J1653$+$2349, is already considered a radio-AGN based on morphology. 

    \item[iii)] Brightness temperature ($T_\mathrm{B}$): high $T_\mathrm{B}$ values are inconsistent with star formation and point to synchrotron emission from an AGN core \citep[e.g.,][]{Condon1992}. A compact unresolved radio core at sub-arcsecond ($< 0\farcs3$) scales with high $T_\mathrm{B} > 10^{4.6}\,$K is indicative of radio AGN. Only one target (J2304-0841; i.e., $\sim3\,$per\,cent) has $T_\mathrm{B} > 10^{4.6}\,$K. This target was already considered a radio-AGN based on a flat spectral index. 
\end{itemize}
Overall, 11/29 targets ($\sim38$\,per\,cent) of QFeedS-2 are classified as radio-AGN using the combined criteria of radio morphology, spectral index and brightness temperature. However, the origin of radio emission in the remaining 18/29 ($\sim62$\,per\,cent) remains uncertain, largely due to our radio data limitations and a lack of complementary multi-wavelength data, particularly in the far infrared. For example, with far infrared emission we could assess if the radio luminosity is above that expected from star formation following the far-infrared radio correlation of star-forming galaxies \citep[e.g.][also see Section~\ref{sec:AGNvsSF}]{DelMoro2013,Delhaize2017,Jarvis2021,Eberhard2025,Wang2024}. Therefore, we consider the 38\,per\,cent of radio-identified AGN in QFeedS-2 to be a lower limit of the true fraction.

\section{Discussion}\label{sec:Discussions}
The full Quasar Feedback Survey (QFeedS), comprises of 71 optically selected, $z < 0.2$ quasars (Section~\ref{sec:sample_selection}). Combining the initial QFeedS-1 sample (\citealt{Jarvis2021}) and QFeedS-2 sample (this work), this spans an extensive radio luminosity range from $L_\mathrm{1.4\,GHz} = 10 ^{22.6} \-- 10^{26.3}$\,W\,Hz$^{-1}$, encompassing nearly four orders of magnitude. The QFeedS-1 sample, covers the higher luminosities, with a radio luminosity range of $L_\mathrm{1.4\,GHz} = 10^{23.45} \-- 10^{26.30}$\,W\,Hz$^{-1}$  and a median of $\sim10^{23.8}$\,W\,Hz$^{-1}$, while the low luminosity QFeedS-2 sample has a radio luminosity range of $L_\mathrm{1.4\,GHz} = 10 ^{22.63} \-- 10^{23.44}$\,W\,Hz$^{-1}$, and a median of $ \sim 10^{23.29}$\,W\,Hz$^{-1}$. 

This combined sample enables a rare opportunity to systematically study how the radio emission in quasars evolves with radio luminosity, and obtain more insight into feedback processes in a representative quasar sample. Here we discuss the radio AGN identification across the full QFeedS sample (Section~\ref{sec:QFS1_QFS2}), before exploring more deeply the potential origins of the radio emission (Section~\ref{sec:radio_origin}), and comparing to other AGN populations (Section~\ref{sec:low_luminosity}).

\subsection{Radio-AGN identification in the full QFeedS sample}\label{sec:QFS1_QFS2}

In Figure~\ref{fig:TB_plot}, we summarise the radio-AGN selection criteria used in this work for QFeedS-2 (green stars), which are: (1) morphology (highlighted with squares; 9 targets, i.e., $31\,$per\,cent); (2) flat core spectral indices of $\alpha>-0.5$ (right of dot-dashed line; 3 targets, i.e., 10\,per\,cent); and, (3) high brightness temperature measurements (above horizontal dashed line; 1 target, i.e., 3\,per\,cent). Based on these combined criteria, 11/29 (38\,per\,cent) of sources are classified as radio-AGN in QFeedS-2. In the same figure we show QFeedS-1 (black stars), which show a higher incidence of flat spectral indices ($17^{+6.5}_{-5.0}$\,per\,cent, compared to $10^{+7.0}_{-4.4}$\,per\,cent in QFeedS-2) and high brightness temperatures ($17^{+6.5}_{-5.0}$\,per\,cent compared to $3^{+5.2}_{-2.1}$\,per\,cent in QFeedS-2). 

Following the same radio morphology criteria applied in this work (i.e., sources with double, triple, or jet-like morphologies), $55^{+7.5}_{-7.7}$\,per\,cent\footnote{{We note that 67\,per\,cent of sources in \cite{Jarvis2021} are described as having extended structures. However, excluding sources showing diffuse/irregular morphologies, only $55\,$per\,cent of the sample is classified as radio-AGN based on morphology alone.} We also note that \cite{Jarvis2021} quoted a lower fraction of $\sim$16\,per\,cent based on much stricter morphology criteria; however, here we apply a looser criteria, following \cite{Njeri2025}, and note that sources not classified as radio-AGN based on morphology in \cite{Jarvis2021}, are identified as radio-AGN based on other diagnostics, reassuring us that this is a reasonable approach.} of targets in the QFeedS-1 sample are classified as radio-AGN based on radio morphology alone, compared to $31^{+9.1}_{-7.8}\,$per\,cent in QFeedS-2. These fractions of identified AGN-like morphologies are broadly consistent with results for comparable samples at similar luminosities and redshifts (see Section~\ref{sec:low_luminosity}). The QFeedS-1 sources span radio sizes of $\sim 0.1 \-- 67\,$kpc \citep[][]{Jarvis2021}, while QFeedS-2 sizes range $\sim 0.1 \-- 20\,$kpc (see Fig.~\ref{fig:luminosity_plot}). Most objects in both samples have radio emission confined to host-galaxy scales (i.e., $\leq$ a few kpc), although a minority of QFeedS-1 sources {and one QFeedS-2 target} exhibit substantially larger sizes (see Section~\ref{sec:linearsizes}). Importantly, this size distribution masks a wide variety of radio morphologies, and even though the targets were selected to have similar [O\,~{\sc iii}] luminosities, we observe compact unresolved cores, small jets, hotspots and diffuse structures across the whole sample (Figs.~\ref{fig:compact_sources}--~\ref{fig:jet_like}).


Using {the consistent and combined} three criteria of radio morphology, brightness temperature and spectral indices, we identify 28/42 ($67^{+6.8}_{-7.6}$\,per\,cent) and 11/29 ($38^{+9.3}_{-8.5}$\,per\,cent) sources as radio-AGN across QFeedS-1 and QFeedS-2, respectively. However, it is already known that additional information can identify more radio-AGN, and these are likely lower limits of the true fractions. For example, using sufficiently high quality far-infrared measurements, that were available for a sub-set of the sample, \cite{Jarvis2021} found another $12\,$per\,cent of QFeedS-1 quasars to be classified as radio-AGN solely based on radio-excess measurements (following e.g., \citealt{Condon1992,Helou1985,DelMoro2013,Marvil2015}), i.e., where their radio morphology, spectral indices and brightness temperature criteria were inadequate \citep[see Fig.~6 in][]{Jarvis2021}. Sufficient quality far-infrared data are not available for QFeedS-2 for a comparable analysis; however, we briefly investigate the likely contribution of star formation in Section~\ref{sec:AGNvsSF}. Overall, this highlights the need for high resolution multi-band radio imaging combined with multi-wavelength information, particularly infrared, as we push further into the radio-quiet regime (also see e.g., \citealt{Hardcastle2025}).  

\begin{figure}
\centerline{\includegraphics[width=1\linewidth]{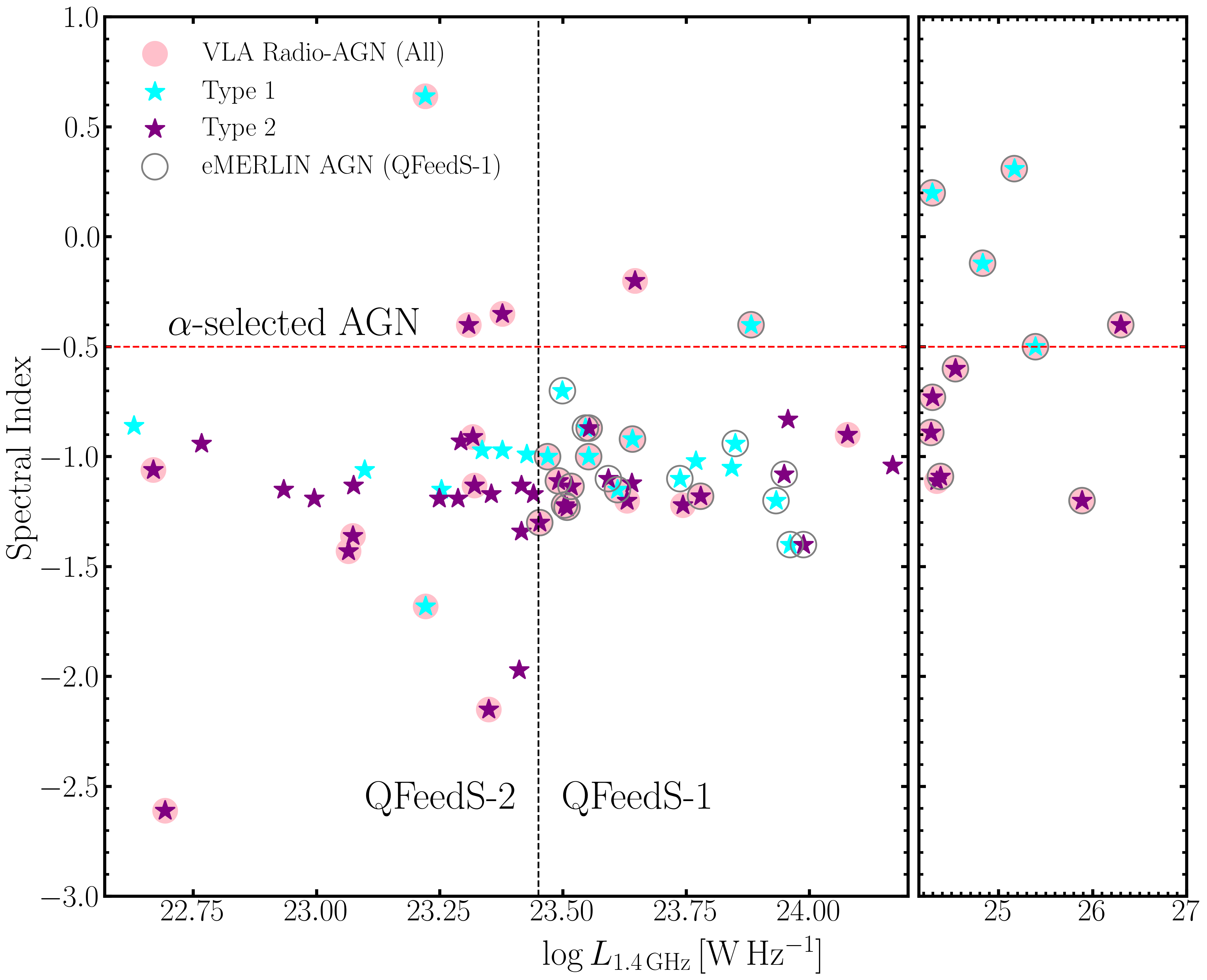}}
\caption[]{Core spectral index ($\sim$5---7\,GHz) versus $1.4\,$GHz (NVSS) radio luminosity for the QFeedS sample. The vertical dashed line separates QFeedS-1 from QFeedS-2 sources. The plot is split into two panels, covering different luminosity ranges for more clarity over the wide parameter space. Cyan stars denote Type 1 quasars and purple stars denote Type 2 quasars. Solid circles highlight quasars identified as radio-AGN based on the criteria used in this work. Both sub-samples are dominated by steep spectrum emission; however, flatter-spectrum sources are more common at higher luminosities (QFeedS-1), for the Type 1 quasars, consistent with an increased contribution from compact AGN-related cores. Open circles mark QFeedS-1 sources identified as radio-AGN based on follow-up higher resolution \emerlin{} radio imaging.} 
\label{fig:Lumin_All}
\end{figure}


In Figure~\ref{fig:Lumin_All} we present a summary of the radio-AGN identification approaches across the whole QFeedS sample, in the form of spectral index versus radio luminosity diagram\footnote{One QFeedS-1 target (J1016+0028) is absent from Figure~\ref{fig:Lumin_All} due to the lack of a core spectral index measurement, as it was undetected in the C-band data \citep[][]{Jarvis2021}. Nevertheless, we note that the source was classified as a radio-AGN based on its morphology.}. Unsurprisingly, for the moderate radio luminosities covered by the QFeedS quasars, only a minority of the sample have strong radio-AGN cores based on our VLA data, as indicated by flat spectral indices or high brightness temperatures. However, as expected, this fraction increases as a function of radio luminosity. On the other hand, radio morphology results in a higher fraction of radio-AGN being identified in QFeedS, which is only possible thanks to our high resolution radio maps. 

Higher-resolution follow-up of QFeedS-2 is needed because a substantial fraction of the radio power in these quasars appears to arise on sub-kiloparsec scales (i.e., $\leq 1\,$kpc to a few hundred parcecs; e.g., \citeauthor{Njeri2025} \citeyear{Njeri2025}). Compact cores, frustrated jets, hotspot knots and shocked regions can all be concentrated within these small spatial extents and therefore remain unresolved or beam-diluted in our VLA maps. Observations with finer angular resolution (e.g., \emerlin{} at $\leq100\,$pc scales or VLBI at parsec scales) both increase brightness temperature sensitivity and resolve compact AGN features, often revealing AGN signatures missed at lower resolution. For example, 6\,GHz \emerlin{} imaging of the QFeedS-1 sample at $\leq100\,$pc resolution increases the fraction of sources confidently classified as radio-AGN to nearly $90\,$per\,cent (see open circles in Fig.~\ref{fig:Lumin_All}), compared with $60\,$per\,cent identified from low resolution data \citep[][]{Njeri2025}. Higher-resolution observations of QFeedS-2 are likely to uncover additional compact AGN components and to provide a more complete census of radio-AGN across the full QFeedS luminosity range.


\subsubsection{Type 1 versus Type 2 quasars}

Since our sample is composed of $\sim 35\,$per\,cent Type 1 and $\sim 65\,$per\,cent Type 2 quasars, we briefly investigate the radio properties across the two classes of AGN in our QFeedS sample. The division of sources into optical Type 1 (cyan stars) and Type 2 (purple stars) in Figure~\ref{fig:Lumin_All} shows no strong offset in spectral index distribution between the two classes. The exception is for the highest radio luminosities ($\gtrsim$10$^{24}$\,W\,Hz$^{-1}$), where the Type 1 systematically have flatter spectral indices. We also note that the extended AGN-like morphologies are more frequently identified in Type 2 sources, with $\sim 52\,$per\,cent compared to only $\sim 32\,$per\,cent for the Type 1. These trends, albeit weak, may reflect observational biases in detecting compact cores against different host galaxy environments \citep[][]{Zakamska2016,Villar-Marti2021}, and are broadly consistent with an orientation-based model. Specifically, for a more directly face on view for Type~1 quasars, we would expect the radio emission to be more dominated by the radio core (with compact, flat spectrum emission).

\subsection{Potential origin of the Radio Emission}\label{sec:radio_origin}

For $\sim$60\% of our QFeedS-2 sample, which covers the modest radio luminosities ($L_\mathrm{1.4\,GHz} \sim 10^{22.63} \-- 10^{23.45}\,$W\,Hz$^{-1}$), the origin of the radio emission remains ambiguous. Although we lack the diagnostics to definitively rule out star formation in individual targets, here we briefly assess the likelihood that star formation is a dominantly contributing factor across the QFeedS sample as a whole.

\subsubsection{Star formation versus radio-quiet AGN}\label{sec:AGNvsSF}

Single-band radio conversions have been used to convert radio luminosity into star formation rates for star-forming galaxies (e.g., \citealt{Yun2001,Murphy2011,Kennicutt2012}). Compact, low-power jets, wind–ISM shocks, and coronal synchrotron all produce steep-spectrum emission that masquerades as star formation when using single-band radio calibrations. Therefore, when AGN may be present these calibrations would only provide upper limits on star-formation rate measurements \citep[e.g.][]{Yun2001,DelMoro2013,Padovani2017}. Based on large sample studies, at $L_\mathrm{1.4\,GHz} \approx 10^{23}\,$W\,Hz$^{-1}$, the radio emission in quasars increasingly overlaps the regime of star-forming galaxies, with non star-forming processes easily supplying a large fraction of the radio power. We place our QFeedS sample into the context of such studies, assessing the likely contribution from star formation to the sample as a whole. 


\begin{figure}
\centerline{\includegraphics[width=1\linewidth]{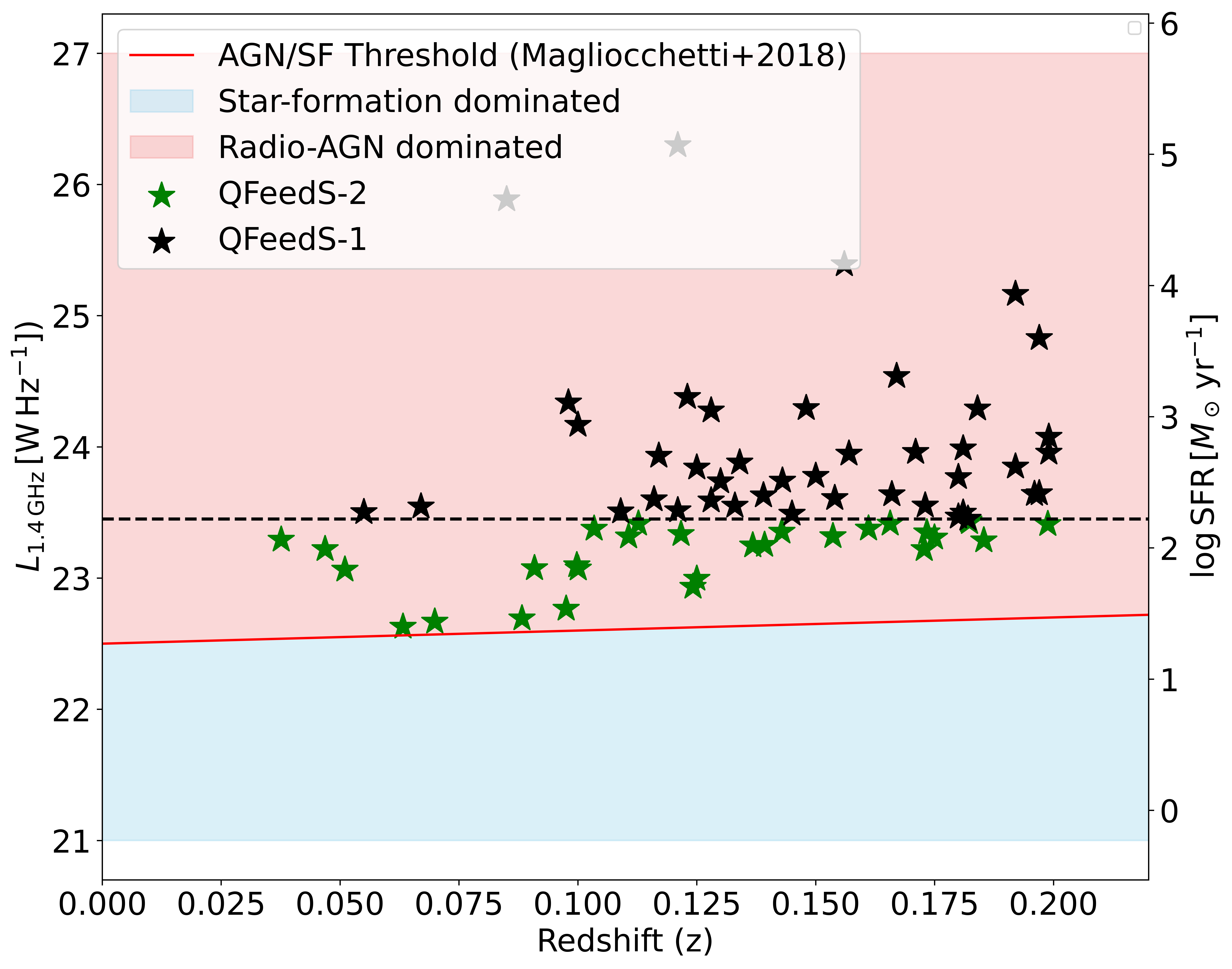}}
\caption[]{Radio luminosity at $1.4\,$GHz (NVSS) versus redshift for QFeedS quasars and the corresponding star-formation rates as derived from the standard star-formation rate calibrators, is shown on the right axis (see Section~\ref{sec:AGNvsSF}). Black stars represent QFeedS-1 and green stars represent QFeedS-2. The solid red line shows the AGN / star formation threshold for the galaxy population from \cite{Magliocchetti2018}, separating the star-formation dominated (blue shaded) and radio-AGN dominated (pink shaded) regimes. The dashed black line marks the QFeedS-1/QFeedS-2 selection limit. Most QFeedS quasars, particularly the luminous QFeedS-1, lie firmly above the star-forming locus, indicating that their radio emission is most likely dominated by AGN processes rather than star formation. The lower-luminosity QFeedS-2 source scatter closer to the threshold but still occupy predominantly AGN-dominated space.}
\label{fig:AGN_SF}
\end{figure}

In Figure~\ref{fig:AGN_SF} we present radio luminosity as a function of redshift for the combined QFeedS samples. As a rough estimate, on the right-hand axis, we convert these radio luminosities into star-formation rates, assuming a 100\% contribution from star formation to the radio emission and averaging multiple conversion calibrations \citep[][]{Yun2001,Murphy2011,Kennicutt2012}. The inferred star formation rates range from $\sim$30\,M$_{\odot}$\,yr$^{-1}$ to $\sim$10$^{5}$\,M$_{\odot}$\,yr$^{-1}$. For context, the typical star-formation rates of radiatively-luminous AGN at $z\sim0$ are found to be 0.1--50\,M$_{\odot}$\,yr$^{-1}$ (e.g., \citealt{Zakamska2016,Shimizu2017,Jarvis2019,Jackson2020}). Using this conversion, we find that 67/71 (94\,per\,cent) of the inferred star-formation rates are greater than 50\,M$_{\odot}$\,yr$^{-1}$, which is an uncomfortably high value for these very low redshift quasars. Therefore, it is most likely that there is a significant contribution to the radio emission from AGN for the majority of the sample. Indeed, these standard single-band calibrations yield implausibly high star-formation rates in AGN hosts due to contamination from compact synchrotron processes. The inflated inferred star-formation rate values we measure are consistent with previous works that report radio-based star-formation rates in quasars and radio-quiet AGN tend to overshoot far-infrared or spectral energy distribution based estimates due to contamination from low-power jets, coronal activity, and wind-driven shocks \citep[e.g.][]{DelMoro2013,Behar2015,Zakamska2016,Delvecchio2017,Padovani2017,Smolcic2017,Bernhard2022,Liao2024,CalistroRivera2024}. 

We further compare to the redshift-dependent separation between star-formation dominated and AGN-dominated radio luminosity regimes as derived by \cite{Magliocchetti2018}, which is represented by the solid red curve in Figure~\ref{fig:AGN_SF}. The QFeedS-1 quasars, at higher radio luminosities, sit firmly within the AGN-dominated region, consistent with their high fraction of radio AGN identification from other methods \citep[][]{Jarvis2021,Njeri2025}. In contrast, QFeedS-2 sources cluster closer to the threshold, with some appearing near the boundary between the two regimes; however, still within the AGN-dominated regime. 

Although we can not rule out a significant contribution from star formation in all individual sources, we conclude that the QFeedS quasars, which are preselected as luminous [O{\sc iii}] and radio detected in FIRST, are likely to have their radio emission predominantly associated with AGN related processes. This supports the idea of an increasing body of work implying a significant contribution of AGN-related emission to the radio output of radio-quiet quasars (e.g., \citealt{Zakamska2016,White2017,Panessa2019}).




\subsubsection{Jets and winds in radio-quiet quasars}

At the high-luminosity end (QFeedS-1), the radio emission is often clearly AGN-dominated, with compact jet-like morphologies, flat to moderate spectral indices and high brightness temperatures ($T_\mathrm{B}\geq 10^{4.6}\,$K), pointing to relativistic jet activity as the primary driver, although shocks due to quasar-driven winds can not be ruled out in all cases (\citealt{Jarvis2021}). These sources align with classical radio-intermediate and radio-loud quasars, where radio-AGN activity is well established \citep[][]{Jarvis2021,Njeri2025}. 

Interestingly, the majority of QFeedS sources show very steep spectral indices of $\approx$-1 (Fig.~\ref{fig:Lumin_All}). Such a steep spectral slope could be an indication of shocks due to quasar driven winds and/or low power radio jets interacting with the ISM (e.g., \citealt{Odea1998,Nims2015,Fawcett2025}). Indeed, spatially-resolved multi-wavelength analysis of subsets of the QFeedS-1 sample show a high incidence of interaction with the ISM (e.g., \citealt{Jarvis2019,Girdhar2022,Girdhar2024}). However, distinguishing between jets and winds as the origin of the radio emission in AGN-dominated systems, is very challenging. Whilst spatially-resolved polarisation-sensitive observations may provide some diagnostics, sensitivity-limited imaging alone can be expected to remain ambiguous based on simulations (e.g., \citealt{Meenaksh2024}). Nonetheless, in sources where `AGN-like' (extended) morphologies are observed in QFeedS-2 (see Fig.~\ref{fig:jet_like}), we discuss their morphologies below, in the context of previous observations. 


The jet-like feature in J0841$+$0101 appears to be bipolar with an end-to-end extent of nearly 20\,kpc. It is important to note the filamentary nature of the jet and its western end lying close to the neighboring galaxy (but not coincident with its optical center). On the eastern end, only a single radio feature is observed, and only in the L-band data. The jet-like features in J1300$+$5454, J1034$+$6001, J0827$+$2233, and J1152$+$1016 display morphologies similar to those observed in jetted radio-loud AGN except for the fact that these jets are much smaller in extent. J1300$+$5454, J1034$+$6001 and J0827$+$2233 appear to show terminal hotspots like FRII radio galaxies, while the jet in J1152$+$1016 shows an S-shaped FRI-like morphology. The more compact extensions in J0232$-$0811 and J1410$+$2233 resemble similar extensions observed in Seyfert galaxies and other radio-quiet quasars \citep[e.g.,][]{Rao2023, Silpa2023}. The radio structures in J1450$+$0713 and J1653$+$2349 resemble a core-hotspot-like structure but the counter-hotspot is not clearly visible. For all radio sources that are compact (Fig.~\ref{fig:compact_sources}), it is possible that jets exist on even smaller sub-kpc scales that would need sub-arcsecond or VLBI observations to resolve \citep[e.g.,][]{Kharb2019, Kharb2021,Njeri2025}. 


Taken together, these diverse jet-like morphologies illustrate that even within radio-quiet quasars, small-scale jets and compact outflows significantly impact the observed radio emission.  Although much smaller than classical FR I/II radio galaxies (Fig.~\ref{fig:luminosity_plot}), the presence of hotspots, bends, and filamentary features strongly suggests jet–ISM interactions, rather than star-formation origins \citep[e.g.][]{Baldi2018,Jarvis2019,Rosario2020}. At the same time, the coexistence of compact, steep-spectrum sources without obvious extended features highlights that multiple mechanisms, ranging from low-power or frustrated jets, coronal emission and outflow-driven shocks, likely contribute across the QFeedS population \citep[e.g.][]{Laor2008,Panessa2019,Silpa2023}. Thus, the radio emission in these quasars is unlikely to be ascribed to a single origin. Instead, it reflects a continuum of AGN-driven processes operating over different spatial and energetic scales, potentially intertwined with the dynamical state of the host galaxy \citep[e.g.][]{Padovani2017,Hardcastle2020}. 


%

\subsection{Comparison to other AGN samples}\label{sec:low_luminosity}

Compact, low-luminosity radio-AGN often exhibit steep spectra and show jet-like structures on kiloparsec (or smaller) scales, if spatially resolved \cite[see review by][]{Baldi2023}. For QFeedS-2, 26/29 sources show steep spectra $\alpha <-0.5$ (Fig.~\ref{fig:TB_plot}), consistent with optically thin synchrotron emission. However, since nearly all remain compact on kiloparsec scales, this compactness likely reflects low-power or frustrated jets, or shocks from quasar-driven winds, confined to small spatial scales \citep[e.g.][Section~\ref{sec:AGNvsSF}]{Baldi2015,Silpa2020,Baldi2023,Njeri2025}. As shown in Figure~\ref{fig:luminosity_plot}, the QFeedS quasars have radio sizes overlapping Compact Steep Spectrum (CSS) and low-size end of FRI sources. Strikingly, $\sim 80$ per cent of the QFeedS-2 sources cluster in the same region as FR0 galaxies (Fig.~\ref{fig:luminosity_plot}; yellow shaded region), which is traditionally associated with low-accretion radio galaxies (\citealt{Baldi2023}). However, the high {\sc [O iii]} luminosities of our sample indicate radiatively efficient quasars, illustrating a decoupling between radiative accretion power and observed radio luminosities and sizes.

The compact, low-luminosity objects in QFeedS may also represent early evolutionary stages of Gigahertz Peaked Spectrum (GPS) and Compact Steep Spectrum (CSS) sources, in which jets remain confined within their host galaxies \citep[particularly for the QFeedS-1 sample; see more discussion in][]{Njeri2025}. \cite{Kunert-Bajraszewska2010} proposed that such low-luminosity compact systems could be young radio sources that later evolve into CSS of FRI galaxies, while \citet{Slob2022} similarly identified compact peaked-spectrum sources in LOFAR consistent with either young or confined jets. 



Recent low-redshift studies provide additional insights into the low-luminosity compact radio AGN population. For instance, \cite{Pierce2020} analysed intermediate-power high excitation radio galaxies (HERGs; $22.5 < \mathrm{log}\,L_\mathrm{1.4\,GHz}< 24.0\,$W\,Hz$^{-1}$) at $z<0.1$ and found a lower incidence of extended radio morphologies, steep-spectrum compact sources in the less powerful half of their sample, compared to classical radio galaxies (see crosses in Fig.~\ref{fig:luminosity_plot}). This may suggest that the low-power AGN can persist in relatively undisturbed systems, contrasting with the merger-driven environments of more powerful radio AGN. The QFeedS-2 sample, while overlapping in radio luminosity with these HERGS, they differ in that they are preselected as luminous {\sc[O iii]} quasars, pointing to strong radiative accretion event with radio output that is relatively compact and weak. 

Similarly, \cite{Chilufya2024} studied LOFAR-selected low-luminosity ($L_\mathrm{150\,MHz} < 10^{25}\,$W\,Hz$^{-1}$) radio-AGN at $0.03 < z < 0.1$ and reported that most sources are unresolved at sub-kpc scales, and often have steep-spectrum, strongly resembling CSS, GPS and FR0-like systems (see plus symbols in Fig.~\ref{fig:luminosity_plot}). This is directly comparable to the QFeedS-2 sources, which also occupy the FR0 region in the luminosity-size plane. The strong overlap highlights that compact, confined radio sources are common manifestation of low-radio power AGN activity across both radio-selected and optically-selected samples. These comparisons underscore that the QFeedS sample maps onto the broader landscape of compact radio-AGN populations. While \citeauthor{Pierce2020} highlights the role of host morphology, \citeauthor{Chilufya2024} emphasizes the compactness and FR0-like character, with the QFeedS connecting these with the quasar regime of high radiative accretion, demonstrating how luminous quasars can host radio emission that mimics low-power radio galaxies. Overall, the QFeedS sample traces a continuum, from luminous quasars with extended jet-like structures, to compact, low-power systems whose radio emission resembles FR0s, thereby mapping the diversity of radio AGN activity across nearly four orders of magnitude in radio luminosity.

\section{Conclusion}\label{sec:conclusion}
We present a new sample of 29 $z<0.2$ quasars (QFeedS-2) with $L_{\rm 1.4GHz} < 10^{23.45}$\,W\,Hz$^{-1}$, as an extension to our Quasar Feedback Survey (QFeedS; see Fig.~\ref{fig:selection}). In this work, we study the gigahertz radio emission using new 1.4\,GHz (L-band) and 6\,GHz (C-band) data from the VLA, which cover a spatial resolution of $\sim1$--3\,kpc. Our main conclusions are:

\begin{itemize}
    \item For the vast majority of sources ($\sim80\,$per\,cent), our L-band VLA imaging recovers the FIRST 1.4\,GHz flux, indicating that the high-resolution maps capture the bulk of the total gigahertz emission.
    \item The gigahertz radio emission is extended on $\sim$0.1–20\,kpc scales. $31^{+9.1}_{-7.8}$\,per\,cent of the sample show extended radio structures indicative of AGN-driven outflows (i.e., jet or wind-driven; Fig.~\ref{fig:jet_like}). This is compared to the $55^{+7.5}_{-7.7}$\,per\,cent AGN-like extended radio structures for the more radio luminous QFeedS-1 quasars, which are also more likely to exhibit the most extended, $\gtrsim$10\,kpc scale emission (Fig.~\ref{fig:luminosity_plot}). 
     \item Nine of the quasars display collimated morphologies indicative of AGN (e.g., bipolar features, hot spots, S-shaped jets), although typically smaller in scale than classical FR I/II galaxies (Fig.~\ref{fig:jet_like}; \citealt{Jarvis2021,Njeri2025}). Others (20/29) remain compact (Fig.~\ref{fig:compact_sources}) and may host sub-kpc jets, thus requiring higher resolution radio imaging to confirm (e.g., \citealt{Njeri2025}). 
    \item The majority (90\,per\,cent) of the sample show very steep spectral indices across 5--7\,GHz with $\alpha\lesssim$-1, potentially indicating a strong contribution from interactions between the AGN and ISM. Only three QFeedS-2 sources show radio emission dominated by AGN-core like emission, based on flat spectral indices or high brightness temperatures (see Fig.~\ref{fig:TB_plot}). 
    \item Combining spectral indices, brightness temperatures, and morphology criteria, 11/29 of the targets (i.e., $38^{+9.3}_{-8.5}$\,per\,cent) are confirmed as radio-AGN. This is compared to 28/42 ($67^{+6.8}_{-7.6}$\,per\,cent) following the same criteria, in the more radio luminous QFeedS-1 sample (Section~\ref{sec:QFS1_QFS2}; Fig.~\ref{fig:Lumin_All}). However, the total fraction of QFeedS-1 confirmed as radio-AGN is 86\,per\,cent based on additional data from \emerlin{} and far-infrared measurements; which are not currently available for QFeedS-2 (Fig.~\ref{fig:Lumin_All}; \citealt{Njeri2025}). Hence, 38\,per\,cent is considered a lower limit of radio-AGN in QFeedS-2. Overall, $66^{+5.8}_{-5.9}\,$per\,cent of the full QFeedS sample can be attributed to a radio-AGN origin.
    \item For the sources that are not identified as a radio-AGN ($\sim 62^{+8.5}_{-9.3}\,$per\,cent) in QFeedS-2, it is not possible to rule out a strong contribution of star formation to the radio emission. However, across the sample as a whole, the radio emission is unlikely to be star-formation dominated, due to the compact emission and very high inferred star formation rates at the radio luminosity range of the sample (Fig.~\ref{fig:AGN_SF}).
    \item Across the whole of QFeedS, the Type 2 sources display a slightly higher incidence of extended radio structures (52\,per\,cent compared to 32\,per\,cent). Furthermore, at the highest radio luminosities, the Type 1 sources are more likely to be core-dominated (with high brightness temperatures and flat spectral indices), compared to the Type 2 sources (Fig.~\ref{fig:Lumin_All}). These observations are broadly consistent with an orientation based model of AGN types. 
    \item With $\sim 80\,$per\,cent of our QFeedS-2 sample falling in the FR0 region (Fig.~\ref{fig:luminosity_plot}), their radio properties would, at face value, be consistent with with the classical FR0 population \citep[][]{Baldi2015,Baldi2023}. However, by construction, our sample is selected to have very high [O~{\sc iii}] luminosities ($L_\mathrm{[O\,III]} \geq 10^{42}\,$erg\,s$^{-1}$), unambiguously pointing to radiatively efficient, high-accretion quasars. This highlights that FR0-like radio morphologies do not necessarily imply radiatively inefficient accretion (Section~\ref{sec:low_luminosity}). 
\end{itemize}
The Quasar Feedback Survey demonstrates that quasars, even those traditionally labeled radio quiet, are not radio silent but span a continuous spectrum of radio activity and feedback modes; from powerful jets in luminous systems to weaker, radio emission in lower-luminosity systems, likely associated with weak radio jets or outflows from quasar winds. These results emphasize that robust identification of radio-AGN in the radio-quiet regime requires high-resolution, multi-band radio data, ideally combined with multi-wavelength constraints.

Our QFeedS-1 sample, which represents the higher radio luminosities ($L_\mathrm{1.4\,GHz}\sim 10^{23.45\--26.30}\,$W\,Hz$^{-1}$), has already revealed that compact, AGN-driven jets and/or winds strongly couple to the ISM, driving significant multi-phase outflows \citep[e.g.][]{Jarvis2021,Girdhar2022,Silpa2022,Girdhar2024}. However, these objects align with the classical `radio-intermediate' population where compact but powerful radio activity is becoming a clear signature of feedback activity (\citealt{Harrison2024}). Follow-up observations of QFeedS-2 will assess the driving mechanisms and impact on the ISM of lower radio power systems. Thus, the combined QFeedS quasars, covering four orders of magnitude in radio luminosity, represent a compelling laboratory for bridging radio-quiet and radio-loud AGN regimes, uncovering the diverse physical channels through which quasars generate radio emission, and how they can influence their galactic environments.

More broadly, current and future radio facilities (e.g., LOFAR-ILT, SKA and ngVLA; \citealt{SKA2009,ngVLA2018,Morabito2025}) will be transformative in understanding radio emission and related feedback processes across the whole AGN population. High-frequency arrays will unveil low power milliarcsecond jets unidentified in current radio maps, while low-frequency, high-sensitivity observations will capture aged, diffuse plasma, which is key to understanding the full lifecycle of AGN outflows. Integrating such radio diagnostics with X-ray observations, sub-mm interferometry and integral-field spectroscopy, will allow us to directly observe how compact jets or quasar-driven winds impact the surrounding ISM and assess their role in feedback processes in large, and representative, AGN samples.

\section*{Acknowledgements}
We thank the referee for their prompt and constructive report. AN, CMH, and VF acknowledge funding from an United Kingdom Research and Innovation grant (code: MR/V022830/1). PK acknowledges the support of the Department of Atomic Energy, Government of India, under the project 12-R\&D-TFR-5.02-0700. DMA thanks the Science Technology Facilities Council (STFC) for support through the Durham consolidated grant (ST/T000244/1). SS acknowledges funding from ANID through Fondecyt Postdoctorado (project code 3250762), Millenium Nucleus NCN23\_002 (TITANs), and Comité Mixto ESO-Chile. We acknowledge the use of the ilifu cloud computing facility - \href{www.ilifu.ac.za}{www.ilifu.ac.za}, a partnership between the University of Cape Town, the University of the Western Cape, the University of Stellenbosch, Sol Plaatje University, the Cape Peninsula University of Technology and the South African Radio Astronomy Observatory.


\section*{Data Availability}
The data underlying this article were accessed from the NRAO Science Data Archive (\href{https://data.nrao.edu/portal/}{https://data.nrao.edu/portal/}) using the proposal id: 23A-214. The spectral index, C-band and L-band maps produced in this work, including the figures for all 29 targets showing the radio contours overlaid on the optical images (in the style of Figure~\ref{fig:compact_sources}) are all available at \url{https://doi.org/10.25405/data.ncl.c.8110946}. All public data related to the Quasar Feedback Survey can also be located via our website: \url{https://blogs.ncl.ac.uk/quasarfeedbacksurvey/data/}.



\bibliographystyle{mnras}
\bibliography{QFS} 

@ARTICLE{Bernhard2022,
       author = {{Bernhard}, E. and {Tadhunter}, C.~N. and {Pierce}, J.~C.~S. and {Dicken}, D. and {Mullaney}, J.~R. and {Morganti}, R. and {Ramos Almeida}, C. and {Daddi}, E.},
        title = "{Quantifying the cool ISM in radio AGNs: evidence for late-time retriggering by galaxy mergers and interactions}",
      journal = {\mnras},
         year = 2022,
        month = may,
       volume = {512},
       number = {1},
        pages = {86-103},
          doi = {10.1093/mnras/stac474},
archivePrefix = {arXiv},
       eprint = {2202.08276},
 primaryClass = {astro-ph.GA},
       adsurl = {https://ui.adsabs.harvard.edu/abs/2022MNRAS.512...86B}
}

@ARTICLE{Ilha2025,
       author = {{Ilha}, Gabriele S. and {Harrison}, C.~M. and {Mainieri}, V. and {Njeri}, Ann and {Bertola}, E. and {Bischetti}, M. and {Circosta}, C. and {Cicone}, C. and {Cresci}, G. and {Fawcett}, V.~A. and {Georgakakis}, A. and {Kakkad}, D. and {Lamperti}, I. and {Marconi}, A. and {Perna}, M. and {Puglisi}, A. and {Rosario}, D. and {Tozzi}, G. and {Vignali}, C. and {Zamorani}, G.},
        title = "{Connecting outflows with radio emission in active galactic nuclei at cosmic noon}",
      journal = {arXiv e-prints},
         year = 2025,
        month = oct,
          eid = {arXiv:2510.14152},
        pages = {arXiv:2510.14152},
          doi = {10.48550/arXiv.2510.14152},
archivePrefix = {arXiv},
       eprint = {2510.14152},
 primaryClass = {astro-ph.GA},
       adsurl = {https://ui.adsabs.harvard.edu/abs/2025arXiv251014152I}
}

@ARTICLE{Roy2025,
       author = {{Roy}, Namrata and {Heckman}, Timothy and {Henry}, Alaina},
        title = "{Mapping Jet-Gas Coupling and energetic ionized outflows in High-Redshift Radio Galaxies with JWST/NIRSpec}",
      journal = {arXiv e-prints},
         year = 2025,
        month = aug,
          eid = {arXiv:2508.06707},
        pages = {arXiv:2508.06707},
          doi = {10.48550/arXiv.2508.06707},
archivePrefix = {arXiv},
       eprint = {2508.06707},
 primaryClass = {astro-ph.GA},
       adsurl = {https://ui.adsabs.harvard.edu/abs/2025arXiv250806707R}
}

@ARTICLE{Hardcastle2025,
       author = {{Hardcastle}, M.~J. and {Pierce}, J.~C.~S. and {Duncan}, K.~J. and {G{\"u}rkan}, G. and {Gong}, Y. and {Horton}, M.~A. and {Mingo}, B. and {R{\"o}ttgering}, H.~J.~A. and {Smith}, D.~J.~B.},
        title = "{Radio AGN selection in LoTSS DR2}",
      journal = {\mnras},
         year = 2025,
        month = may,
       volume = {539},
       number = {2},
        pages = {1856-1878},
          doi = {10.1093/mnras/staf622},
archivePrefix = {arXiv},
       eprint = {2504.09303},
 primaryClass = {astro-ph.GA},
       adsurl = {https://ui.adsabs.harvard.edu/abs/2025MNRAS.539.1856H}
}

@ARTICLE{Mingo2022,
       author = {{Mingo}, B. and {Croston}, J.~H. and {Best}, P.~N. and {Duncan}, K.~J. and {Hardcastle}, M.~J. and {Kondapally}, R. and {Prandoni}, I. and {Sabater}, J. and {Shimwell}, T.~W. and {Williams}, W.~L. and {Baldi}, R.~D. and {Bonato}, M. and {Bondi}, M. and {Dabhade}, P. and {G{\"u}rkan}, G. and {Ineson}, J. and {Magliocchetti}, M. and {Miley}, G. and {Pierce}, J.~C.~S. and {R{\"o}ttgering}, H.~J.~A.},
        title = "{Accretion mode versus radio morphology in the LOFAR Deep Fields}",
      journal = {\mnras},
         year = 2022,
        month = apr,
       volume = {511},
       number = {3},
        pages = {3250-3271},
          doi = {10.1093/mnras/stac140},
archivePrefix = {arXiv},
       eprint = {2201.04433},
 primaryClass = {astro-ph.GA},
       adsurl = {https://ui.adsabs.harvard.edu/abs/2022MNRAS.511.3250M}
}

@ARTICLE{Ulivi2024,
       author = {{Ulivi}, L. and {Venturi}, G. and {Cresci}, G. and {Marconi}, A. and {Marconcini}, C. and {Amiri}, A. and {Belfiore}, F. and {Bertola}, E. and {Carniani}, S. and {D'Amato}, Q. and {Di Teodoro}, E. and {Ginolfi}, M. and {Girdhar}, A. and {Harrison}, C. and {Maiolino}, R. and {Mannucci}, F. and {Mingozzi}, M. and {Perna}, M. and {Scialpi}, M. and {Tomicic}, N. and {Tozzi}, G. and {Treister}, E.},
        title = "{Feedback and ionized gas outflows in four low-radio power AGN at z {\ensuremath{\sim}} 0.15}",
      journal = {\aap},
         year = 2024,
        month = may,
       volume = {685},
          eid = {A122},
        pages = {A122},
          doi = {10.1051/0004-6361/202347436},
archivePrefix = {arXiv},
       eprint = {2403.01258},
 primaryClass = {astro-ph.GA},
       adsurl = {https://ui.adsabs.harvard.edu/abs/2024A&A...685A.122U}
}

@ARTICLE{Morganti2023,
       author = {{Morganti}, Raffaella and {Murthy}, Suma and {Guillard}, Pierre and {Oosterloo}, Tom and {Garcia-Burillo}, Santiago},
        title = "{Young Radio Sources Expanding in Gas-Rich ISM: Using Cold Molecular Gas to Trace Their Impact}",
      journal = {Galaxies},
         year = 2023,
        month = feb,
       volume = {11},
       number = {1},
          eid = {24},
        pages = {24},
          doi = {10.3390/galaxies11010024},
archivePrefix = {arXiv},
       eprint = {2302.14095},
 primaryClass = {astro-ph.GA},
       adsurl = {https://ui.adsabs.harvard.edu/abs/2023Galax..11...24M}
}

@ARTICLE{Cresci2023,
       author = {{Cresci}, G. and {Tozzi}, G. and {Perna}, M. and {Brusa}, M. and {Marconcini}, C. and {Marconi}, A. and {Carniani}, S. and {Brienza}, M. and {Giroletti}, M. and {Belfiore}, F. and {Ginolfi}, M. and {Mannucci}, F. and {Ulivi}, L. and {Scholtz}, J. and {Venturi}, G. and {Arribas}, S. and {{\"U}bler}, H. and {D'Eugenio}, F. and {Mingozzi}, M. and {Balmaverde}, B. and {Capetti}, A. and {Parlanti}, E. and {Zana}, T.},
        title = "{Bubbles and outflows: The novel JWST/NIRSpec view of the z = 1.59 obscured quasar XID2028}",
      journal = {\aap},
         year = 2023,
        month = apr,
       volume = {672},
          eid = {A128},
        pages = {A128},
          doi = {10.1051/0004-6361/202346001},
archivePrefix = {arXiv},
       eprint = {2301.11060},
 primaryClass = {astro-ph.GA},
       adsurl = {https://ui.adsabs.harvard.edu/abs/2023A&A...672A.128C}
}

@ARTICLE{Jackson2020,
       author = {{Jackson}, Thomas M. and {Rosario}, D.~J. and {Alexander}, D.~M. and {Scholtz}, J. and {McAlpine}, Stuart and {Bower}, R.~G.},
        title = "{The star formation properties of the observed and simulated AGN Universe: BAT versus EAGLE}",
      journal = {\mnras},
         year = 2020,
        month = oct,
       volume = {498},
       number = {2},
        pages = {2323-2338},
          doi = {10.1093/mnras/staa2414},
archivePrefix = {arXiv},
       eprint = {2008.05115},
 primaryClass = {astro-ph.GA},
       adsurl = {https://ui.adsabs.harvard.edu/abs/2020MNRAS.498.2323J}
}

@ARTICLE{Shimizu2017,
       author = {{Shimizu}, T. Taro and {Mushotzky}, Richard F. and {Mel{\'e}ndez}, Marcio and {Koss}, Michael J. and {Barger}, Amy J. and {Cowie}, Lennox L.},
        title = "{Herschel far-infrared photometry of the Swift Burst Alert Telescope active galactic nuclei sample of the local universe - III. Global star-forming properties and the lack of a connection to nuclear activity}",
      journal = {\mnras},
         year = 2017,
        month = apr,
       volume = {466},
       number = {3},
        pages = {3161-3183},
          doi = {10.1093/mnras/stw3268},
archivePrefix = {arXiv},
       eprint = {1612.03941},
 primaryClass = {astro-ph.GA},
       adsurl = {https://ui.adsabs.harvard.edu/abs/2017MNRAS.466.3161S}
}

@ARTICLE{Odea1998,
       author = {{O'Dea}, Christopher P.},
        title = "{The Compact Steep-Spectrum and Gigahertz Peaked-Spectrum Radio Sources}",
      journal = {\pasp},
         year = 1998,
        month = may,
       volume = {110},
       number = {747},
        pages = {493-532},
          doi = {10.1086/316162},
       adsurl = {https://ui.adsabs.harvard.edu/abs/1998PASP..110..493O}
}

@ARTICLE{Marvil2015,
       author = {{Marvil}, Joshua and {Owen}, Frazer and {Eilek}, Jean},
        title = "{Integrated Radio Continuum Spectra of Galaxies}",
      journal = {\aj},
         year = 2015,
        month = jan,
       volume = {149},
       number = {1},
          eid = {32},
        pages = {32},
          doi = {10.1088/0004-6256/149/1/32},
archivePrefix = {arXiv},
       eprint = {1408.6296},
 primaryClass = {astro-ph.GA},
       adsurl = {https://ui.adsabs.harvard.edu/abs/2015AJ....149...32M}
}

@ARTICLE{Helou1985,
       author = {{Helou}, G. and {Soifer}, B.~T. and {Rowan-Robinson}, M.},
        title = "{Thermal infrared and nonthermal radio : remarkable correlation in disks of galaxies.}",
      journal = {\apjl},
         year = 1985,
        month = nov,
       volume = {298},
        pages = {L7-L11},
          doi = {10.1086/184556},
       adsurl = {https://ui.adsabs.harvard.edu/abs/1985ApJ...298L...7H}
}

@ARTICLE{SKA2009,
       author = {{Dewdney}, P.~E. and {Hall}, P.~J. and {Schilizzi}, R.~T. and {Lazio}, T.~J.~L.~W.},
        title = "{The Square Kilometre Array}",
      journal = {IEEE Proceedings},
         year = 2009,
        month = aug,
       volume = {97},
       number = {8},
        pages = {1482-1496},
          doi = {10.1109/JPROC.2009.2021005},
       adsurl = {https://ui.adsabs.harvard.edu/abs/2009IEEEP..97.1482D}
}

@ARTICLE{Morabito2025,
       author = {{Morabito}, Leah K. and {Jackson}, Neal and {de Jong}, Jurjen and {Escott}, Emmy and {Groeneveld}, Christian and {Mahatma}, Vijay and {Petley}, James and {Sweijen}, Frits and {Timmerman}, Roland and {van Weeren}, Reinout J.},
        title = "{A decade of sub-arcsecond imaging with the International LOFAR Telescope}",
      journal = {\apss},
         year = 2025,
        month = feb,
       volume = {370},
       number = {2},
          eid = {19},
        pages = {19},
          doi = {10.1007/s10509-025-04406-x},
archivePrefix = {arXiv},
       eprint = {2502.06946},
 primaryClass = {astro-ph.IM},
       adsurl = {https://ui.adsabs.harvard.edu/abs/2025Ap&SS.370...19M}
}

@INPROCEEDINGS{ngVLA2018,
       author = {{Murphy}, E.~J. and {Bolatto}, A. and {Chatterjee}, S. and {Casey}, C.~M. and {Chomiuk}, L. and {Dale}, D. and {de Pater}, I. and {Dickinson}, M. and {Francesco}, J.~D. and {Hallinan}, G. and {Isella}, A. and {Kohno}, K. and {Kulkarni}, S.~R. and {Lang}, C. and {Lazio}, T.~J.~W. and {Leroy}, A.~K. and {Loinard}, L. and {Maccarone}, T.~J. and {Matthews}, B.~C. and {Osten}, R.~A. and {Reid}, M.~J. and {Riechers}, D. and {Sakai}, N. and {Walter}, F. and {Wilner}, D.},
        title = "{The ngVLA Science Case and Associated Science Requirements}",
    booktitle = {Science with a Next Generation Very Large Array},
         year = 2018,
       editor = {{Murphy}, Eric},
       series = {Astronomical Society of the Pacific Conference Series},
       volume = {517},
        month = dec,
        pages = {3},
          doi = {10.48550/arXiv.1810.07524},
archivePrefix = {arXiv},
       eprint = {1810.07524},
 primaryClass = {astro-ph.IM},
       adsurl = {https://ui.adsabs.harvard.edu/abs/2018ASPC..517....3M}
}

@ARTICLE{Tanner2022,
       author = {{Tanner}, Ryan and {Weaver}, Kimberly A.},
        title = "{Simulations of AGN-driven Galactic Outflow Morphology and Content}",
      journal = {\aj},
         year = 2022,
        month = mar,
       volume = {163},
       number = {3},
          eid = {134},
        pages = {134},
          doi = {10.3847/1538-3881/ac4d23},
archivePrefix = {arXiv},
       eprint = {2201.08360},
 primaryClass = {astro-ph.GA},
       adsurl = {https://ui.adsabs.harvard.edu/abs/2022AJ....163..134T}
}

@ARTICLE{White2017,
       author = {{White}, Sarah V. and {Jarvis}, Matt J. and {Kalfountzou}, Eleni and {Hardcastle}, Martin J. and {Verma}, Aprajita and {Cao Orjales}, Jos{\'e} M. and {Stevens}, Jason},
        title = "{Evidence that the AGN dominates the radio emission in z {\ensuremath{\sim}} 1 radio-quiet quasars}",
      journal = {\mnras},
         year = 2017,
        month = jun,
       volume = {468},
       number = {1},
        pages = {217-238},
          doi = {10.1093/mnras/stx284},
archivePrefix = {arXiv},
       eprint = {1702.00904},
 primaryClass = {astro-ph.GA},
       adsurl = {https://ui.adsabs.harvard.edu/abs/2017MNRAS.468..217W}
}

@ARTICLE{White2015,
       author = {{White}, Sarah V. and {Jarvis}, Matt J. and {H{\"a}u{\ss}ler}, Boris and {Maddox}, Natasha},
        title = "{Radio-quiet quasars in the VIDEO survey: evidence for AGN-powered radio emission at S\_\{1.4 GHz < 1\} mJy}",
      journal = {\mnras},
         year = 2015,
        month = apr,
       volume = {448},
       number = {3},
        pages = {2665-2686},
          doi = {10.1093/mnras/stv134},
archivePrefix = {arXiv},
       eprint = {1410.3892},
 primaryClass = {astro-ph.GA},
       adsurl = {https://ui.adsabs.harvard.edu/abs/2015MNRAS.448.2665W}
}

@ARTICLE{Karouzos2016,
       author = {{Karouzos}, Marios and {Woo}, Jong-Hak and {Bae}, Hyun-Jin},
        title = "{Unraveling the Complex Structure of AGN-driven Outflows. I. Kinematics and Sizes}",
      journal = {\apj},
         year = 2016,
        month = mar,
       volume = {819},
       number = {2},
          eid = {148},
        pages = {148},
          doi = {10.3847/0004-637X/819/2/148},
archivePrefix = {arXiv},
       eprint = {1601.02621},
 primaryClass = {astro-ph.GA},
       adsurl = {https://ui.adsabs.harvard.edu/abs/2016ApJ...819..148K}
}

@ARTICLE{Fluetsch2021,
       author = {{Fluetsch}, A. and {Maiolino}, R. and {Carniani}, S. and {Arribas}, S. and {Belfiore}, F. and {Bellocchi}, E. and {Cazzoli}, S. and {Cicone}, C. and {Cresci}, G. and {Fabian}, A.~C. and {Gallagher}, R. and {Ishibashi}, W. and {Mannucci}, F. and {Marconi}, A. and {Perna}, M. and {Sturm}, E. and {Venturi}, G.},
        title = "{Properties of the multiphase outflows in local (ultra)luminous infrared galaxies}",
      journal = {\mnras},
         year = 2021,
        month = aug,
       volume = {505},
       number = {4},
        pages = {5753-5783},
          doi = {10.1093/mnras/stab1666},
archivePrefix = {arXiv},
       eprint = {2006.13232},
 primaryClass = {astro-ph.GA},
       adsurl = {https://ui.adsabs.harvard.edu/abs/2021MNRAS.505.5753F}
}

@ARTICLE{Morganti2005,
       author = {{Morganti}, R. and {Tadhunter}, C.~N. and {Oosterloo}, T.~A.},
        title = "{Fast neutral outflows in powerful radio galaxies: a major source of feedback in massive galaxies}",
      journal = {\aap},
         year = 2005,
        month = dec,
       volume = {444},
       number = {1},
        pages = {L9-L13},
          doi = {10.1051/0004-6361:200500197},
archivePrefix = {arXiv},
       eprint = {astro-ph/0510263},
 primaryClass = {astro-ph},
       adsurl = {https://ui.adsabs.harvard.edu/abs/2005A&A...444L...9M}
}

@ARTICLE{Santoro2018,
       author = {{Santoro}, F. and {Tadhunter}, C. and {Baron}, D. and {Morganti}, R. and {Holt}, J.},
        title = "{AGN-driven outflows and the AGN feedback efficiency in young radio galaxies}",
      journal = {\aap},
         year = 2020,
        month = dec,
       volume = {644},
          eid = {A54},
        pages = {A54},
          doi = {10.1051/0004-6361/202039077},
archivePrefix = {arXiv},
       eprint = {2009.11175},
 primaryClass = {astro-ph.GA},
       adsurl = {https://ui.adsabs.harvard.edu/abs/2020A&A...644A..54S}
}

@ARTICLE{Speranza2021,
       author = {{Speranza}, G. and {Balmaverde}, B. and {Capetti}, A. and {Massaro}, F. and {Tremblay}, G. and {Marconi}, A. and {Venturi}, G. and {Chiaberge}, M. and {Baldi}, R.~D. and {Baum}, S. and {Grandi}, P. and {Meyer}, E.~T. and {O'Dea}, C. and {Sparks}, W. and {Terrazas}, B.~A. and {Torresi}, E.},
        title = "{The MURALES survey. IV. Searching for nuclear outflows in 3C radio galaxies at z < 0.3 with MUSE observations}",
      journal = {\aap},
         year = 2021,
        month = sep,
       volume = {653},
          eid = {A150},
        pages = {A150},
          doi = {10.1051/0004-6361/202140686},
archivePrefix = {arXiv},
       eprint = {2106.09743},
 primaryClass = {astro-ph.GA},
       adsurl = {https://ui.adsabs.harvard.edu/abs/2021A&A...653A.150S}
}

@ARTICLE{Rose2018,
       author = {{Rose}, Marvin and {Tadhunter}, Clive and {Ramos Almeida}, Cristina and {Rodr{\'\i}guez Zaur{\'\i}n}, Javier and {Santoro}, Francesco and {Spence}, Robert},
        title = "{Quantifying the AGN-driven outflows in ULIRGs (QUADROS) - I: VLT/Xshooter observations of nine nearby objects}",
      journal = {\mnras},
         year = 2018,
        month = feb,
       volume = {474},
       number = {1},
        pages = {128-156},
          doi = {10.1093/mnras/stx2590},
archivePrefix = {arXiv},
       eprint = {1710.06600},
 primaryClass = {astro-ph.GA},
       adsurl = {https://ui.adsabs.harvard.edu/abs/2018MNRAS.474..128R}
}

@ARTICLE{Riffel2017,
       author = {{Riffel}, Rogemar A. and {Storchi-Bergmann}, Thaisa and {Riffel}, Rogerio and {Dahmer-Hahn}, Luis G. and {Diniz}, Marlon R. and {Sch{\"o}nell}, Astor J. and {Dametto}, Natacha Z.},
        title = "{Gemini NIFS survey of feeding and feedback processes in nearby active galaxies - I. Stellar kinematics}",
      journal = {\mnras},
         year = 2017,
        month = sep,
       volume = {470},
       number = {1},
        pages = {992-1016},
          doi = {10.1093/mnras/stx1308},
archivePrefix = {arXiv},
       eprint = {1705.06949},
 primaryClass = {astro-ph.GA},
       adsurl = {https://ui.adsabs.harvard.edu/abs/2017MNRAS.470..992R}
}

@ARTICLE{Davies2020,
       author = {{Davies}, R. and {Baron}, D. and {Shimizu}, T. and {Netzer}, H. and {Burtscher}, L. and {de Zeeuw}, P.~T. and {Genzel}, R. and {Hicks}, E.~K.~S. and {Koss}, M. and {Lin}, M. -Y. and {Lutz}, D. and {Maciejewski}, W. and {M{\"u}ller-S{\'a}nchez}, F. and {Orban de Xivry}, G. and {Ricci}, C. and {Riffel}, R. and {Riffel}, R.~A. and {Rosario}, D. and {Schartmann}, M. and {Schnorr-M{\"u}ller}, A. and {Shangguan}, J. and {Sternberg}, A. and {Sturm}, E. and {Storchi-Bergmann}, T. and {Tacconi}, L. and {Veilleux}, S.},
        title = "{Ionized outflows in local luminous AGN: what are the real densities and outflow rates?}",
      journal = {\mnras},
         year = 2020,
        month = nov,
       volume = {498},
       number = {3},
        pages = {4150-4177},
          doi = {10.1093/mnras/staa2413},
archivePrefix = {arXiv},
       eprint = {2003.06153},
 primaryClass = {astro-ph.GA},
       adsurl = {https://ui.adsabs.harvard.edu/abs/2020MNRAS.498.4150D}
}

@ARTICLE{GarciaBurillo2021,
       author = {{Garc{\'\i}a-Burillo}, S. and {Alonso-Herrero}, A. and {Ramos Almeida}, C. and {Gonz{\'a}lez-Mart{\'\i}n}, O. and {Combes}, F. and {Usero}, A. and {H{\"o}nig}, S. and {Querejeta}, M. and {Hicks}, E.~K.~S. and {Hunt}, L.~K. and {Rosario}, D. and {Davies}, R. and {Boorman}, P.~G. and {Bunker}, A.~J. and {Burtscher}, L. and {Colina}, L. and {D{\'\i}az-Santos}, T. and {Gandhi}, P. and {Garc{\'\i}a-Bernete}, I. and {Garc{\'\i}a-Lorenzo}, B. and {Ichikawa}, K. and {Imanishi}, M. and {Izumi}, T. and {Labiano}, A. and {Levenson}, N.~A. and {L{\'o}pez-Rodr{\'\i}guez}, E. and {Packham}, C. and {Pereira-Santaella}, M. and {Ricci}, C. and {Rigopoulou}, D. and {Rouan}, D. and {Shimizu}, T. and {Stalevski}, M. and {Wada}, K. and {Williamson}, D.},
        title = "{The Galaxy Activity, Torus, and Outflow Survey (GATOS). I. ALMA images of dusty molecular tori in Seyfert galaxies}",
      journal = {\aap},
         year = 2021,
        month = aug,
       volume = {652},
          eid = {A98},
        pages = {A98},
          doi = {10.1051/0004-6361/202141075},
archivePrefix = {arXiv},
       eprint = {2104.10227},
 primaryClass = {astro-ph.GA},
       adsurl = {https://ui.adsabs.harvard.edu/abs/2021A&A...652A..98G}
}

@ARTICLE{Circosta2018,
       author = {{Circosta}, C. and {Mainieri}, V. and {Padovani}, P. and {Lanzuisi}, G. and {Salvato}, M. and {Harrison}, C.~M. and {Kakkad}, D. and {Puglisi}, A. and {Vietri}, G. and {Zamorani}, G. and {Cicone}, C. and {Husemann}, B. and {Vignali}, C. and {Balmaverde}, B. and {Bischetti}, M. and {Bongiorno}, A. and {Brusa}, M. and {Carniani}, S. and {Civano}, F. and {Comastri}, A. and {Cresci}, G. and {Feruglio}, C. and {Fiore}, F. and {Fotopoulou}, S. and {Karim}, A. and {Lamastra}, A. and {Magnelli}, B. and {Mannucci}, F. and {Marconi}, A. and {Merloni}, A. and {Netzer}, H. and {Perna}, M. and {Piconcelli}, E. and {Rodighiero}, G. and {Schinnerer}, E. and {Schramm}, M. and {Schulze}, A. and {Silverman}, J. and {Zappacosta}, L.},
        title = "{SUPER. I. Toward an unbiased study of ionized outflows in z {\ensuremath{\sim}} 2 active galactic nuclei: survey overview and sample characterization}",
      journal = {\aap},
         year = 2018,
        month = nov,
       volume = {620},
          eid = {A82},
        pages = {A82},
          doi = {10.1051/0004-6361/201833520},
archivePrefix = {arXiv},
       eprint = {1809.04858},
 primaryClass = {astro-ph.GA},
       adsurl = {https://ui.adsabs.harvard.edu/abs/2018A&A...620A..82C}
}

@ARTICLE{Nesvadba2017,
       author = {{Nesvadba}, N.~P.~H. and {De Breuck}, C. and {Lehnert}, M.~D. and {Best}, P.~N. and {Collet}, C.},
        title = "{The SINFONI survey of powerful radio galaxies at z 2: Jet-driven AGN feedback during the Quasar Era}",
      journal = {\aap},
         year = 2017,
        month = mar,
       volume = {599},
          eid = {A123},
        pages = {A123},
          doi = {10.1051/0004-6361/201528040},
archivePrefix = {arXiv},
       eprint = {1610.02057},
 primaryClass = {astro-ph.GA},
       adsurl = {https://ui.adsabs.harvard.edu/abs/2017A&A...599A.123N}
}

@ARTICLE{Mukherjee2025,
       author = {{Mukherjee}, Dipanjan},
        title = "{Jet-Feedback on kpc scales: a review}",
      journal = {arXiv e-prints},
         year = 2025,
        month = jun,
          eid = {arXiv:2506.03888},
        pages = {arXiv:2506.03888},
          doi = {10.48550/arXiv.2506.03888},
archivePrefix = {arXiv},
       eprint = {2506.03888},
 primaryClass = {astro-ph.GA},
       adsurl = {https://ui.adsabs.harvard.edu/abs/2025arXiv250603888M}
}

@ARTICLE{Rao2023,
       author = {{Rao}, Vaishnav V. and {Kharb}, P. and {Rubinur}, K. and {Silpa}, S. and {Roy}, N. and {Sebastian}, B. and {Singh}, V. and {Baghel}, J. and {Manna}, S. and {Ishwara-Chandra}, C.~H.},
        title = "{AGN feedback through multiple jet cycles in the Seyfert galaxy NGC 2639}",
      journal = {\mnras},
         year = 2023,
        month = sep,
       volume = {524},
       number = {2},
        pages = {1615-1624},
          doi = {10.1093/mnras/stad1901},
archivePrefix = {arXiv},
       eprint = {2301.01610},
 primaryClass = {astro-ph.GA},
       adsurl = {https://ui.adsabs.harvard.edu/abs/2023MNRAS.524.1615R}
}

@ARTICLE{Silpa2023,
       author = {{Silpa}, S. and {Kharb}, P. and {Ho}, Luis C. and {Harrison}, C.~M.},
        title = "{Probing the Interplay between Jets, Winds, and Multi-phase Gas in 11 Radio-quiet PG Quasars: A uGMRT-VLA Study}",
      journal = {\apj},
         year = 2023,
        month = nov,
       volume = {958},
       number = {1},
          eid = {47},
        pages = {47},
          doi = {10.3847/1538-4357/acf7c9},
archivePrefix = {arXiv},
       eprint = {2301.07929},
 primaryClass = {astro-ph.GA},
       adsurl = {https://ui.adsabs.harvard.edu/abs/2023ApJ...958...47S}
}

@ARTICLE{Kharb2019,
       author = {{Kharb}, P. and {Vaddi}, S. and {Sebastian}, B. and {Subramanian}, S. and {Das}, M. and {Paragi}, Z.},
        title = "{A Curved 150 pc Long Jet in the Double-peaked Emission-line AGN KISSR 434}",
      journal = {\apj},
         year = 2019,
        month = feb,
       volume = {871},
       number = {2},
          eid = {249},
        pages = {249},
          doi = {10.3847/1538-4357/aafad7},
archivePrefix = {arXiv},
       eprint = {1812.11074},
 primaryClass = {astro-ph.GA},
       adsurl = {https://ui.adsabs.harvard.edu/abs/2019ApJ...871..249K}
}

@ARTICLE{Kharb2021,
       author = {{Kharb}, P. and {Subramanian}, S. and {Das}, M. and {Vaddi}, S. and {Paragi}, Z.},
        title = "{The Nature of Jets in Double-peaked Emission-line AGN in the KISSR Sample}",
      journal = {\apj},
         year = 2021,
        month = oct,
       volume = {919},
       number = {2},
          eid = {108},
        pages = {108},
          doi = {10.3847/1538-4357/ac0c82},
archivePrefix = {arXiv},
       eprint = {2106.09304},
 primaryClass = {astro-ph.GA},
       adsurl = {https://ui.adsabs.harvard.edu/abs/2021ApJ...919..108K}
}

@ARTICLE{Jarvis2021,
       author = {{Jarvis}, M.~E. and {Harrison}, C.~M. and {Mainieri}, V. and {Alexander}, D.~M. and {Arrigoni Battaia}, F. and {Calistro Rivera}, G. and {Circosta}, C. and {Costa}, T. and {De Breuck}, C. and {Edge}, A.~C. and {Girdhar}, A. and {Kakkad}, D. and {Kharb}, P. and {Lansbury}, G.~B. and {Molyneux}, S.~J. and {Mukherjee}, D. and {Mullaney}, J.~R. and {Farina}, E.~P. and {Silpa}, S. and {Thomson}, A.~P. and {Ward}, S.~R.},
        title = "{The quasar feedback survey: discovering hidden Radio-AGN and their connection to the host galaxy ionized gas}",
      journal = {\mnras},
         year = 2021,
        month = may,
       volume = {503},
       number = {2},
        pages = {1780-1797},
          doi = {10.1093/mnras/stab549},
archivePrefix = {arXiv},
       eprint = {2103.00014},
 primaryClass = {astro-ph.GA},
       adsurl = {https://ui.adsabs.harvard.edu/abs/2021MNRAS.503.1780J}
}

@ARTICLE{Njeri2025,
       author = {{Njeri}, Ann and {Harrison}, Chris M. and {Kharb}, Preeti and {Beswick}, Robert and {Calistro-Rivera}, Gabriela and {Circosta}, Chiara and {Mainieri}, Vincenzo and {Molyneux}, Stephen and {Mullaney}, James and {Silpa}, Sasikumar},
        title = "{The Quasar Feedback Survey: zooming into the origin of radio emission with e-MERLIN}",
      journal = {\mnras},
         year = 2025,
        month = feb,
       volume = {537},
       number = {2},
        pages = {705-722},
          doi = {10.1093/mnras/staf020},
archivePrefix = {arXiv},
       eprint = {2501.03433},
 primaryClass = {astro-ph.GA},
       adsurl = {https://ui.adsabs.harvard.edu/abs/2025MNRAS.537..705N}
}

@ARTICLE{Mullaney2013,
       author = {{Mullaney}, J.~R. and {Alexander}, D.~M. and {Fine}, S. and {Goulding}, A.~D. and {Harrison}, C.~M. and {Hickox}, R.~C.},
        title = "{Narrow-line region gas kinematics of 24 264 optically selected AGN: the radio connection}",
      journal = {\mnras},
         year = 2013,
        month = jul,
       volume = {433},
       number = {1},
        pages = {622-638},
          doi = {10.1093/mnras/stt751},
archivePrefix = {arXiv},
       eprint = {1305.0263},
 primaryClass = {astro-ph.CO},
       adsurl = {https://ui.adsabs.harvard.edu/abs/2013MNRAS.433..622M}
}

@ARTICLE{Jarvis2020,
       author = {{Jarvis}, M.~E. and {Harrison}, C.~M. and {Mainieri}, V. and {Calistro Rivera}, G. and {Jethwa}, P. and {Zhang}, Z. -Y. and {Alexander}, D.~M. and {Circosta}, C. and {Costa}, T. and {De Breuck}, C. and {Kakkad}, D. and {Kharb}, P. and {Lansbury}, G.~B. and {Thomson}, A.~P.},
        title = "{High molecular gas content and star formation rates in local galaxies that host quasars, outflows, and jets}",
      journal = {\mnras},
         year = 2020,
        month = oct,
       volume = {498},
       number = {2},
        pages = {1560-1575},
          doi = {10.1093/mnras/staa2196},
archivePrefix = {arXiv},
       eprint = {2007.10351},
 primaryClass = {astro-ph.GA},
       adsurl = {https://ui.adsabs.harvard.edu/abs/2020MNRAS.498.1560J}
}

@ARTICLE{Becker1995,
       author = {{Becker}, Robert H. and {White}, Richard L. and {Helfand}, David J.},
        title = "{The FIRST Survey: Faint Images of the Radio Sky at Twenty Centimeters}",
      journal = {\apj},
         year = 1995,
        month = sep,
       volume = {450},
        pages = {559},
          doi = {10.1086/176166},
       adsurl = {https://ui.adsabs.harvard.edu/abs/1995ApJ...450..559B}
}

@ARTICLE{Middelberg2007,
       author = {{Middelberg}, E. and {Agudo}, I. and {Roy}, A.~L. and {Krichbaum}, T.~P.},
        title = "{Jet-cloud collisions in the jet of the Seyfert galaxy NGC3079}",
      journal = {\mnras},
         year = 2007,
        month = may,
       volume = {377},
       number = {2},
        pages = {731-740},
          doi = {10.1111/j.1365-2966.2007.11639.x},
archivePrefix = {arXiv},
       eprint = {astro-ph/0702481},
 primaryClass = {astro-ph},
       adsurl = {https://ui.adsabs.harvard.edu/abs/2007MNRAS.377..731M}
}

@ARTICLE{Odea2021,
       author = {{O'Dea}, Christopher P. and {Saikia}, D.~J.},
        title = "{Compact steep-spectrum and peaked-spectrum radio sources}",
      journal = {\aapr},
         year = 2021,
        month = dec,
       volume = {29},
       number = {1},
          eid = {3},
        pages = {3},
          doi = {10.1007/s00159-021-00131-w},
archivePrefix = {arXiv},
       eprint = {2009.02750},
 primaryClass = {astro-ph.GA},
       adsurl = {https://ui.adsabs.harvard.edu/abs/2021A&ARv..29....3O}
}

@ARTICLE{Baldi2018,
       author = {{Baldi}, R.~D. and {Williams}, D.~R.~A. and {McHardy}, I.~M. and {Beswick}, R.~J. and {Argo}, M.~K. and {Dullo}, B.~T. and {Knapen}, J.~H. and {Brinks}, E. and {Muxlow}, T.~W.~B. and {Aalto}, S. and {Alberdi}, A. and {Bendo}, G.~J. and {Corbel}, S. and {Evans}, R. and {Fenech}, D.~M. and {Green}, D.~A. and {Kl{\"o}ckner}, H. -R. and {K{\"o}rding}, E. and {Kharb}, P. and {Maccarone}, T.~J. and {Mart{\'\i}-Vidal}, I. and {Mundell}, C.~G. and {Panessa}, F. and {Peck}, A.~B. and {P{\'e}rez-Torres}, M.~A. and {Saikia}, D.~J. and {Saikia}, P. and {Shankar}, F. and {Spencer}, R.~E. and {Stevens}, I.~R. and {Uttley}, P. and {Westcott}, J.},
        title = "{LeMMINGs - I. The eMERLIN legacy survey of nearby galaxies. 1.5-GHz parsec-scale radio structures and cores}",
      journal = {\mnras},
         year = 2018,
        month = may,
       volume = {476},
       number = {3},
        pages = {3478-3522},
          doi = {10.1093/mnras/sty342},
archivePrefix = {arXiv},
       eprint = {1802.02162},
 primaryClass = {astro-ph.GA},
       adsurl = {https://ui.adsabs.harvard.edu/abs/2018MNRAS.476.3478B}
}

@ARTICLE{Harrison2024,
       author = {{Harrison}, Chris M. and {Ramos Almeida}, Cristina},
        title = "{Observational Tests of Active Galactic Nuclei Feedback: An Overview of Approaches and Interpretation}",
      journal = {Galaxies},
         year = 2024,
        month = apr,
       volume = {12},
       number = {2},
          eid = {17},
        pages = {17},
          doi = {10.3390/galaxies12020017},
archivePrefix = {arXiv},
       eprint = {2404.08050},
 primaryClass = {astro-ph.GA},
       adsurl = {https://ui.adsabs.harvard.edu/abs/2024Galax..12...17H}
}

@ARTICLE{Begelman1984,
       author = {{Begelman}, Mitchell C. and {Blandford}, Roger D. and {Rees}, Martin J.},
        title = "{Theory of extragalactic radio sources}",
      journal = {Reviews of Modern Physics},
         year = 1984,
        month = apr,
       volume = {56},
       number = {2},
        pages = {255-351},
          doi = {10.1103/RevModPhys.56.255},
       adsurl = {https://ui.adsabs.harvard.edu/abs/1984RvMP...56..255B}
}

@ARTICLE{Laing1980,
       author = {{Laing}, R.~A. and {Peacock}, J.~A.},
        title = "{The relation between radio luminosity and spectrum for extended extragalactic radio sources.}",
      journal = {\mnras},
         year = 1980,
        month = mar,
       volume = {190},
        pages = {903-924},
          doi = {10.1093/mnras/190.4.903},
       adsurl = {https://ui.adsabs.harvard.edu/abs/1980MNRAS.190..903L}
}

@ARTICLE{Hardcastle2020,
       author = {{Hardcastle}, M.~J. and {Croston}, J.~H.},
        title = "{Radio galaxies and feedback from AGN jets}",
      journal = {\nar},
         year = 2020,
        month = jun,
       volume = {88},
          eid = {101539},
        pages = {101539},
          doi = {10.1016/j.newar.2020.101539},
archivePrefix = {arXiv},
       eprint = {2003.06137},
 primaryClass = {astro-ph.HE},
       adsurl = {https://ui.adsabs.harvard.edu/abs/2020NewAR..8801539H}
}

@ARTICLE{Harwood2022,
       author = {{Harwood}, J.~J. and {Mooney}, S. and {Morabito}, L.~K. and {Quinn}, J. and {Sweijen}, F. and {Groeneveld}, C. and {Bonnassieux}, E. and {Kappes}, A. and {Moldon}, J.},
        title = "{The resolved jet of 3C 273 at 150 MHz. Sub-arcsecond imaging with the LOFAR international baselines}",
      journal = {\aap},
         year = 2022,
        month = feb,
       volume = {658},
          eid = {A8},
        pages = {A8},
          doi = {10.1051/0004-6361/202141579},
       adsurl = {https://ui.adsabs.harvard.edu/abs/2022A&A...658A...8H}
}

@ARTICLE{Condon1982,
       author = {{Condon}, J.~J. and {Condon}, M.~A. and {Gisler}, G. and
         {Puschell}, J.~J.},
        title = "{Strong radio sources in bright spiral galaxies. II. Rapid star formation and galaxy-galaxy interactions.}",
      journal = {\apj},
         year = 1982,
        month = jan,
       volume = {252},
        pages = {102-124},
          doi = {10.1086/159538},
       adsurl = {https://ui.adsabs.harvard.edu/abs/1982ApJ...252..102C}
}

@ARTICLE{Ulvestad2005,
       author = {{Ulvestad}, James S. and {Antonucci}, Robert R.~J. and {Barvainis}, Richard},
        title = "{VLBA Imaging of Central Engines in Radio-Quiet Quasars}",
      journal = {\apj},
         year = 2005,
        month = mar,
       volume = {621},
       number = {1},
        pages = {123-129},
          doi = {10.1086/427426},
archivePrefix = {arXiv},
       eprint = {astro-ph/0411678},
 primaryClass = {astro-ph},
       adsurl = {https://ui.adsabs.harvard.edu/abs/2005ApJ...621..123U}
}

@ARTICLE{Morabito2022,
       author = {{Morabito}, Leah K. and {Sweijen}, F. and {Radcliffe}, J.~F. and {Best}, P.~N. and {Kondapally}, Rohit and {Bondi}, Marco and {Bonato}, Matteo and {Duncan}, K.~J. and {Prandoni}, Isabella and {Shimwell}, T.~W. and {Williams}, W.~L. and {van Weeren}, R.~J. and {Conway}, J.~E. and {Calistro Rivera}, G.},
        title = "{Identifying active galactic nuclei via brightness temperature with sub-arcsecond international LOFAR telescope observations}",
      journal = {\mnras},
         year = 2022,
        month = oct,
       volume = {515},
       number = {4},
        pages = {5758-5774},
          doi = {10.1093/mnras/stac2129},
archivePrefix = {arXiv},
       eprint = {2207.13096},
 primaryClass = {astro-ph.GA},
       adsurl = {https://ui.adsabs.harvard.edu/abs/2022MNRAS.515.5758M}
}

@ARTICLE{Kellerman1989,
       author = {{Kellermann}, K.~I. and {Sramek}, R. and {Schmidt}, M. and {Shaffer}, D.~B. and {Green}, R.},
        title = "{VLA Observations of Objects in the Palomar Bright Quasar Survey}",
      journal = {\aj},
         year = 1989,
        month = oct,
       volume = {98},
        pages = {1195},
          doi = {10.1086/115207},
       adsurl = {https://ui.adsabs.harvard.edu/abs/1989AJ.....98.1195K}
}

@ARTICLE{Ivezic2002,
       author = {{Ivezi{\'c}}, {\v{Z}}eljko and {Menou}, Kristen and {Knapp}, Gillian R. and {Strauss}, Michael A. and {Lupton}, Robert H. and {Vanden Berk}, Daniel E. and {Richards}, Gordon T. and {Tremonti}, Christy and {Weinstein}, Michael A. and {Anderson}, Scott and {Bahcall}, Neta A. and {Becker}, Robert H. and {Bernardi}, Mariangela and {Blanton}, Michael and {Eisenstein}, Daniel and {Fan}, Xiaohui and {Finkbeiner}, Douglas and {Finlator}, Kristian and {Frieman}, Joshua and {Gunn}, James E. and {Hall}, Pat B. and {Kim}, Rita S.~J. and {Kinkhabwala}, Ali and {Narayanan}, Vijay K. and {Rockosi}, Constance M. and {Schlegel}, David and {Schneider}, Donald P. and {Strateva}, Iskra and {SubbaRao}, Mark and {Thakar}, Aniruddha R. and {Voges}, Wolfgang and {White}, Richard L. and {Yanny}, Brian and {Brinkmann}, Jonathan and {Doi}, Mamoru and {Fukugita}, Masataka and {Hennessy}, Gregory S. and {Munn}, Jeffrey A. and {Nichol}, Robert C. and {York}, Donald G.},
        title = "{Optical and Radio Properties of Extragalactic Sources Observed by the FIRST Survey and the Sloan Digital Sky Survey}",
      journal = {\aj},
         year = 2002,
        month = nov,
       volume = {124},
       number = {5},
        pages = {2364-2400},
          doi = {10.1086/344069},
archivePrefix = {arXiv},
       eprint = {astro-ph/0202408},
 primaryClass = {astro-ph},
       adsurl = {https://ui.adsabs.harvard.edu/abs/2002AJ....124.2364I}
}

@ARTICLE{Xu1999,
       author = {{Xu}, Chun and {Livio}, Mario and {Baum}, Stefi},
        title = "{Radio-loud and Radio-quiet Active Galactic Nuclei}",
      journal = {\aj},
         year = 1999,
        month = sep,
       volume = {118},
       number = {3},
        pages = {1169-1176},
          doi = {10.1086/301007},
archivePrefix = {arXiv},
       eprint = {astro-ph/9905322},
 primaryClass = {astro-ph},
       adsurl = {https://ui.adsabs.harvard.edu/abs/1999AJ....118.1169X}
}

@ARTICLE{Jarvis2019,
       author = {{Jarvis}, M.~E. and {Harrison}, C.~M. and {Thomson}, A.~P. and {Circosta}, C. and {Mainieri}, V. and {Alexander}, D.~M. and {Edge}, A.~C. and {Lansbury}, G.~B. and {Molyneux}, S.~J. and {Mullaney}, J.~R.},
        title = "{Prevalence of radio jets associated with galactic outflows and feedback from quasars}",
      journal = {\mnras},
         year = 2019,
        month = may,
       volume = {485},
       number = {2},
        pages = {2710-2730},
          doi = {10.1093/mnras/stz556},
archivePrefix = {arXiv},
       eprint = {1902.07727},
 primaryClass = {astro-ph.GA},
       adsurl = {https://ui.adsabs.harvard.edu/abs/2019MNRAS.485.2710J}
}

@ARTICLE{Zakamska2014,
       author = {{Zakamska}, Nadia L. and {Greene}, Jenny E.},
        title = "{Quasar feedback and the origin of radio emission in radio-quiet quasars}",
      journal = {\mnras},
         year = 2014,
        month = jul,
       volume = {442},
       number = {1},
        pages = {784-804},
          doi = {10.1093/mnras/stu842},
archivePrefix = {arXiv},
       eprint = {1402.6736},
 primaryClass = {astro-ph.GA},
       adsurl = {https://ui.adsabs.harvard.edu/abs/2014MNRAS.442..784Z}
}

@ARTICLE{Girdhar2024,
       author = {{Girdhar}, A. and {Harrison}, C.~M. and {Mainieri}, V. and {Fern{\'a}ndez Aranda}, R. and {Alexander}, D.~M. and {Arrigoni Battaia}, F. and {Bianchin}, M. and {Calistro Rivera}, G. and {Circosta}, C. and {Costa}, T. and {Edge}, A.~C. and {Farina}, E.~P. and {Kakkad}, D. and {Kharb}, P. and {Molyneux}, S.~J. and {Mukherjee}, D. and {Njeri}, A. and {Silpa}, S. and {Venturi}, G. and {Ward}, S.~R.},
        title = "{Quasar feedback survey: molecular gas affected by central outflows and by  10-kpc radio lobes reveal dual feedback effects in 'radio quiet' quasars}",
      journal = {\mnras},
         year = 2024,
        month = jan,
       volume = {527},
       number = {3},
        pages = {9322-9342},
          doi = {10.1093/mnras/stad3453},
archivePrefix = {arXiv},
       eprint = {2311.03453},
 primaryClass = {astro-ph.GA},
       adsurl = {https://ui.adsabs.harvard.edu/abs/2024MNRAS.527.9322G}
}

@ARTICLE{Girdhar2022,
       author = {{Girdhar}, A. and {Harrison}, C.~M. and {Mainieri}, V. and {Bittner}, A. and {Costa}, T. and {Kharb}, P. and {Mukherjee}, D. and {Arrigoni Battaia}, F. and {Alexander}, D.~M. and {Calistro Rivera}, G. and {Circosta}, C. and {De Breuck}, C. and {Edge}, A.~C. and {Farina}, E.~P. and {Kakkad}, D. and {Lansbury}, G.~B. and {Molyneux}, S.~J. and {Mullaney}, J.~R. and {Silpa}, S. and {Thomson}, A.~P. and {Ward}, S.~R.},
        title = "{Quasar feedback survey: multiphase outflows, turbulence, and evidence for feedback caused by low power radio jets inclined into the galaxy disc}",
      journal = {\mnras},
         year = 2022,
        month = may,
       volume = {512},
       number = {2},
        pages = {1608-1628},
          doi = {10.1093/mnras/stac073},
archivePrefix = {arXiv},
       eprint = {2201.02208},
 primaryClass = {astro-ph.GA},
       adsurl = {https://ui.adsabs.harvard.edu/abs/2022MNRAS.512.1608G}
}

@ARTICLE{Kennicutt2012,
       author = {{Kennicutt}, Robert C. and {Evans}, Neal J.},
        title = "{Star Formation in the Milky Way and Nearby Galaxies}",
      journal = {\araa},
         year = 2012,
        month = sep,
       volume = {50},
        pages = {531-608},
          doi = {10.1146/annurev-astro-081811-125610},
archivePrefix = {arXiv},
       eprint = {1204.3552},
 primaryClass = {astro-ph.GA},
       adsurl = {https://ui.adsabs.harvard.edu/abs/2012ARA&A..50..531K}
}

@ARTICLE{Panessa2019,
       author = {{Panessa}, Francesca and {Baldi}, Ranieri Diego and {Laor}, Ari and {Padovani}, Paolo and {Behar}, Ehud and {McHardy}, Ian},
        title = "{The origin of radio emission from radio-quiet active galactic nuclei}",
      journal = {Nature Astronomy},
         year = 2019,
        month = apr,
       volume = {3},
        pages = {387-396},
          doi = {10.1038/s41550-019-0765-4},
archivePrefix = {arXiv},
       eprint = {1902.05917},
 primaryClass = {astro-ph.GA},
       adsurl = {https://ui.adsabs.harvard.edu/abs/2019NatAs...3..387P}
}

@ARTICLE{Woo2016,
       author = {{Woo}, Jong-Hak and {Bae}, Hyun-Jin and {Son}, Donghoon and {Karouzos}, Marios},
        title = "{The Prevalence of Gas Outflows in Type 2 AGNs}",
      journal = {\apj},
         year = 2016,
        month = feb,
       volume = {817},
       number = {2},
          eid = {108},
        pages = {108},
          doi = {10.3847/0004-637X/817/2/108},
archivePrefix = {arXiv},
       eprint = {1511.05142},
 primaryClass = {astro-ph.GA},
       adsurl = {https://ui.adsabs.harvard.edu/abs/2016ApJ...817..108W}
}

@ARTICLE{Laor2008,
       author = {{Laor}, Ari and {Behar}, Ehud},
        title = "{On the origin of radio emission in radio-quiet quasars}",
      journal = {\mnras},
         year = 2008,
        month = oct,
       volume = {390},
       number = {2},
        pages = {847-862},
          doi = {10.1111/j.1365-2966.2008.13806.x},
archivePrefix = {arXiv},
       eprint = {0808.0637},
 primaryClass = {astro-ph},
       adsurl = {https://ui.adsabs.harvard.edu/abs/2008MNRAS.390..847L}
}

@ARTICLE{Kunert-Bajraszewska2010,
       author = {{Kunert-Bajraszewska}, M. and {Gawro{\'n}ski}, M.~P. and {Labiano}, A. and {Siemiginowska}, A.},
        title = "{A survey of low-luminosity compact sources and its implication for the evolution of radio-loud active galactic nuclei - I. Radio data}",
      journal = {\mnras},
         year = 2010,
        month = nov,
       volume = {408},
       number = {4},
        pages = {2261-2278},
          doi = {10.1111/j.1365-2966.2010.17271.x},
archivePrefix = {arXiv},
       eprint = {1009.5235},
 primaryClass = {astro-ph.CO},
       adsurl = {https://ui.adsabs.harvard.edu/abs/2010MNRAS.408.2261K}
}

@ARTICLE{Slob2022,
       author = {{Slob}, M.~M. and {Callingham}, J.~R. and {R{\"o}ttgering}, H.~J.~A. and {Williams}, W.~L. and {Duncan}, K.~J. and {de Gasperin}, F. and {Hardcastle}, M.~J. and {Miley}, G.~K.},
        title = "{Extragalactic peaked-spectrum radio sources at low frequencies are young radio galaxies}",
      journal = {\aap},
         year = 2022,
        month = dec,
       volume = {668},
          eid = {A186},
        pages = {A186},
          doi = {10.1051/0004-6361/202244651},
archivePrefix = {arXiv},
       eprint = {2210.16570},
 primaryClass = {astro-ph.GA},
       adsurl = {https://ui.adsabs.harvard.edu/abs/2022A&A...668A.186S}
}

@ARTICLE{Baldi2023,
       author = {{Baldi}, Ranieri D.},
        title = "{The nature of compact radio sources: the case of FR 0 radio galaxies}",
      journal = {\aapr},
         year = 2023,
        month = dec,
       volume = {31},
       number = {1},
          eid = {3},
        pages = {3},
          doi = {10.1007/s00159-023-00148-3},
archivePrefix = {arXiv},
       eprint = {2307.08379},
 primaryClass = {astro-ph.GA},
       adsurl = {https://ui.adsabs.harvard.edu/abs/2023A&ARv..31....3B}
}

@ARTICLE{Baldi2015,
       author = {{Baldi}, Ranieri D. and {Capetti}, Alessandro and {Giovannini}, Gabriele},
        title = "{Pilot study of the radio-emitting AGN population: the emerging new class of FR 0 radio-galaxies}",
      journal = {\aap},
         year = 2015,
        month = apr,
       volume = {576},
          eid = {A38},
        pages = {A38},
          doi = {10.1051/0004-6361/201425426},
archivePrefix = {arXiv},
       eprint = {1502.00427},
 primaryClass = {astro-ph.GA},
       adsurl = {https://ui.adsabs.harvard.edu/abs/2015A&A...576A..38B}
}

@ARTICLE{Dey2019,
       author = {{Dey}, Arjun and {Schlegel}, David J. and {Lang}, Dustin and {Blum}, Robert and {Burleigh}, Kaylan and {Fan}, Xiaohui and {Findlay}, Joseph R. and {Finkbeiner}, Doug and {Herrera}, David and {Juneau}, St{\'e}phanie and {Landriau}, Martin and {Levi}, Michael and {McGreer}, Ian and {Meisner}, Aaron and {Myers}, Adam D. and {Moustakas}, John and {Nugent}, Peter and {Patej}, Anna and {Schlafly}, Edward F. and {Walker}, Alistair R. and {Valdes}, Francisco and {Weaver}, Benjamin A. and {Y{\`e}che}, Christophe and {Zou}, Hu and {Zhou}, Xu and {Abareshi}, Behzad and {Abbott}, T.~M.~C. and {Abolfathi}, Bela and {Aguilera}, C. and {Alam}, Shadab and {Allen}, Lori and {Alvarez}, A. and {Annis}, James and {Ansarinejad}, Behzad and {Aubert}, Marie and {Beechert}, Jacqueline and {Bell}, Eric F. and {BenZvi}, Segev Y. and {Beutler}, Florian and {Bielby}, Richard M. and {Bolton}, Adam S. and {Brice{\~n}o}, C{\'e}sar and {Buckley-Geer}, Elizabeth J. and {Butler}, Karen and {Calamida}, Annalisa and {Carlberg}, Raymond G. and {Carter}, Paul and {Casas}, Ricard and {Castander}, Francisco J. and {Choi}, Yumi and {Comparat}, Johan and {Cukanovaite}, Elena and {Delubac}, Timoth{\'e}e and {DeVries}, Kaitlin and {Dey}, Sharmila and {Dhungana}, Govinda and {Dickinson}, Mark and {Ding}, Zhejie and {Donaldson}, John B. and {Duan}, Yutong and {Duckworth}, Christopher J. and {Eftekharzadeh}, Sarah and {Eisenstein}, Daniel J. and {Etourneau}, Thomas and {Fagrelius}, Parker A. and {Farihi}, Jay and {Fitzpatrick}, Mike and {Font-Ribera}, Andreu and {Fulmer}, Leah and {G{\"a}nsicke}, Boris T. and {Gaztanaga}, Enrique and {George}, Koshy and {Gerdes}, David W. and {Gontcho}, Satya Gontcho A. and {Gorgoni}, Claudio and {Green}, Gregory and {Guy}, Julien and {Harmer}, Diane and {Hernandez}, M. and {Honscheid}, Klaus and {Huang}, Lijuan Wendy and {James}, David J. and {Jannuzi}, Buell T. and {Jiang}, Linhua and {Joyce}, Richard and {Karcher}, Armin and {Karkar}, Sonia and {Kehoe}, Robert and {Kneib}, Jean-Paul and {Kueter-Young}, Andrea and {Lan}, Ting-Wen and {Lauer}, Tod R. and {Le Guillou}, Laurent and {Le Van Suu}, Auguste and {Lee}, Jae Hyeon and {Lesser}, Michael and {Perreault Levasseur}, Laurence and {Li}, Ting S. and {Mann}, Justin L. and {Marshall}, Robert and {Mart{\'\i}nez-V{\'a}zquez}, C.~E. and {Martini}, Paul and {du Mas des Bourboux}, H{\'e}lion and {McManus}, Sean and {Meier}, Tobias Gabriel and {M{\'e}nard}, Brice and {Metcalfe}, Nigel and {Mu{\~n}oz-Guti{\'e}rrez}, Andrea and {Najita}, Joan and {Napier}, Kevin and {Narayan}, Gautham and {Newman}, Jeffrey A. and {Nie}, Jundan and {Nord}, Brian and {Norman}, Dara J. and {Olsen}, Knut A.~G. and {Paat}, Anthony and {Palanque-Delabrouille}, Nathalie and {Peng}, Xiyan and {Poppett}, Claire L. and {Poremba}, Megan R. and {Prakash}, Abhishek and {Rabinowitz}, David and {Raichoor}, Anand and {Rezaie}, Mehdi and {Robertson}, A.~N. and {Roe}, Natalie A. and {Ross}, Ashley J. and {Ross}, Nicholas P. and {Rudnick}, Gregory and {Safonova}, Sasha and {Saha}, Abhijit and {S{\'a}nchez}, F. Javier and {Savary}, Elodie and {Schweiker}, Heidi and {Scott}, Adam and {Seo}, Hee-Jong and {Shan}, Huanyuan and {Silva}, David R. and {Slepian}, Zachary and {Soto}, Christian and {Sprayberry}, David and {Staten}, Ryan and {Stillman}, Coley M. and {Stupak}, Robert J. and {Summers}, David L. and {Sien Tie}, Suk and {Tirado}, H. and {Vargas-Maga{\~n}a}, Mariana and {Vivas}, A. Katherina and {Wechsler}, Risa H. and {Williams}, Doug and {Yang}, Jinyi and {Yang}, Qian and {Yapici}, Tolga and {Zaritsky}, Dennis and {Zenteno}, A. and {Zhang}, Kai and {Zhang}, Tianmeng and {Zhou}, Rongpu and {Zhou}, Zhimin},
        title = "{Overview of the DESI Legacy Imaging Surveys}",
      journal = {\aj},
         year = 2019,
        month = may,
       volume = {157},
       number = {5},
          eid = {168},
        pages = {168},
          doi = {10.3847/1538-3881/ab089d},
archivePrefix = {arXiv},
       eprint = {1804.08657},
 primaryClass = {astro-ph.IM},
       adsurl = {https://ui.adsabs.harvard.edu/abs/2019AJ....157..168D}
}

@ARTICLE{Zakamska2004,
       author = {{Zakamska}, Nadia L. and {Strauss}, Michael A. and {Heckman}, Timothy M. and {Ivezi{\'c}}, {\v{Z}}eljko and {Krolik}, Julian H.},
        title = "{Candidate Type II Quasars from the Sloan Digital Sky Survey. II. From Radio to X-Rays}",
      journal = {\aj},
         year = 2004,
        month = sep,
       volume = {128},
       number = {3},
        pages = {1002-1016},
          doi = {10.1086/423220},
archivePrefix = {arXiv},
       eprint = {astro-ph/0406248},
 primaryClass = {astro-ph},
       adsurl = {https://ui.adsabs.harvard.edu/abs/2004AJ....128.1002Z}
}

@ARTICLE{Harrison2015,
       author = {{Harrison}, C.~M. and {Thomson}, A.~P. and {Alexander}, D.~M. and {Bauer}, F.~E. and {Edge}, A.~C. and {Hogan}, M.~T. and {Mullaney}, J.~R. and {Swinbank}, A.~M.},
        title = "{Storm in a ``Teacup'': A Radio-quiet Quasar with {\ensuremath{\approx}}10 kpc Radio-emitting Bubbles and Extreme Gas Kinematics}",
      journal = {\apj},
         year = 2015,
        month = feb,
       volume = {800},
       number = {1},
          eid = {45},
        pages = {45},
          doi = {10.1088/0004-637X/800/1/45},
archivePrefix = {arXiv},
       eprint = {1410.4198},
 primaryClass = {astro-ph.GA},
       adsurl = {https://ui.adsabs.harvard.edu/abs/2015ApJ...800...45H}
}

@ARTICLE{Venturi2021,
       author = {{Venturi}, G. and {Cresci}, G. and {Marconi}, A. and {Mingozzi}, M. and {Nardini}, E. and {Carniani}, S. and {Mannucci}, F. and {Marasco}, A. and {Maiolino}, R. and {Perna}, M. and {Treister}, E. and {Bland-Hawthorn}, J. and {Gallimore}, J.},
        title = "{MAGNUM survey: Compact jets causing large turmoil in galaxies. Enhanced line widths perpendicular to radio jets as tracers of jet-ISM interaction}",
      journal = {\aap},
         year = 2021,
        month = apr,
       volume = {648},
          eid = {A17},
        pages = {A17},
          doi = {10.1051/0004-6361/202039869},
archivePrefix = {arXiv},
       eprint = {2011.04677},
 primaryClass = {astro-ph.GA},
       adsurl = {https://ui.adsabs.harvard.edu/abs/2021A&A...648A..17V}
}

@ARTICLE{Kellerman2016,
       author = {{Kellermann}, K.~I. and {Condon}, J.~J. and {Kimball}, A.~E. and {Perley}, R.~A. and {Ivezi{\'c}}, {\v{Z}}eljko},
        title = "{Radio-loud and Radio-quiet QSOs}",
      journal = {\apj},
         year = 2016,
        month = nov,
       volume = {831},
       number = {2},
          eid = {168},
        pages = {168},
          doi = {10.3847/0004-637X/831/2/168},
archivePrefix = {arXiv},
       eprint = {1608.04586},
 primaryClass = {astro-ph.GA},
       adsurl = {https://ui.adsabs.harvard.edu/abs/2016ApJ...831..168K}
}

@ARTICLE{Condon1998,
       author = {{Condon}, J.~J. and {Cotton}, W.~D. and {Greisen}, E.~W. and {Yin}, Q.~F. and {Perley}, R.~A. and {Taylor}, G.~B. and {Broderick}, J.~J.},
        title = "{The NRAO VLA Sky Survey}",
      journal = {\aj},
         year = 1998,
        month = may,
       volume = {115},
       number = {5},
        pages = {1693-1716},
          doi = {10.1086/300337},
       adsurl = {https://ui.adsabs.harvard.edu/abs/1998AJ....115.1693C}
}

@ARTICLE{Molyneux2024,
       author = {{Molyneux}, S.~J. and {Calistro Rivera}, G. and {De Breuck}, C. and {Harrison}, C.~M. and {Mainieri}, V. and {Lundgren}, A. and {Kakkad}, D. and {Circosta}, C. and {Girdhar}, A. and {Costa}, T. and {Mullaney}, J.~R. and {Kharb}, P. and {Arrigoni Battaia}, F. and {Farina}, E.~P. and {Alexander}, D.~M. and {Ward}, S.~R. and {Silpa}, S. and {Smit}, R.},
        title = "{The Quasar Feedback Survey: characterizing CO excitation in quasar host galaxies}",
      journal = {\mnras},
         year = 2024,
        month = jan,
       volume = {527},
       number = {3},
        pages = {4420-4439},
          doi = {10.1093/mnras/stad3133},
archivePrefix = {arXiv},
       eprint = {2310.10235},
 primaryClass = {astro-ph.GA},
       adsurl = {https://ui.adsabs.harvard.edu/abs/2024MNRAS.527.4420M}
}

@ARTICLE{Silpa2022,
       author = {{Silpa}, S. and {Kharb}, P. and {Harrison}, C.~M. and {Girdhar}, A. and {Mukherjee}, D. and {Mainieri}, V. and {Jarvis}, M.~E.},
        title = "{The Quasar Feedback Survey: revealing the interplay of jets, winds, and emission-line gas in type 2 quasars with radio polarization}",
      journal = {\mnras},
         year = 2022,
        month = jul,
       volume = {513},
       number = {3},
        pages = {4208-4223},
          doi = {10.1093/mnras/stac1044},
archivePrefix = {arXiv},
       eprint = {2204.05613},
 primaryClass = {astro-ph.GA},
       adsurl = {https://ui.adsabs.harvard.edu/abs/2022MNRAS.513.4208S}
}

@ARTICLE{An_Baan2012,
       author = {{An}, Tao and {Baan}, Willem A.},
        title = "{The Dynamic Evolution of Young Extragalactic Radio Sources}",
      journal = {\apj},
         year = 2012,
        month = nov,
       volume = {760},
       number = {1},
          eid = {77},
        pages = {77},
          doi = {10.1088/0004-637X/760/1/77},
archivePrefix = {arXiv},
       eprint = {1211.1760},
 primaryClass = {astro-ph.CO},
       adsurl = {https://ui.adsabs.harvard.edu/abs/2012ApJ...760...77A}
}

@ARTICLE{Mukherjee2018,
       author = {{Mukherjee}, Dipanjan and {Bicknell}, Geoffrey V. and {Wagner}, Alexander Y. and {Sutherland}, Ralph S. and {Silk}, Joseph},
        title = "{Relativistic jet feedback - III. Feedback on gas discs}",
      journal = {\mnras},
         year = 2018,
        month = oct,
       volume = {479},
       number = {4},
        pages = {5544-5566},
          doi = {10.1093/mnras/sty1776},
archivePrefix = {arXiv},
       eprint = {1803.08305},
 primaryClass = {astro-ph.HE},
       adsurl = {https://ui.adsabs.harvard.edu/abs/2018MNRAS.479.5544M}
}

@ARTICLE{Meenakshi2022,
       author = {{Meenakshi}, Moun and {Mukherjee}, Dipanjan and {Wagner}, Alexander Y. and {Nesvadba}, Nicole P.~H. and {Bicknell}, Geoffrey V. and {Morganti}, Raffaella and {Janssen}, Reinier M.~J. and {Sutherland}, Ralph S. and {Mandal}, Ankush},
        title = "{Modelling observable signatures of jet-ISM interaction: thermal emission and gas kinematics}",
      journal = {\mnras},
         year = 2022,
        month = oct,
       volume = {516},
       number = {1},
        pages = {766-786},
          doi = {10.1093/mnras/stac2251},
archivePrefix = {arXiv},
       eprint = {2203.10251},
 primaryClass = {astro-ph.GA},
       adsurl = {https://ui.adsabs.harvard.edu/abs/2022MNRAS.516..766M}
}

@ARTICLE{Meenaksh2024,
       author = {{Meenakshi}, Moun and {Mukherjee}, Dipanjan and {Bodo}, Gianluigi and {Rossi}, Paola and {Harrison}, Chris M.},
        title = "{A comparative study of radio signatures from winds and jets: modelling synchrotron emission and polarization}",
      journal = {\mnras},
         year = 2024,
        month = sep,
       volume = {533},
       number = {2},
        pages = {2213-2231},
          doi = {10.1093/mnras/stae1890},
archivePrefix = {arXiv},
       eprint = {2408.00099},
 primaryClass = {astro-ph.HE},
       adsurl = {https://ui.adsabs.harvard.edu/abs/2024MNRAS.533.2213M}
}

@ARTICLE{Baldi2021a,
       author = {{Baldi}, Ranieri D. and {Giovannini}, Gabriele and {Capetti}, Alessandro},
        title = "{The eMERLIN and EVN View of FR 0 Radio Galaxies}",
      journal = {Galaxies},
         year = 2021,
        month = nov,
       volume = {9},
       number = {4},
          eid = {106},
        pages = {106},
          doi = {10.3390/galaxies9040106},
archivePrefix = {arXiv},
       eprint = {2111.09899},
 primaryClass = {astro-ph.GA},
       adsurl = {https://ui.adsabs.harvard.edu/abs/2021Galax...9..106B}
}

@article{HeckmanBest2014,
author = {Heckman, Timothy M. and Best, Philip N.},
title = {The Coevolution of Galaxies and Supermassive Black Holes: Insights from Surveys of the Contemporary Universe},
journal = {Annual Review of Astronomy and Astrophysics},
volume = {52},
number = {1},
pages = {589-660},
year = {2014},
doi = {10.1146/annurev-astro-081913-035722},

URL = { 
        https://doi.org/10.1146/annurev-astro-081913-035722},
eprint = { 
        https://doi.org/10.1146/annurev-astro-081913-035722}
}

@ARTICLE{Ahumada2020,
       author = {{Ahumada}, Romina and {Allende Prieto}, Carlos and {Almeida}, Andr{\'e}s and {Anders}, Friedrich and {Anderson}, Scott F. and {Andrews}, Brett H. and {Anguiano}, Borja and {Arcodia}, Riccardo and {Armengaud}, Eric and {Aubert}, Marie and {Avila}, Santiago and {Avila-Reese}, Vladimir and {Badenes}, Carles and {Balland}, Christophe and {Barger}, Kat and {Barrera-Ballesteros}, Jorge K. and {Basu}, Sarbani and {Bautista}, Julian and {Beaton}, Rachael L. and {Beers}, Timothy C. and {Benavides}, B. Izamar T. and {Bender}, Chad F. and {Bernardi}, Mariangela and {Bershady}, Matthew and {Beutler}, Florian and {Bidin}, Christian Moni and {Bird}, Jonathan and {Bizyaev}, Dmitry and {Blanc}, Guillermo A. and {Blanton}, Michael R. and {Boquien}, M{\'e}d{\'e}ric and {Borissova}, Jura and {Bovy}, Jo and {Brandt}, W.~N. and {Brinkmann}, Jonathan and {Brownstein}, Joel R. and {Bundy}, Kevin and {Bureau}, Martin and {Burgasser}, Adam and {Burtin}, Etienne and {Cano-D{\'\i}az}, Mariana and {Capasso}, Raffaella and {Cappellari}, Michele and {Carrera}, Ricardo and {Chabanier}, Sol{\`e}ne and {Chaplin}, William and {Chapman}, Michael and {Cherinka}, Brian and {Chiappini}, Cristina and {Doohyun Choi}, Peter and {Chojnowski}, S. Drew and {Chung}, Haeun and {Clerc}, Nicolas and {Coffey}, Damien and {Comerford}, Julia M. and {Comparat}, Johan and {da Costa}, Luiz and {Cousinou}, Marie-Claude and {Covey}, Kevin and {Crane}, Jeffrey D. and {Cunha}, Katia and {Ilha}, Gabriele da Silva and {Dai}, Yu Sophia and {Damsted}, Sanna B. and {Darling}, Jeremy and {Davidson}, Jr., James W. and {Davies}, Roger and {Dawson}, Kyle and {De}, Nikhil and {de la Macorra}, Axel and {De Lee}, Nathan and {Queiroz}, Anna B{\'a}rbara de Andrade and {Deconto Machado}, Alice and {de la Torre}, Sylvain and {Dell'Agli}, Flavia and {du Mas des Bourboux}, H{\'e}lion and {Diamond-Stanic}, Aleksandar M. and {Dillon}, Sean and {Donor}, John and {Drory}, Niv and {Duckworth}, Chris and {Dwelly}, Tom and {Ebelke}, Garrett and {Eftekharzadeh}, Sarah and {Davis Eigenbrot}, Arthur and {Elsworth}, Yvonne P. and {Eracleous}, Mike and {Erfanianfar}, Ghazaleh and {Escoffier}, Stephanie and {Fan}, Xiaohui and {Farr}, Emily and {Fern{\'a}ndez-Trincado}, Jos{\'e} G. and {Feuillet}, Diane and {Finoguenov}, Alexis and {Fofie}, Patricia and {Fraser-McKelvie}, Amelia and {Frinchaboy}, Peter M. and {Fromenteau}, Sebastien and {Fu}, Hai and {Galbany}, Llu{\'\i}s and {Garcia}, Rafael A. and {Garc{\'\i}a-Hern{\'a}ndez}, D.~A. and {Garma Oehmichen}, Luis Alberto and {Ge}, Junqiang and {Geimba Maia}, Marcio Antonio and {Geisler}, Doug and {Gelfand}, Joseph and {Goddy}, Julian and {Gonzalez-Perez}, Violeta and {Grabowski}, Kathleen and {Green}, Paul and {Grier}, Catherine J. and {Guo}, Hong and {Guy}, Julien and {Harding}, Paul and {Hasselquist}, Sten and {Hawken}, Adam James and {Hayes}, Christian R. and {Hearty}, Fred and {Hekker}, S. and {Hogg}, David W. and {Holtzman}, Jon A. and {Horta}, Danny and {Hou}, Jiamin and {Hsieh}, Bau-Ching and {Huber}, Daniel and {Hunt}, Jason A.~S. and {Ider Chitham}, J. and {Imig}, Julie and {Jaber}, Mariana and {Jimenez Angel}, Camilo Eduardo and {Johnson}, Jennifer A. and {Jones}, Amy M. and {J{\"o}nsson}, Henrik and {Jullo}, Eric and {Kim}, Yerim and {Kinemuchi}, Karen and {Kirkpatrick}, IV, Charles C. and {Kite}, George W. and {Klaene}, Mark and {Kneib}, Jean-Paul and {Kollmeier}, Juna A. and {Kong}, Hui and {Kounkel}, Marina and {Krishnarao}, Dhanesh and {Lacerna}, Ivan and {Lan}, Ting-Wen and {Lane}, Richard R. and {Law}, David R. and {Le Goff}, Jean-Marc and {Leung}, Henry W. and {Lewis}, Hannah and {Li}, Cheng and {Lian}, Jianhui and {Lin}, Lihwai and {Long}, Dan and {Longa-Pe{\~n}a}, Pen{\'e}lope and {Lundgren}, Britt and {Lyke}, Brad W. and {Mackereth}, J. Ted and {MacLeod}, Chelsea L. and {Majewski}, Steven R. and {Manchado}, Arturo and {Maraston}, Claudia and {Martini}, Paul and {Masseron}, Thomas and {Masters}, Karen L. and {Mathur}, Savita and {McDermid}, Richard M. and {Merloni}, Andrea and {Merrifield}, Michael and {M{\'e}sz{\'a}ros}, Szabolcs and {Miglio}, Andrea and {Minniti}, Dante and {Minsley}, Rebecca and {Miyaji}, Takamitsu and {Mohammad}, Faizan Gohar and {Mosser}, Benoit and {Mueller}, Eva-Maria and {Muna}, Demitri and {Mu{\~n}oz-Guti{\'e}rrez}, Andrea and {Myers}, Adam D. and {Nadathur}, Seshadri and {Nair}, Preethi and {Nandra}, Kirpal and {Correa do Nascimento}, Janaina and {Nevin}, Rebecca Jean and {Newman}, Jeffrey A. and {Nidever}, David L. and {Nitschelm}, Christian and {Noterdaeme}, Pasquier and {O'Connell}, Julia E. and {Olmstead}, Matthew D. and {Oravetz}, Daniel and {Oravetz}, Audrey and {Osorio}, Yeisson and {Pace}, Zachary J. and {Padilla}, Nelson and {Palanque-Delabrouille}, Nathalie and {Palicio}, Pedro A.},
        title = "{The 16th Data Release of the Sloan Digital Sky Surveys: First Release from the APOGEE-2 Southern Survey and Full Release of eBOSS Spectra}",
      journal = {\apjs},
         year = 2020,
        month = jul,
       volume = {249},
       number = {1},
          eid = {3},
        pages = {3},
          doi = {10.3847/1538-4365/ab929e},
archivePrefix = {arXiv},
       eprint = {1912.02905},
 primaryClass = {astro-ph.GA},
       adsurl = {https://ui.adsabs.harvard.edu/abs/2020ApJS..249....3A}
}

@ARTICLE{Padovani2016,
       author = {{Padovani}, Paolo},
        title = "{The faint radio sky: radio astronomy becomes mainstream}",
      journal = {\aapr},
         year = 2016,
        month = sep,
       volume = {24},
       number = {1},
          eid = {13},
        pages = {13},
          doi = {10.1007/s00159-016-0098-6},
archivePrefix = {arXiv},
       eprint = {1609.00499},
 primaryClass = {astro-ph.GA},
       adsurl = {https://ui.adsabs.harvard.edu/abs/2016A&ARv..24...13P}
}

@ARTICLE{Condon1992,
       author = {{Condon}, J.~J.},
        title = "{Radio emission from normal galaxies.}",
      journal = {\araa},
         year = 1992,
        month = jan,
       volume = {30},
        pages = {575-611},
          doi = {10.1146/annurev.aa.30.090192.003043},
       adsurl = {https://ui.adsabs.harvard.edu/abs/1992ARA&A..30..575C}
}

@ARTICLE{Tadhunter2016,
       author = {{Tadhunter}, C.},
        title = "{The impact of compact radio sources on their host galaxies: observations}",
      journal = {Astronomische Nachrichten},
         year = 2016,
        month = feb,
       volume = {337},
       number = {1-2},
        pages = {159},
          doi = {10.1002/asna.201512286},
       adsurl = {https://ui.adsabs.harvard.edu/abs/2016AN....337..159T}
}

@ARTICLE{McCaffrey2022,
       author = {{McCaffrey}, Trevor V. and {Kimball}, Amy E. and {Momjian}, Emmanuel and {Richards}, Gordon T.},
        title = "{Kiloparsec-scale Radio Structure in z   0.25 Radio-quiet QSOs}",
      journal = {\aj},
         year = 2022,
        month = oct,
       volume = {164},
       number = {4},
          eid = {122},
        pages = {122},
          doi = {10.3847/1538-3881/ac853e},
archivePrefix = {arXiv},
       eprint = {2207.13792},
 primaryClass = {astro-ph.GA},
       adsurl = {https://ui.adsabs.harvard.edu/abs/2022AJ....164..122M}
}

@ARTICLE{Lansbury2018,
       author = {{Lansbury}, George B. and {Jarvis}, Miranda E. and {Harrison}, Chris M. and {Alexander}, David M. and {Del Moro}, Agnese and {Edge}, Alastair C. and {Mullaney}, James R. and {Thomson}, Alasdair P.},
        title = "{Storm in a Teacup: X-Ray View of an Obscured Quasar and Superbubble}",
      journal = {\apjl},
         year = 2018,
        month = mar,
       volume = {856},
       number = {1},
          eid = {L1},
        pages = {L1},
          doi = {10.3847/2041-8213/aab357},
archivePrefix = {arXiv},
       eprint = {1803.00009},
 primaryClass = {astro-ph.GA},
       adsurl = {https://ui.adsabs.harvard.edu/abs/2018ApJ...856L...1L}
}

@ARTICLE{Chilufya2024,
       author = {{Chilufya}, J. and {Hardcastle}, M.~J. and {Pierce}, J.~C.~S. and {Croston}, J.~H. and {Mingo}, B. and {Zheng}, X. and {Baldi}, R.~D. and {R{\"o}ttgering}, H.~J.~A.},
        title = "{The nature of compact radio-loud AGN: a systematic look at the LOFAR AGN population}",
      journal = {\mnras},
         year = 2024,
        month = apr,
       volume = {529},
       number = {2},
        pages = {1472-1492},
          doi = {10.1093/mnras/stae658},
archivePrefix = {arXiv},
       eprint = {2402.19424},
 primaryClass = {astro-ph.GA},
       adsurl = {https://ui.adsabs.harvard.edu/abs/2024MNRAS.529.1472C}
}

@ARTICLE{Fawcett2025,
       author = {{Fawcett}, V.~A. and {Harrison}, C.~M. and {Alexander}, D.~M. and {Morabito}, L.~K. and {Kharb}, P. and {Rosario}, D.~J. and {Baghel}, Janhavi and {Ghosh}, Salmoli and {Silpa}, Sasikumar and {Petley}, J. and {Sargent}, C. and {Calistro Rivera}, G.},
        title = "{Connection between steep radio spectral slopes and dust extinction in QSOs: evidence for outflow-driven shocks in dusty QSOs}",
      journal = {\mnras},
         year = 2025,
        month = feb,
       volume = {537},
       number = {2},
        pages = {2003-2023},
          doi = {10.1093/mnras/staf105},
archivePrefix = {arXiv},
       eprint = {2501.10501},
 primaryClass = {astro-ph.GA},
       adsurl = {https://ui.adsabs.harvard.edu/abs/2025MNRAS.537.2003F}
}

@ARTICLE{SDSS72009,
       author = {{Abazajian}, Kevork N. and {Adelman-McCarthy}, Jennifer K. and {Ag{\"u}eros}, Marcel A. and {Allam}, Sahar S. and {Allende Prieto}, Carlos and {An}, Deokkeun and {Anderson}, Kurt S.~J. and {Anderson}, Scott F. and {Annis}, James and {Bahcall}, Neta A. and {Bailer-Jones}, C.~A.~L. and {Barentine}, J.~C. and {Bassett}, Bruce A. and {Becker}, Andrew C. and {Beers}, Timothy C. and {Bell}, Eric F. and {Belokurov}, Vasily and {Berlind}, Andreas A. and {Berman}, Eileen F. and {Bernardi}, Mariangela and {Bickerton}, Steven J. and {Bizyaev}, Dmitry and {Blakeslee}, John P. and {Blanton}, Michael R. and {Bochanski}, John J. and {Boroski}, William N. and {Brewington}, Howard J. and {Brinchmann}, Jarle and {Brinkmann}, J. and {Brunner}, Robert J. and {Budav{\'a}ri}, Tam{\'a}s and {Carey}, Larry N. and {Carliles}, Samuel and {Carr}, Michael A. and {Castander}, Francisco J. and {Cinabro}, David and {Connolly}, A.~J. and {Csabai}, Istv{\'a}n and {Cunha}, Carlos E. and {Czarapata}, Paul C. and {Davenport}, James R.~A. and {de Haas}, Ernst and {Dilday}, Ben and {Doi}, Mamoru and {Eisenstein}, Daniel J. and {Evans}, Michael L. and {Evans}, N.~W. and {Fan}, Xiaohui and {Friedman}, Scott D. and {Frieman}, Joshua A. and {Fukugita}, Masataka and {G{\"a}nsicke}, Boris T. and {Gates}, Evalyn and {Gillespie}, Bruce and {Gilmore}, G. and {Gonzalez}, Belinda and {Gonzalez}, Carlos F. and {Grebel}, Eva K. and {Gunn}, James E. and {Gy{\"o}ry}, Zsuzsanna and {Hall}, Patrick B. and {Harding}, Paul and {Harris}, Frederick H. and {Harvanek}, Michael and {Hawley}, Suzanne L. and {Hayes}, Jeffrey J.~E. and {Heckman}, Timothy M. and {Hendry}, John S. and {Hennessy}, Gregory S. and {Hindsley}, Robert B. and {Hoblitt}, J. and {Hogan}, Craig J. and {Hogg}, David W. and {Holtzman}, Jon A. and {Hyde}, Joseph B. and {Ichikawa}, Shin-ichi and {Ichikawa}, Takashi and {Im}, Myungshin and {Ivezi{\'c}}, {\v{Z}}eljko and {Jester}, Sebastian and {Jiang}, Linhua and {Johnson}, Jennifer A. and {Jorgensen}, Anders M. and {Juri{\'c}}, Mario and {Kent}, Stephen M. and {Kessler}, R. and {Kleinman}, S.~J. and {Knapp}, G.~R. and {Konishi}, Kohki and {Kron}, Richard G. and {Krzesinski}, Jurek and {Kuropatkin}, Nikolay and {Lampeitl}, Hubert and {Lebedeva}, Svetlana and {Lee}, Myung Gyoon and {Lee}, Young Sun and {French Leger}, R. and {L{\'e}pine}, S{\'e}bastien and {Li}, Nolan and {Lima}, Marcos and {Lin}, Huan and {Long}, Daniel C. and {Loomis}, Craig P. and {Loveday}, Jon and {Lupton}, Robert H. and {Magnier}, Eugene and {Malanushenko}, Olena and {Malanushenko}, Viktor and {Mandelbaum}, Rachel and {Margon}, Bruce and {Marriner}, John P. and {Mart{\'\i}nez-Delgado}, David and {Matsubara}, Takahiko and {McGehee}, Peregrine M. and {McKay}, Timothy A. and {Meiksin}, Avery and {Morrison}, Heather L. and {Mullally}, Fergal and {Munn}, Jeffrey A. and {Murphy}, Tara and {Nash}, Thomas and {Nebot}, Ada and {Neilsen}, Jr., Eric H. and {Newberg}, Heidi Jo and {Newman}, Peter R. and {Nichol}, Robert C. and {Nicinski}, Tom and {Nieto-Santisteban}, Maria and {Nitta}, Atsuko and {Okamura}, Sadanori and {Oravetz}, Daniel J. and {Ostriker}, Jeremiah P. and {Owen}, Russell and {Padmanabhan}, Nikhil and {Pan}, Kaike and {Park}, Changbom and {Pauls}, George and {Peoples}, Jr., John and {Percival}, Will J. and {Pier}, Jeffrey R. and {Pope}, Adrian C. and {Pourbaix}, Dimitri and {Price}, Paul A. and {Purger}, Norbert and {Quinn}, Thomas and {Raddick}, M. Jordan and {Re Fiorentin}, Paola and {Richards}, Gordon T. and {Richmond}, Michael W. and {Riess}, Adam G. and {Rix}, Hans-Walter and {Rockosi}, Constance M. and {Sako}, Masao and {Schlegel}, David J. and {Schneider}, Donald P. and {Scholz}, Ralf-Dieter and {Schreiber}, Matthias R. and {Schwope}, Axel D. and {Seljak}, Uro{\v{s}} and {Sesar}, Branimir and {Sheldon}, Erin and {Shimasaku}, Kazu and {Sibley}, Valena C. and {Simmons}, A.~E. and {Sivarani}, Thirupathi and {Allyn Smith}, J. and {Smith}, Martin C. and {Smol{\v{c}}i{\'c}}, Vernesa and {Snedden}, Stephanie A. and {Stebbins}, Albert and {Steinmetz}, Matthias and {Stoughton}, Chris and {Strauss}, Michael A. and {SubbaRao}, Mark and {Suto}, Yasushi and {Szalay}, Alexander S. and {Szapudi}, Istv{\'a}n and {Szkody}, Paula and {Tanaka}, Masayuki and {Tegmark}, Max and {Teodoro}, Luis F.~A. and {Thakar}, Aniruddha R. and {Tremonti}, Christy A. and {Tucker}, Douglas L. and {Uomoto}, Alan and {Vanden Berk}, Daniel E. and {Vandenberg}, Jan and {Vidrih}, S. and {Vogeley}, Michael S. and {Voges}, Wolfgang and {Vogt}, Nicole P. and {Wadadekar}, Yogesh and {Watters}, Shannon and {Weinberg}, David H. and {West}, Andrew A. and {White}, Simon D.~M. and {Wilhite}, Brian C. and {Wonders}, Alainna C. and {Yanny}, Brian and {Yocum}, D.~R.},
        title = "{The Seventh Data Release of the Sloan Digital Sky Survey}",
      journal = {\apjs},
         year = 2009,
        month = jun,
       volume = {182},
       number = {2},
        pages = {543-558},
          doi = {10.1088/0067-0049/182/2/543},
archivePrefix = {arXiv},
       eprint = {0812.0649},
 primaryClass = {astro-ph},
       adsurl = {https://ui.adsabs.harvard.edu/abs/2009ApJS..182..543A}
}

@ARTICLE{Kharb2023,
       author = {{Kharb}, Preeti and {Silpa}, Sasikumar},
        title = "{Looking for Signatures of AGN Feedback in Radio-Quiet AGN}",
      journal = {Galaxies},
         year = 2023,
        month = feb,
       volume = {11},
       number = {1},
          eid = {27},
        pages = {27},
          doi = {10.3390/galaxies11010027},
archivePrefix = {arXiv},
       eprint = {2302.06954},
 primaryClass = {astro-ph.GA},
       adsurl = {https://ui.adsabs.harvard.edu/abs/2023Galax..11...27K}
}

@ARTICLE{Silpa2020,
       author = {{Silpa}, S. and {Kharb}, P. and {Ho}, L.~C. and {Ishwara-Chandra}, C.~H. and {Jarvis}, M.~E. and {Harrison}, C.},
        title = "{Probing the origin of low-frequency radio emission in PG quasars with the uGMRT - I}",
      journal = {\mnras},
         year = 2020,
        month = dec,
       volume = {499},
       number = {4},
        pages = {5826-5839},
          doi = {10.1093/mnras/staa2970},
archivePrefix = {arXiv},
       eprint = {2009.12264},
 primaryClass = {astro-ph.GA},
       adsurl = {https://ui.adsabs.harvard.edu/abs/2020MNRAS.499.5826S}
}

@ARTICLE{Fawcett2020,
       author = {{Fawcett}, V.~A. and {Alexander}, D.~M. and {Rosario}, D.~J. and {Klindt}, L. and {Fotopoulou}, S. and {Lusso}, E. and {Morabito}, L.~K. and {Calistro Rivera}, G.},
        title = "{Fundamental differences in the radio properties of red and blue quasars: enhanced compact AGN emission in red quasars}",
      journal = {\mnras},
         year = 2020,
        month = jun,
       volume = {494},
       number = {4},
        pages = {4802-4818},
          doi = {10.1093/mnras/staa954},
archivePrefix = {arXiv},
       eprint = {2004.01197},
 primaryClass = {astro-ph.GA},
       adsurl = {https://ui.adsabs.harvard.edu/abs/2020MNRAS.494.4802F}
}

@ARTICLE{Klindt2019,
       author = {{Klindt}, L. and {Alexander}, D.~M. and {Rosario}, D.~J. and {Lusso}, E. and {Fotopoulou}, S.},
        title = "{Fundamental differences in the radio properties of red and blue quasars: evolution strongly favoured over orientation}",
      journal = {\mnras},
         year = 2019,
        month = sep,
       volume = {488},
       number = {3},
        pages = {3109-3128},
          doi = {10.1093/mnras/stz1771},
archivePrefix = {arXiv},
       eprint = {1905.12108},
 primaryClass = {astro-ph.GA},
       adsurl = {https://ui.adsabs.harvard.edu/abs/2019MNRAS.488.3109K}
}

@ARTICLE{Radcliffe2018,
       author = {{Radcliffe}, J.~F. and {Garrett}, M.~A. and {Muxlow}, T.~W.~B. and {Beswick}, R.~J. and {Barthel}, P.~D. and {Deller}, A.~T. and {Keimpema}, A. and {Campbell}, R.~M. and {Wrigley}, N.},
        title = "{Nowhere to Hide: Radio-faint AGN in GOODS-N field. I. Initial catalogue and radio properties}",
      journal = {\aap},
         year = 2018,
        month = nov,
       volume = {619},
          eid = {A48},
        pages = {A48},
          doi = {10.1051/0004-6361/201833399},
archivePrefix = {arXiv},
       eprint = {1808.04296},
 primaryClass = {astro-ph.GA},
       adsurl = {https://ui.adsabs.harvard.edu/abs/2018A&A...619A..48R}
}

@ARTICLE{Njeri2023,
       author = {{Njeri}, Ann and {Beswick}, Robert J. and {Radcliffe}, Jack F. and {Thomson}, A.~P. and {Wrigley}, N. and {Muxlow}, T.~W.~B. and {Garrett}, M.~A. and {Deane}, Roger P. and {Moldon}, Javier and {Norris}, Ray P. and {Kothes}, Roland},
        title = "{SPARCS-North Wide-field VLBI Survey: exploring the resolved {\ensuremath{\mu}}Jy extragalactic radio source population with EVN + e-MERLIN}",
      journal = {\mnras},
         year = 2023,
        month = feb,
       volume = {519},
       number = {2},
        pages = {1732-1744},
          doi = {10.1093/mnras/stac3569},
archivePrefix = {arXiv},
       eprint = {2212.01874},
 primaryClass = {astro-ph.GA},
       adsurl = {https://ui.adsabs.harvard.edu/abs/2023MNRAS.519.1732N}
}

@ARTICLE{Wang2024,
       author = {{Wang}, Yijun and {Wang}, Tao and {Liu}, Daizhong and {Sargent}, Mark T. and {Gao}, Fangyou and {Alexander}, David M. and {Rujopakarn}, Wiphu and {Zhou}, Luwenjia and {Daddi}, Emanuele and {Xu}, Ke and {Kohno}, Kotaro and {Jin}, Shuowen},
        title = "{Cosmic evolution of radio-excess active galactic nuclei in quiescent and star-forming galaxies across 0 < z < 4}",
      journal = {\aap},
         year = 2024,
        month = may,
       volume = {685},
          eid = {A79},
        pages = {A79},
          doi = {10.1051/0004-6361/202347787},
archivePrefix = {arXiv},
       eprint = {2401.04924},
 primaryClass = {astro-ph.GA},
       adsurl = {https://ui.adsabs.harvard.edu/abs/2024A&A...685A..79W}
}

@ARTICLE{DelMoro2013,
       author = {{Del Moro}, A. and {Alexander}, D.~M. and {Mullaney}, J.~R. and {Daddi}, E. and {Pannella}, M. and {Bauer}, F.~E. and {Pope}, A. and {Dickinson}, M. and {Elbaz}, D. and {Barthel}, P.~D. and {Garrett}, M.~A. and {Brandt}, W.~N. and {Charmandaris}, V. and {Chary}, R.~R. and {Dasyra}, K. and {Gilli}, R. and {Hickox}, R.~C. and {Hwang}, H.~S. and {Ivison}, R.~J. and {Juneau}, S. and {Le Floc'h}, E. and {Luo}, B. and {Morrison}, G.~E. and {Rovilos}, E. and {Sargent}, M.~T. and {Xue}, Y.~Q.},
        title = "{GOODS-Herschel: radio-excess signature of hidden AGN activity in distant star-forming galaxies}",
      journal = {\aap},
         year = 2013,
        month = jan,
       volume = {549},
          eid = {A59},
        pages = {A59},
          doi = {10.1051/0004-6361/201219880},
archivePrefix = {arXiv},
       eprint = {1210.2521},
 primaryClass = {astro-ph.CO},
       adsurl = {https://ui.adsabs.harvard.edu/abs/2013A&A...549A..59D}
}

@ARTICLE{Delhaize2017,
       author = {{Delhaize}, J. and {Smol{\v{c}}i{\'c}}, V. and {Delvecchio}, I. and {Novak}, M. and {Sargent}, M. and {Baran}, N. and {Magnelli}, B. and {Zamorani}, G. and {Schinnerer}, E. and {Murphy}, E.~J. and {Aravena}, M. and {Berta}, S. and {Bondi}, M. and {Capak}, P. and {Carilli}, C. and {Ciliegi}, P. and {Civano}, F. and {Ilbert}, O. and {Karim}, A. and {Laigle}, C. and {Le F{\`e}vre}, O. and {Marchesi}, S. and {McCracken}, H.~J. and {Salvato}, M. and {Seymour}, N. and {Tasca}, L.},
        title = "{The VLA-COSMOS 3 GHz Large Project: The infrared-radio correlation of star-forming galaxies and AGN to z {\ensuremath{\lesssim}} 6}",
      journal = {\aap},
         year = 2017,
        month = jun,
       volume = {602},
          eid = {A4},
        pages = {A4},
          doi = {10.1051/0004-6361/201629430},
archivePrefix = {arXiv},
       eprint = {1703.09723},
 primaryClass = {astro-ph.GA},
       adsurl = {https://ui.adsabs.harvard.edu/abs/2017A&A...602A...4D}
}

@ARTICLE{Eberhard2025,
       author = {{Eberhard}, John-Michael and {Reines}, Amy E. and {Gim}, Hansung B. and {Darling}, Jeremy and {Greene}, Jenny E.},
        title = "{Dwarf Galaxies with Radio-excess Active Galactic Nuclei in the Very Large Array Sky Survey}",
      journal = {\apj},
         year = 2025,
        month = jan,
       volume = {978},
       number = {2},
          eid = {158},
        pages = {158},
          doi = {10.3847/1538-4357/ad9584},
       adsurl = {https://ui.adsabs.harvard.edu/abs/2025ApJ...978..158E}
}

@ARTICLE{Yun2001,
       author = {{Yun}, Min S. and {Reddy}, Naveen A. and {Condon}, J.~J.},
        title = "{Radio Properties of Infrared-selected Galaxies in the IRAS 2 Jy Sample}",
      journal = {\apj},
         year = 2001,
        month = jun,
       volume = {554},
       number = {2},
        pages = {803-822},
          doi = {10.1086/323145},
archivePrefix = {arXiv},
       eprint = {astro-ph/0102154},
 primaryClass = {astro-ph},
       adsurl = {https://ui.adsabs.harvard.edu/abs/2001ApJ...554..803Y}
}

@ARTICLE{Padovani2017,
       author = {{Padovani}, Paolo},
        title = "{On the two main classes of active galactic nuclei}",
      journal = {Nature Astronomy},
         year = 2017,
        month = aug,
       volume = {1},
          eid = {0194},
        pages = {0194},
          doi = {10.1038/s41550-017-0194},
archivePrefix = {arXiv},
       eprint = {1707.08069},
 primaryClass = {astro-ph.GA},
       adsurl = {https://ui.adsabs.harvard.edu/abs/2017NatAs...1E.194P}
}

@ARTICLE{Godfrey2016,
       author = {{Godfrey}, L.~E.~H. and {Shabala}, S.~S.},
        title = "{Mutual distance dependence drives the observed jet-power-radio-luminosity scaling relations in radio galaxies}",
      journal = {\mnras},
         year = 2016,
        month = feb,
       volume = {456},
       number = {2},
        pages = {1172-1184},
          doi = {10.1093/mnras/stv2712},
archivePrefix = {arXiv},
       eprint = {1511.06007},
 primaryClass = {astro-ph.GA},
       adsurl = {https://ui.adsabs.harvard.edu/abs/2016MNRAS.456.1172G}
}

@ARTICLE{RamosAlmeida2022,
       author = {{Ramos Almeida}, C. and {Bischetti}, M. and {Garc{\'\i}a-Burillo}, S. and {Alonso-Herrero}, A. and {Audibert}, A. and {Cicone}, C. and {Feruglio}, C. and {Tadhunter}, C.~N. and {Pierce}, J.~C.~S. and {Pereira-Santaella}, M. and {Bessiere}, P.~S.},
        title = "{The diverse cold molecular gas contents, morphologies, and kinematics of type-2 quasars as seen by ALMA}",
      journal = {\aap},
         year = 2022,
        month = feb,
       volume = {658},
          eid = {A155},
        pages = {A155},
          doi = {10.1051/0004-6361/202141906},
archivePrefix = {arXiv},
       eprint = {2111.13578},
 primaryClass = {astro-ph.GA},
       adsurl = {https://ui.adsabs.harvard.edu/abs/2022A&A...658A.155R}
}

@ARTICLE{Audibert2023,
       author = {{Audibert}, A. and {Ramos Almeida}, C. and {Garc{\'\i}a-Burillo}, S. and {Combes}, F. and {Bischetti}, M. and {Meenakshi}, M. and {Mukherjee}, D. and {Bicknell}, G. and {Wagner}, A.~Y.},
        title = "{Jet-induced molecular gas excitation and turbulence in the Teacup}",
      journal = {\aap},
         year = 2023,
        month = mar,
       volume = {671},
          eid = {L12},
        pages = {L12},
          doi = {10.1051/0004-6361/202345964},
archivePrefix = {arXiv},
       eprint = {2302.13884},
 primaryClass = {astro-ph.GA},
       adsurl = {https://ui.adsabs.harvard.edu/abs/2023A&A...671L..12A}
}

@ARTICLE{Bessiere2024,
       author = {{Bessiere}, P.~S. and {Ramos Almeida}, C. and {Holden}, L.~R. and {Tadhunter}, C.~N. and {Canalizo}, G.},
        title = "{QSOFEED: Relationship between star formation and active galactic nuclei feedback}",
      journal = {\aap},
         year = 2024,
        month = sep,
       volume = {689},
          eid = {A271},
        pages = {A271},
          doi = {10.1051/0004-6361/202348795},
archivePrefix = {arXiv},
       eprint = {2405.06421},
 primaryClass = {astro-ph.GA},
       adsurl = {https://ui.adsabs.harvard.edu/abs/2024A&A...689A.271B}
}

@ARTICLE{Harrison2014,
       author = {{Harrison}, C.~M. and {Alexander}, D.~M. and {Mullaney}, J.~R. and {Swinbank}, A.~M.},
        title = "{Kiloparsec-scale outflows are prevalent among luminous AGN: outflows and feedback in the context of the overall AGN population}",
      journal = {\mnras},
         year = 2014,
        month = jul,
       volume = {441},
       number = {4},
        pages = {3306-3347},
          doi = {10.1093/mnras/stu515},
archivePrefix = {arXiv},
       eprint = {1403.3086},
 primaryClass = {astro-ph.GA},
       adsurl = {https://ui.adsabs.harvard.edu/abs/2014MNRAS.441.3306H}
}

@ARTICLE{Murphy2011,
       author = {{Murphy}, E.~J. and {Condon}, J.~J. and {Schinnerer}, E. and {Kennicutt}, R.~C. and {Calzetti}, D. and {Armus}, L. and {Helou}, G. and {Turner}, J.~L. and {Aniano}, G. and {Beir{\~a}o}, P. and {Bolatto}, A.~D. and {Brandl}, B.~R. and {Croxall}, K.~V. and {Dale}, D.~A. and {Donovan Meyer}, J.~L. and {Draine}, B.~T. and {Engelbracht}, C. and {Hunt}, L.~K. and {Hao}, C. -N. and {Koda}, J. and {Roussel}, H. and {Skibba}, R. and {Smith}, J. -D.~T.},
        title = "{Calibrating Extinction-free Star Formation Rate Diagnostics with 33 GHz Free-free Emission in NGC 6946}",
      journal = {\apj},
         year = 2011,
        month = aug,
       volume = {737},
       number = {2},
          eid = {67},
        pages = {67},
          doi = {10.1088/0004-637X/737/2/67},
archivePrefix = {arXiv},
       eprint = {1105.4877},
 primaryClass = {astro-ph.CO},
       adsurl = {https://ui.adsabs.harvard.edu/abs/2011ApJ...737...67M}
}

@ARTICLE{Delvecchio2017,
       author = {{Delvecchio}, I. and {Smol{\v{c}}i{\'c}}, V. and {Zamorani}, G. and {Lagos}, C. Del P. and {Berta}, S. and {Delhaize}, J. and {Baran}, N. and {Alexander}, D.~M. and {Rosario}, D.~J. and {Gonzalez-Perez}, V. and {Ilbert}, O. and {Lacey}, C.~G. and {Le F{\`e}vre}, O. and {Miettinen}, O. and {Aravena}, M. and {Bondi}, M. and {Carilli}, C. and {Ciliegi}, P. and {Mooley}, K. and {Novak}, M. and {Schinnerer}, E. and {Capak}, P. and {Civano}, F. and {Fanidakis}, N. and {Herrera Ruiz}, N. and {Karim}, A. and {Laigle}, C. and {Marchesi}, S. and {McCracken}, H.~J. and {Middleberg}, E. and {Salvato}, M. and {Tasca}, L.},
        title = "{The VLA-COSMOS 3 GHz Large Project: AGN and host-galaxy properties out to z {\ensuremath{\lesssim}} 6}",
      journal = {\aap},
         year = 2017,
        month = jun,
       volume = {602},
          eid = {A3},
        pages = {A3},
          doi = {10.1051/0004-6361/201629367},
archivePrefix = {arXiv},
       eprint = {1703.09720},
 primaryClass = {astro-ph.GA},
       adsurl = {https://ui.adsabs.harvard.edu/abs/2017A&A...602A...3D}
}

@ARTICLE{Smolcic2017,
       author = {{Smol{\v{c}}i{\'c}}, V. and {Novak}, M. and {Delvecchio}, I. and {Ceraj}, L. and {Bondi}, M. and {Delhaize}, J. and {Marchesi}, S. and {Murphy}, E. and {Schinnerer}, E. and {Vardoulaki}, E. and {Zamorani}, G.},
        title = "{The VLA-COSMOS 3 GHz Large Project: Cosmic evolution of radio AGN and implications for radio-mode feedback since z   5}",
      journal = {\aap},
         year = 2017,
        month = jun,
       volume = {602},
          eid = {A6},
        pages = {A6},
          doi = {10.1051/0004-6361/201730685},
archivePrefix = {arXiv},
       eprint = {1705.07090},
 primaryClass = {astro-ph.GA},
       adsurl = {https://ui.adsabs.harvard.edu/abs/2017A&A...602A...6S}
}

@ARTICLE{Zakamska2016,
       author = {{Zakamska}, Nadia L. and {Lampayan}, Kelly and {Petric}, Andreea and {Dicken}, Daniel and {Greene}, Jenny E. and {Heckman}, Timothy M. and {Hickox}, Ryan C. and {Ho}, Luis C. and {Krolik}, Julian H. and {Nesvadba}, Nicole P.~H. and {Strauss}, Michael A. and {Geach}, James E. and {Oguri}, Masamune and {Strateva}, Iskra V.},
        title = "{Star formation in quasar hosts and the origin of radio emission in radio-quiet quasars}",
      journal = {\mnras},
         year = 2016,
        month = feb,
       volume = {455},
       number = {4},
        pages = {4191-4211},
          doi = {10.1093/mnras/stv2571},
archivePrefix = {arXiv},
       eprint = {1511.00013},
 primaryClass = {astro-ph.GA},
       adsurl = {https://ui.adsabs.harvard.edu/abs/2016MNRAS.455.4191Z}
}

@ARTICLE{Liao2024,
       author = {{Liao}, Mai and {Wang}, Junxian and {Ren}, Wenke and {Zhou}, Minhua},
        title = "{Outflow-related radio emission in radio-quiet quasars}",
      journal = {\mnras},
         year = 2024,
        month = feb,
       volume = {528},
       number = {2},
        pages = {3696-3704},
          doi = {10.1093/mnras/stae126},
archivePrefix = {arXiv},
       eprint = {2401.05212},
 primaryClass = {astro-ph.GA},
       adsurl = {https://ui.adsabs.harvard.edu/abs/2024MNRAS.528.3696L}
}

@ARTICLE{Behar2015,
       author = {{Behar}, Ehud and {Baldi}, Ranieri D. and {Laor}, Ari and {Horesh}, Assaf and {Stevens}, Jamie and {Tzioumis}, Tasso},
        title = "{Discovery of millimetre-wave excess emission in radio-quiet active galactic nuclei}",
      journal = {\mnras},
         year = 2015,
        month = jul,
       volume = {451},
       number = {1},
        pages = {517-526},
          doi = {10.1093/mnras/stv988},
archivePrefix = {arXiv},
       eprint = {1504.01226},
 primaryClass = {astro-ph.GA},
       adsurl = {https://ui.adsabs.harvard.edu/abs/2015MNRAS.451..517B}
}

@ARTICLE{CalistroRivera2024,
       author = {{Calistro Rivera}, G. and {Alexander}, D.~M. and {Harrison}, C.~M. and {Fawcett}, V.~A. and {Best}, P.~N. and {Williams}, W.~L. and {Hardcastle}, M.~J. and {Rosario}, D.~J. and {Smith}, D.~J.~B. and {Arnaudova}, M.~I. and {Escott}, E. and {G{\"u}rkan}, G. and {Kondapally}, R. and {Miley}, G. and {Morabito}, L.~K. and {Petley}, J. and {Prandoni}, I. and {R{\"o}ttgering}, H.~J.~A. and {Yue}, B. -H.},
        title = "{Ubiquitous radio emission in quasars: Predominant AGN origin and a connection to jets, dust, and winds}",
      journal = {\aap},
         year = 2024,
        month = nov,
       volume = {691},
          eid = {A191},
        pages = {A191},
          doi = {10.1051/0004-6361/202348982},
archivePrefix = {arXiv},
       eprint = {2312.10177},
 primaryClass = {astro-ph.GA},
       adsurl = {https://ui.adsabs.harvard.edu/abs/2024A&A...691A.191C}
}

@ARTICLE{Rosario2020,
       author = {{Rosario}, D.~J. and {Fawcett}, V.~A. and {Klindt}, L. and {Alexander}, D.~M. and {Morabito}, L.~K. and {Fotopoulou}, S. and {Lusso}, E. and {Calistro Rivera}, G.},
        title = "{Fundamental differences in the radio properties of red and blue quasars: insight from the LOFAR Two-metre Sky Survey (LoTSS)}",
      journal = {\mnras},
         year = 2020,
        month = may,
       volume = {494},
       number = {3},
        pages = {3061-3079},
          doi = {10.1093/mnras/staa866},
archivePrefix = {arXiv},
       eprint = {2004.01196},
 primaryClass = {astro-ph.GA},
       adsurl = {https://ui.adsabs.harvard.edu/abs/2020MNRAS.494.3061R}
}

@ARTICLE{Magliocchetti2018,
       author = {{Magliocchetti}, M. and {Popesso}, P. and {Brusa}, M. and {Salvato}, M.},
        title = "{A census of radio-selected AGNs on the COSMOS field and of their FIR properties}",
      journal = {\mnras},
         year = 2018,
        month = jan,
       volume = {473},
       number = {2},
        pages = {2493-2505},
          doi = {10.1093/mnras/stx2424},
archivePrefix = {arXiv},
       eprint = {1709.07230},
 primaryClass = {astro-ph.GA},
       adsurl = {https://ui.adsabs.harvard.edu/abs/2018MNRAS.473.2493M}
}

@ARTICLE{Villar-Marti2021,
       author = {{Villar Mart{\'\i}n}, M. and {Emonts}, B.~H.~C. and {Cabrera Lavers}, A. and {Bellocchi}, E. and {Alonso Herrero}, A. and {Humphrey}, A. and {Dall'Agnol de Oliveira}, B. and {Storchi-Bergmann}, T.},
        title = "{Interactions between large-scale radio structures and gas in a sample of optically selected type 2 quasars}",
      journal = {\aap},
         year = 2021,
        month = jun,
       volume = {650},
          eid = {A84},
        pages = {A84},
          doi = {10.1051/0004-6361/202039642},
archivePrefix = {arXiv},
       eprint = {2103.06805},
 primaryClass = {astro-ph.GA},
       adsurl = {https://ui.adsabs.harvard.edu/abs/2021A&A...650A..84V}
}

@ARTICLE{Pierce2020,
       author = {{Pierce}, J.~C.~S. and {Tadhunter}, C.~N. and {Morganti}, R.},
        title = "{The radio properties of high-excitation radio galaxies with intermediate radio powers}",
      journal = {\mnras},
         year = 2020,
        month = may,
       volume = {494},
       number = {2},
        pages = {2053-2067},
          doi = {10.1093/mnras/staa531},
archivePrefix = {arXiv},
       eprint = {2002.07820},
 primaryClass = {astro-ph.GA},
       adsurl = {https://ui.adsabs.harvard.edu/abs/2020MNRAS.494.2053P}
}

@ARTICLE{Fabian2012,
       author = {{Fabian}, A.~C.},
        title = "{Observational Evidence of Active Galactic Nuclei Feedback}",
      journal = {\araa},
         year = 2012,
        month = sep,
       volume = {50},
        pages = {455-489},
          doi = {10.1146/annurev-astro-081811-125521},
archivePrefix = {arXiv},
       eprint = {1204.4114},
 primaryClass = {astro-ph.CO},
       adsurl = {https://ui.adsabs.harvard.edu/abs/2012ARA&A..50..455F}
}

@ARTICLE{Harrison2017,
       author = {{Harrison}, C.~M.},
        title = "{Impact of supermassive black hole growth on star formation}",
      journal = {Nature Astronomy},
         year = 2017,
        month = jul,
       volume = {1},
          eid = {0165},
        pages = {0165},
          doi = {10.1038/s41550-017-0165},
archivePrefix = {arXiv},
       eprint = {1703.06889},
 primaryClass = {astro-ph.GA},
       adsurl = {https://ui.adsabs.harvard.edu/abs/2017NatAs...1E.165H}
}

@ARTICLE{Sabater2019,
       author = {{Sabater}, J. and {Best}, P.~N. and {Hardcastle}, M.~J. and {Shimwell}, T.~W. and {Tasse}, C. and {Williams}, W.~L. and {Br{\"u}ggen}, M. and {Cochrane}, R.~K. and {Croston}, J.~H. and {de Gasperin}, F. and {Duncan}, K.~J. and {G{\"u}rkan}, G. and {Mechev}, A.~P. and {Morabito}, L.~K. and {Prandoni}, I. and {R{\"o}ttgering}, H.~J.~A. and {Smith}, D.~J.~B. and {Harwood}, J.~J. and {Mingo}, B. and {Mooney}, S. and {Saxena}, A.},
        title = "{The LoTSS view of radio AGN in the local Universe. The most massive galaxies are always switched on}",
      journal = {\aap},
         year = 2019,
        month = feb,
       volume = {622},
          eid = {A17},
        pages = {A17},
          doi = {10.1051/0004-6361/201833883},
archivePrefix = {arXiv},
       eprint = {1811.05528},
 primaryClass = {astro-ph.GA},
       adsurl = {https://ui.adsabs.harvard.edu/abs/2019A&A...622A..17S}
}

@ARTICLE{Nims2015,
       author = {{Nims}, Jesse and {Quataert}, Eliot and {Faucher-Gigu{\`e}re}, Claude-Andr{\'e}},
        title = "{Observational signatures of galactic winds powered by active galactic nuclei}",
      journal = {\mnras},
         year = 2015,
        month = mar,
       volume = {447},
       number = {4},
        pages = {3612-3622},
          doi = {10.1093/mnras/stu2648},
archivePrefix = {arXiv},
       eprint = {1408.5141},
 primaryClass = {astro-ph.GA},
       adsurl = {https://ui.adsabs.harvard.edu/abs/2015MNRAS.447.3612N}
}

@ARTICLE{mcmullin2007,
  author  = {McMullin, J.~P. and Waters, B. and Schiebel, D. and Young, W. and Golap, K.},
  title   = {CASA Architecture and Applications},
  journal = {Astronomical Data Analysis Software and Systems XVI ASP Conference Series},
  year    = {2007},
  volume  = {376},
  pages   = {127},
}

@manual{casa_imfit_2022,
  title        = {{IMFIT} -- Image analysis task (image fitting)},
  author       = {{CASA Team}},
  year         = {2022},
  organization = {National Radio Astronomy Observatory (NRAO) / CASA Development Team},
  note         = {CASA Documentation, CASA v6.5},
  url          = {https://casa.nrao.edu/casadocs},
}

@ARTICLE{condon1991,
       author = {{Condon}, J.~J. and {Huang}, Z.-P. and {Yin}, Q.~F. and {Thuan}, T.~X.},
        title = "{Compact Starbursts in Ultraluminous Infrared Galaxies}",
      journal = {\apj},
         year = 1991,
        month = sep,
       volume = {378},
        pages = {65},
          doi = {10.1086/170407},
       adsurl = {https://ui.adsabs.harvard.edu/abs/1991ApJ...378...65C}
}

@ARTICLE{Kewley2000,
       author = {{Kewley}, L.~J. and {Heisler}, C.~A. and {Dopita}, M.~A. and {Sutherland}, R. and {Norris}, R.~P. and {Reynolds}, J. and {Lumsden}, S.},
        title = "{Compact Radio Emission from Warm Infrared Galaxies}",
      journal = {\apj},
         year = 2000,
        month = feb,
       volume = {530},
       number = {2},
        pages = {704-718},
          doi = {10.1086/308397},
       adsurl = {https://ui.adsabs.harvard.edu/abs/2000ApJ...530..704K}
}

@ARTICLE{Morabito2025b,
       author = {{Morabito}, Leah K. and {Kondapally}, R. and {Best}, P.~N. and {Yue}, B.-H. and {de Jong}, J.~M.~G.~H.~J. and {Sweijen}, F. and {Bondi}, Marco and {Schwarz}, Dominik J. and {Smith}, D.~J.~B. and {van Weeren}, R.~J. and {R{\"o}ttgering}, H.~J.~A. and {Shimwell}, T.~W. and {Prandoni}, Isabella},
        title = "{A hidden active galactic nucleus population: the first radio luminosity functions constructed by physical process}",
      journal = {\mnras},
         year = 2025,
        month = jan,
       volume = {536},
       number = {1},
        pages = {L32-L37},
          doi = {10.1093/mnrasl/slae104},
archivePrefix = {arXiv},
       eprint = {2411.05069},
 primaryClass = {astro-ph.GA},
       adsurl = {https://ui.adsabs.harvard.edu/abs/2025MNRAS.536L..32M}
}

@ARTICLE{Vries2004,
       author = {{de Vries}, W.~H. and {Becker}, R.~H. and {White}, R.~L. and {Helfand}, D.~J.},
        title = "{Optical Properties of faint FIRST Variable Radio Sources}",
      journal = {\aj},
         year = 2004,
        month = may,
       volume = {127},
       number = {5},
        pages = {2565-2578},
          doi = {10.1086/383550},
archivePrefix = {arXiv},
       eprint = {astro-ph/0402117},
 primaryClass = {astro-ph},
       adsurl = {https://ui.adsabs.harvard.edu/abs/2004AJ....127.2565D}
}

@ARTICLE{Ofek2011,
       author = {{Ofek}, Eran O. and {Frail}, Dale A.},
        title = "{The Structure Function of Variable 1.4 GHz Radio Sources Based on NVSS and FIRST Observations}",
      journal = {\apj},
         year = 2011,
        month = aug,
       volume = {737},
       number = {1},
          eid = {45},
        pages = {45},
          doi = {10.1088/0004-637X/737/1/45},
archivePrefix = {arXiv},
       eprint = {1105.3479},
 primaryClass = {astro-ph.GA},
       adsurl = {https://ui.adsabs.harvard.edu/abs/2011ApJ...737...45O}
}

@ARTICLE{Mooley2016,
       author = {{Mooley}, K.~P. and {Hallinan}, G. and {Bourke}, S. and {Horesh}, A. and {Myers}, S.~T. and {Frail}, D.~A. and {Kulkarni}, S.~R. and {Levitan}, D.~B. and {Kasliwal}, M.~M. and {Cenko}, S.~B. and {Cao}, Y. and {Bellm}, E. and {Laher}, R.~R.},
        title = "{The Caltech-NRAO Stripe 82 Survey (CNSS). I. The Pilot Radio Transient Survey In 50 deg$^{2}$}",
      journal = {\apj},
         year = 2016,
        month = feb,
       volume = {818},
       number = {2},
          eid = {105},
        pages = {105},
          doi = {10.3847/0004-637X/818/2/105},
archivePrefix = {arXiv},
       eprint = {1601.01693},
 primaryClass = {astro-ph.HE},
       adsurl = {https://ui.adsabs.harvard.edu/abs/2016ApJ...818..105M}
}

@ARTICLE{Perley2013,
       author = {{Perley}, R.~A. and {Butler}, B.~J.},
        title = "{An Accurate Flux Density Scale from 1 to 50 GHz}",
      journal = {\apjs},
         year = 2013,
        month = feb,
       volume = {204},
       number = {2},
          eid = {19},
        pages = {19},
          doi = {10.1088/0067-0049/204/2/19},
archivePrefix = {arXiv},
       eprint = {1211.1300},
 primaryClass = {astro-ph.IM},
       adsurl = {https://ui.adsabs.harvard.edu/abs/2013ApJS..204...19P}
}

@ARTICLE{Radcliffe2019,
       author = {{Radcliffe}, Jack F. and {Beswick}, Robert J. and {Thomson}, A.~P. and {Garrett}, Michael A. and {Barthel}, Peter D. and {Muxlow}, Thomas W.~B.},
        title = "{An insight into the extragalactic transient and variable microJy radio sky across multiple decades}",
      journal = {\mnras},
         year = 2019,
        month = dec,
       volume = {490},
       number = {3},
        pages = {4024-4039},
          doi = {10.1093/mnras/stz2748},
archivePrefix = {arXiv},
       eprint = {1909.12588},
 primaryClass = {astro-ph.GA},
       adsurl = {https://ui.adsabs.harvard.edu/abs/2019MNRAS.490.4024R}
}






\bsp	
\label{lastpage}
\end{document}